\documentclass[aps,prb,nolongbibliography,notitlepage,twocolumn,superscriptaddress]{revtex4-2}

\usepackage{amsmath}
\usepackage{amssymb}
\usepackage{bm}

\usepackage[dvipsnames]{xcolor}
\usepackage{braket}
\usepackage{graphicx}
\usepackage{CJKutf8}
\usepackage{mathrsfs}
\usepackage{empheq}
\usepackage{etoolbox}
\usepackage{color}
\usepackage{epstopdf}
\usepackage{upgreek} 
\usepackage{mathtools}
\usepackage{soul}
\usepackage[colorlinks=true]{hyperref}
\hypersetup{
    colorlinks=true,
    linkcolor=blue,
    citecolor=blue,
    urlcolor=blue
}

\def\T{{\mathcal T}}

\def\bk{{\bm k}}

\def\bq{{\bm q}}

\newcommand{\ov}{\overline}
\DeclarePairedDelimiter\abs{\lvert}{\rvert}%

\newcommand{\ep}{\varepsilon}

\begin{document}

\title{Controlled expansion for correlated electrons with concentrated kinematics}

\author{Pavel A. Nosov}
\affiliation{Department of Physics, Harvard University, Cambridge, MA 02138, USA}

\author{Eslam Khalaf}
\affiliation{Department of Physics, Harvard University, Cambridge, MA 02138, USA}

\author{Patrick Ledwith}
\affiliation{Department of Physics, Massachusetts Institute of Technology, Cambridge, MA 02139, USA}

\date{\today}
\begin{abstract}

We introduce a systematic expansion tailored to systems with strong local interactions and capable of computing response functions, including finite DC transport, analytically. The expansion is controlled by a small parameter ``$s^2$'' that measures the area of the momentum space where kinematics of the theory is concentrated. In real space, this corresponds to single-particle or correlated hopping terms with amplitudes that decay over a length scale $1/s$ and scale in magnitude as $s^2$.
In the limit $s^2\ll 1$ long, self-avoiding tunneling paths dominate over paths revisiting the same site. This enables systematic controlled calculations of various physical quantities. 
We illustrate the method with three applications. (i) A Hubbard model with concentrated dispersion: we analytically obtain spectral broadening which scales as $s^2$ and identify a high-temperature bad metal with 
$T$-linear resistivity coexisting with parametrically long-lived quasiparticles, as well as an intermediate-temperature "thermal FL*" with a small hole pocket that coexists with thermally disordered fluctuating local moments, all within a single controlled framework. (ii) A correlated-hopping model with interesting electron-trion dynamics. (iii) A model of Chern bands with concentrated Berry curvature, motivated by twisted bilayer graphene, which realizes a Mott semimetal where we compute the broadening for the electron and trion spectral functions. At the end, we discuss how our approach paves the way to addressing various challenging questions in strongly correlated systems and outline its various generalizations.
\end{abstract}
\maketitle

\section{Introduction}
Describing electron correlations in two-dimensional systems with narrow, isolated bands is notoriously challenging. This problem has a long history in the study of strongly correlated systems described by variants of the Hubbard model \cite{Arovas2022_Hubbard,QinHubbard}. More recently, it has become highly relevant for understanding quantum phases in 2D heterostructures where flat bands can be realized \cite{CaoIns,CaoSC,Yankowitz2019,Lu2019,Sharpe2019,Serlin2020}. A natural starting point for describing such systems is the strong-coupling limit, where the kinetic energy is quenched and the interactions dominate. In the Hubbard model, this limit yields a highly degenerate manifold of states corresponding to fluctuating local moments at each site. 
Introducing a small hopping amplitude $t$ couples these sites, eventually leading to ordering at sufficiently low temperatures near half-filling. 

This degeneracy makes electron dynamics in the strong coupling regime ($t \ll U$, where $U$ is the local repulsion strength) a difficult problem, especially away from half filling or at elevated temperatures, where fluctuations of moments are qualitatively important.
Although $t/U$ is a small parameter, dynamical observables such as spectral broadening and incoherent transport generally receive contributions from arbitrarily long hopping sequences, so that a direct low-order expansion in $t/U$ is generally insufficient \cite{Metzner1991_strongU,pairault2000strong}. For instance, there are known examples where the spectral broadening depends non-analytically on the hopping, $\Gamma \sim \abs{t}$, making it invisible to any finite-order expansion in $t$ around the single-site limit. Physically, the difficulty can be traced to the proliferation of long self-returning paths: even when the hopping amplitude is small, the number of such processes grows rapidly with perturbative order. 

One of the most widely used approaches to tackle this problem is dynamical mean-field theory (DMFT) \cite{Georges1996_DMFT}. DMFT is formally exact in the limit of infinite dimensions, where one considers a $d$-dimensional lattice with hopping rescaled as $t \mapsto t/\sqrt{d}$ and takes $d \to \infty$. In this limit, the probability of leaving a site remains of order $t^2$, as the rescaling compensates for the increasing number of neighbors, while the probability of returning to the same site is suppressed by a factor of $1/d$.  Consequently, the self-energy becomes strictly local, and the lattice problem can be solved self-consistently, where one step involves the numerical solution of an impurity problem. DMFT has provided major insight into local correlation effects such as the Mott transition, Hubbard band formation, and bad-metal transport \cite{Georges1996_DMFT,Deng2013_resistivity,Ciuchi2023_local_moments,Millis2022,Rammal2025_2-level-systems}. 
However, its relevance to 2D systems is not always clear. In particular, effects associated with spatially nonlocal correlations (e.g. antiferromagnetic fluctuations and unconventional pairing) are often incompatible with the local self-energy inherited from the $d\to \infty$ limit.
Cluster and diagrammatic extensions \cite{Hafermann2012,rohringerDiagrammaticRoutesNonlocal2018} can capture some of this physics but are generally uncontrolled due to the absence of a small parameter. Controlled $1/d$ corrections involve solving successively more complex impurity problems \cite{schillerSystematic1Corrections1995}; it is not clear if even the leading order corrections are tractable for the Hubbard model.

Unbiased numerical methods such as determinantal quantum
Monte Carlo have provided valuable benchmarks, including evidence for $T$-linear temperature dependence of resistivity in the 2D Hubbard model \cite{Huang2019_Hubbard_DQMC}, but do not yield a simple analytical picture. On the other hand, analytical high-temperature
series expansions \cite{HighT_transport} are controlled for $T \gg t$ but cannot access the intermediate regime $T \ll t$, where local Mott physics and itinerant motion coexist. 

\begin{figure}[t!]
    \centering
\includegraphics[width=0.82\linewidth]{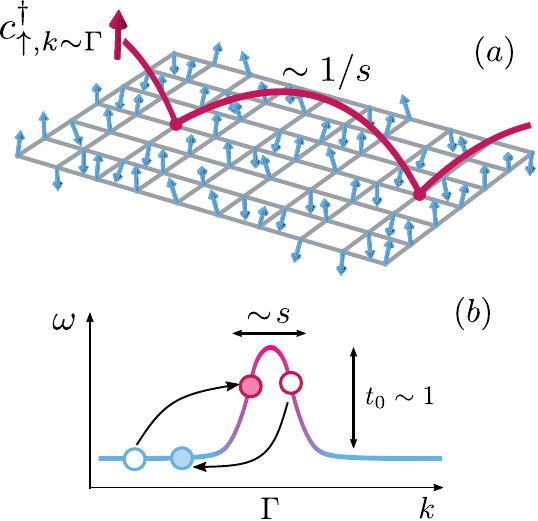}
    \caption{a) Cartoon of the particle with momenta $s\ll 1$ near the $\Gamma$ point propagating on the background of fluctuating local moments.  b) Cartoon of the single-particle dispersion that varies only in the $\sim s$ region near the $\Gamma$ point and has a finite bandwidth $t_0\sim 1$. Arrows show an example of an on-shell scattering process that mixes itinerant degrees of freedom near $\Gamma$ and the "heavy" states in the rest of the Brillouin zone. }
    \label{fig:fig0}
\end{figure}

The recent study of flat bands in moiré platforms has brought a fresh perspective to this problem by introducing new ingredients such as topology and quantum geometry. In some cases, these flat bands resemble Landau levels with approximately uniform quantum geometry. Here, there is no clear separation between charge-ordering (governed by local Hubbard interactions) and spin-ordering (set by exchange interactions between neighboring sites) energy scales. 
However, as noted by two of the present authors \cite{Non-local-moments}, realistic models\cite{Bistritzer2011,carrDerivationWannierOrbitals2019,BernevigTHF,Non-local-moments} of the twisted bilayer graphene flat bands lead to a wide Mott-like regime for temperatures $Us^2 \ll T \ll U$ despite the absence of local moments in the flat band Hilbert space. Here, $s \ll 1$ characterizes the linear size of a small momentum-space region around the $\Gamma$ point where the Berry curvature and quantum geometry is concentrated. When viewed in a local Wannier basis, the concentration of topology in momentum space on the small scale $s$ effectively translates to correlated hopping terms in real space with amplitude $\sim U s^2$ and range $\sim 1/s$\footnote{For Chern bands, topology-enforced Wannier tails lead to $1/r^2$ power-law decay at large distances \cite{Qingchen2024_tails}, producing
logarithmic corrections to certain momentum integrals. Our analysis accommodates such tails.}.

In this work, we introduce a general and systematic strong-coupling expansion inspired by these ideas and apply it to two models: a Hubbard model with concentrated dispersion, and a model of Chern bands with concentrated Berry curvature. The key feature which enables analytic tractability in both cases is concentrated \emph{electron kinematics}; either through the concentration of dispersion or Berry curvature in a small momentum region. The expansion is naturally carried out in real space, where concentrated kinematics corresponds to (correlated) hopping processes with small amplitudes $\sim s^2\ll 1$ but extended range $\sim 1/s$. 
The crucial property of such hopping is that the small amplitude of an individual hopping process is compensated by the large effective coordination volume of available hopping states, so diagrams can be organized systematically in powers of $s$. This structure is illustrated schematically in Fig.~\ref{fig:fig0} for the simplest case of a Hubbard-like model with concentrated dispersion. Panel (b) depicts this dispersion, which is concentrated in a width $\sim s$ neighborhood of the $\Gamma$ point. Fig.~\ref{fig:fig0}(a) shows the corresponding real-space interpretation: spin moments propagate through their self-consistent background through weak, but spatially extended, hopping processes. 
Such dynamics enable electrons of a single species to exhibit characteristics of localized states (large magnetic entropy) and delocalized states (nearly coherent transport) in a self-consistent and analytically tractable way in the fluctuating moment regime, $T \gtrsim t_0 s^2$, where $t_0$ is the bandwidth.

Our method enables controlled, and systematically improvable, analytic calculations of finite DC resistivities in microscopic lattice models with strong local interactions. To our knowledge, such calculations were out of reach until now. As DC resistivity is central to enduring conceptual mysteries like strange metallicity, we anticipate that our method could ultimately provide crucial insights into such puzzles. For concreteness and simplicity, we present leading order DC transport results for the Hubbard model with concentrated dispersion. The DC resistivity of the doped system exhibits a rich hierarchy of parametrically broad regimes, including ``bad metal'' $T$-linear resistivity that persists well below the bandwidth. Remarkably, this $T$-linear scaling coexists with well-defined quasiparticles
with lifetime $\propto 1/s^2$. For lower temperature holes doped relative to half filling (but still in the fluctuating moment regime $T\gtrsim t_0 s^2$), we obtain a Fermi surface with small broadening $\Gamma \sim \abs{t_0}s^2$. This Fermi surface encloses a hole pocket of volume $\delta \ll 1$, where the electron density is $n=1-\delta$, such that it violates the conventional Luttinger's theorem but satisfies that of a (thermal) ``FL$^*$ state.'' Here, doped holes propagate through an environment of one fluctuating moment per site and scatter off them as (annealed) disorder. This yields a small nonzero DC resistivity $\rho \propto s^4/\delta$. We also present transport results for larger hole doping $\delta \sim 1$ and electron doping, assess ordering tendencies through calculating the spin susceptibility, and present single particle spectra for models with concentrated topology.

Although our method was motivated by models with concentrated topology, here we have emphasized its more general role as a controlled deformation of lattice models with local interactions (such as the standard Hubbard model). 
In this sense, it is similar in spirit to large-$N$ approaches: the parameter $s$ provides analytical control while preserving key features of the underlying 2D physics. Large $N$ approaches have provided valuable insight to strongly correlated lattice models, including an analytically controlled theory of a Mott transition\cite{florensSlaverotorMeanfieldTheories2004,zhaoSelfconsistentSlaveRotor2007,podolskyMottTransitionSpinLiquid2009} and insight into low temperature ordered phases\cite{affleckLargeLimitHeisenbergHubbard1988,marstonLargeLimitHubbardHeisenberg1989,arovasFunctionalIntegralTheories1988,readLargeExpansionFrustrated1991,auerbachInteractingElectronsQuantum1994,Arovas2022_Hubbard}. The small $s^2$ expansion can provide another viewpoint on such questions, but it is especially useful for quantities like the DC resistance. Indeed, DC transport in clean systems is elusive in other approximation schemes but tractable through the small-$s$ expansion presented here. We expect that at least part of the structure revealed by this expansion (e.g. aspects of the temperature dependence of the resistivity and some low-temperature ordering tendencies) could remain relevant even for the standard Hubbard model, which corresponds to $s \sim 1$ in our language. More broadly, we see potential applications across the wide class of models and systems that feature strong local correlations. This includes heavy-fermion and mixed-valence systems, recent models\cite{soejima2025jellium,TrithepAHC,tan2025ideallimitrhombohedralgraphene,bernevig2025berry,desrochers2025electronic,han2025exact} for topological electron crystals\cite{HalperinHallCrystal,DongAHC,SenthilAHC,YahuiAHC,guo2024fractional,moirerhomboII,moirerhomboIII}, and of course twisted bilayer graphene where $s^2\ll1$ is furthermore a good approximation quantitatively.

The structure of the paper is as follows.
In Sec.~\ref{sec:CDE_general}, we present a general overview of our method, introducing the class
of models under consideration and the crucial scaling estimates
that underlie the expansion. This section presents the broad idea of our perturbation theory.
In Sec.~\ref{sec:systematics}, we develop the systematic diagrammatic rules and discuss the technicalities of the
concentrated kinematics expansion. This section is more technically involved, and a reader only interested in physical results can skip it. 
In Sec.~\ref{sec:Hubbard}, we summarize our main physical results for the modified Hubbard model with extended and weak single-particle hopping. We focus especially on the variety of transport regimes, including the thermal FL$^*$ phase described above.
In Sec.~\ref{sec:correlated_hop}, we present results for a modified Hubbard model with correlated hopping terms.
In Sec. \ref{sec:concenttratedtopology} we study a model with concentrated topology where, for example, the Berry curvature integrates to $2\pi$ but only has support in a region of momentum space of area $\sim s^2$ motivated by recent work on twisted bilayer graphene\cite{Non-local-moments,ledwithExoticCarriersConcentrated2025,herzogarbeitman2024topologicalheavyfermionprinciple,huProjectedSolvableTopological2025,calugaruObtainingSpectralFunction2025,zhaoTopologicalMottLocalization2025,zhao2025mixedvalencemottinsulator}. 
The nontrivial and concentrated topology leads to a momentum-dependent Mott-like self energy that vanishes at $k=0$\cite{Non-local-moments}. This results in a gapless ``Mott semimetal'' state at charge neutrality and, more broadly, arbitrarily light three-particle ``Dirac trion'' bound states that are exactly orthogonal to the electron at $\Gamma$\cite{ledwithExoticCarriersConcentrated2025,zhaoTopologicalMottLocalization2025,zhao2025mixedvalencemottinsulator}. 
A discussion of future directions is given in Sec.~\ref{sec:Discussion},
and additional computational details are provided in the Appendices.

\section{Small-$s^2$ expansion: Overview}\label{sec:CDE_general}
In this section, we present a general formulation of the small $s^2$ expansion. To this end, we consider a class of fermionic models on a two-dimensional lattice described by the Hamiltonian
\begin{equation}\label{eq:model_general}
    \mathcal{H}=\sum\limits_i \mathcal{H}_{{\rm loc}, i}+s^2\sum\limits_{i\neq j} t(s|\bm{r}_i-\bm{r}_j|) \bm{\gamma}^\dagger_{i}  \bm{M}\bm{\gamma}_j +...
\end{equation}
 The first term, $\mathcal{H}_{{\rm loc}, i}$, captures local on-site interactions (such as the Hubbard $U$, Holstein-type electron-phonon coupling, or interactions with other local degrees of freedom) which we treat non-perturbatively. The second term is non-local and describes fermionic hopping processes. 
 The vector $\bm{\gamma}_i$ consists of local composite operators built from the fundamental fermion operators $\bm{c}_i$. We assume that each component of $\bm{\gamma}_i$ carries nonzero charge under the conserved $U(1)$ symmetry, so that the nonlocal term describes either single-particle or multiparticle hopping. Typical examples of $\gamma$'s include  $c_{i\sigma}$,  $c_{i\uparrow}c_{i\downarrow}$, and  $n_{i\bar{\sigma}}c_{i\sigma}$ (where $\sigma$ is the spin index). The last two cases correspond to pair hopping and correlated hopping, respectively.  The ``flavor" matrix $\bm{M}$ is Hermitian and fixed. The function $t(r)$ determines the hopping profile, $|\bm{r}_i-\bm{r}_j|$ is the distance between $i$ and $j$ sites, and $s$ is a dimensionless parameter whose role will become clear shortly. The ellipsis in Eq.~\eqref{eq:model_general} may include terms that couple three or more sites — for example, terms of the form $s^4 \sum_{ijk} v(s|\bm{r}_i -\bm{r}_j|) v(s|\bm{r}_i - \bm{r}_k|) f_{abc}\tilde{\gamma}'^\dagger_{ia} \tilde{\gamma}_{jb} \tilde{\gamma}_{kc} +h.c.$, etc. We also allow for an arbitrary chemical potential.

  The simplest model that falls into the class described by Eq.~\eqref{eq:model_general} is a modified repulsive Hubbard model on a square lattice with $\bm{\gamma}_i=(c_{i\uparrow},c_{i\downarrow})$ and $\bm{M}=\operatorname{diag}\{1,1\}$ (i.e. just a single-particle hopping term): 
\begin{equation}\label{eq:Hubbard_model}
    \mathcal{H}= U \sum\limits_{i} n_{i\uparrow}n_{i\downarrow}-\mu\sum\limits_{i}n_i+s^2\sum\limits_{i\neq j,\sigma} t(s|\bm{r}_i-\bm{r}_j|)c_{i\sigma}^\dagger c_{j\sigma} .
\end{equation}
 Here $n_i=n_{i\uparrow}+n_{i\downarrow}$ is the total density operator, $n_{i\sigma}=c^\dagger_{i\sigma}c_{i\sigma}$, and $c^\dagger_{i\sigma}$ ($c_{i\sigma}$) are the local fermion creation (annihilation) operators with spin $\sigma=\uparrow/\downarrow$. We will provide specific results for this model in Sec.~\ref{sec:Hubbard}. 

\begin{figure*}[t!]
    \centering
\includegraphics[width=0.95\linewidth]{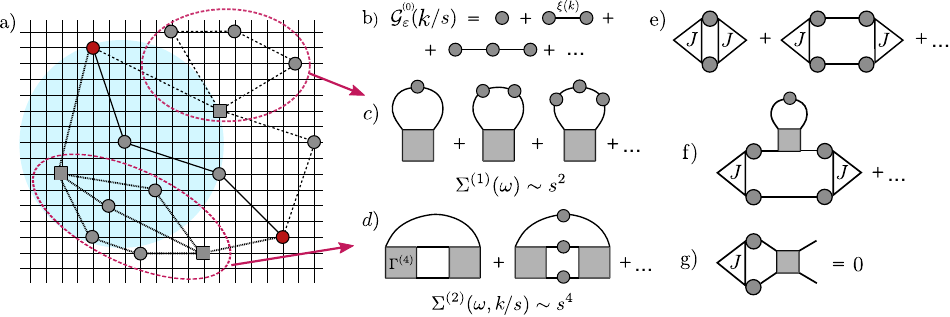}
    \caption{ a) Real-space depiction of various processes contributing to the single-particle Green's function between two distant points (marked as red circles).  The shaded blue region indicates the coordination volume accessible from a fixed site with volume $1/s^2$.  Solid, dashed, and dotted lines represent different paths with zero, one, and two repeated sites, respectively. The sum over paths without repeated site visits is diagrammatically depicted in (b). The detour on the dashed path leads to a local self-energy correction $\Sigma^{(1)}(\omega)\sim s^2$ (shown diagrammatically in (c)). The dotted path involves two repeated site visits and contributes to $\Sigma^{(2)}(\omega,k/s)\sim s^4$ (depicted in (d)), which depends on momentum through the ratio $k/s$. Gray circles denote two-point correlation functions of local operators (i.e. $g_\tau=-\langle \gamma_i(\tau) \bar{\gamma}_i \rangle_0$), and gray squares represent local four-point cumulants (i.e. $\Gamma^{(4)}=-\langle \gamma_i(\tau_1)\gamma_i(\tau_2)  \bar{\gamma}_i(\tau_3) \bar{\gamma}_i \rangle_{0,c}$). e) Resummation of the diagrams contributing to the $\mathcal{O}(1)$ optical conductivity. White triangles represent current vertices. f) Resummation of the self-energy insertions which together with the diagrams in (e) lead to the $\mathcal{O}(1/s^2)$ DC conductivity. g) The lowest-order vertex correction to the current operator vanishes by parity (see discussion in the text).  }  
   
    \label{fig:fig1}
\end{figure*}

A key feature of Eq.~\eqref{eq:model_general} is the presence of a dimensionless parameter $s$. This parameter enters the hopping term such that the hopping amplitude decays over a spatial scale of $1/s$ and is multiplied by an overall factor $s^2$ to ensure finite normalization in the limit $s\rightarrow 0$. 
This spatial dependence is implemented through a suitable choice of the function $t(r)$ which is assumed to have an effective range of order 1; this can be a soft decay, e.g. exponential $t(r) = e^{-r} $, and hard cutoff, e.g. $t(r) = \theta(1-r)$, or even a power law which decays sufficiently quickly as in the case of correlated hopping generated in models of concentrated topology \cite{Non-local-moments, Li2024constraints}. If additional hopping terms coupling $n\geq 3$ sites are included (denoted as ellipsis in Eq.~\eqref{eq:model_general}), they should be accompanied by an overall factor $s^{2(n-1)}$.

The introduction of the parameter $s$ enables a systematic perturbative expansion in powers of $s^2 \ll 1$. At first glance, one might expect this to resemble a conventional weak-hopping expansion, since the hopping amplitudes vanish as $s^2 \to 0$. However, the hopping also becomes increasingly long-ranged in this limit, such that the effective coordination volume for a single hop scales as $1/s^2$. This results in the following crucial estimates:
\begin{equation}
        s^2 \sum_{i}t(s|\bm{r}_i|) = t_0 \sim 1,\quad  s^4 \sum_{i}t^2(s|\bm{r}_i|) \sim s^2 t_0^2\;.\label{eq:count_prob}
\end{equation}
The first quantity $t_0$ denotes the amplitude to leave a site; it is independent of $s$ and corresponds to the bandwidth in momentum space. We will typically use units with $t_0 = 1$ but occasionally restore it for conceptual clarity. The second estimate in \eqref{eq:count_prob} implies that the probability of return is parametrically small. In real space perturbation theory, this delicate balance between suppressed amplitudes and enhanced available phase space leads to a situation where long, self-avoiding spatial loops produce leading contributions, while diagrams with repeated site visits are suppressed by powers of $s^2$ (see Fig.~\ref{fig:fig1}). Since the interaction term $\mathcal{H}_{{\rm loc},i}$ is local, each order in $s^2$ then only involves a finite number of irreducible multi-time correlators (e.g., $\langle \gamma_{\tau_1}\gamma_{\tau_2}  \gamma^\dagger_{\tau_3}\gamma^\dagger_{\tau_4} \rangle_{\rm loc}$, etc.) in the single-site problem defined by $\mathcal{H}_{{\rm loc},i}$. This simplification allows for explicit analytical calculations of all observables (we provide detailed perturbation theory rules in Sec.~\ref{sec:systematics} and Appendix A). 

In momentum space, the limit of weak but extended single-particle hopping (as in Eq.~\eqref{eq:Hubbard_model}) corresponds to a dispersion relation which strongly varies in the $\mathcal{O}(s)$ patch near the $\Gamma$ point and is nearly flat elsewhere in the Brillouin zone (this situation is schematically depicted in Fig.~\ref{fig:fig0}(b)).
More general examples beyond single-particle hopping naturally arise when local density--density interactions are projected onto a narrow Chern band with a nontrivial distribution of Berry curvature. This projection typically leads to the effective Hamiltonian of the form
\begin{equation}
\begin{aligned}
    \mathcal{H} &= \sum\limits_{\bm{q}\sigma}\epsilon(\bm{q}) c_{\bm{q}\sigma}^\dagger c_{\bm{q}\sigma} +\frac{1}{2A}\sum_{\bm q} V_\bq\bigl(\rho_{\bm{q}}-\xi_{\bm{q}}\bigr)\bigl(\rho_{-\bm{q}}-\xi_{-\bm{q}}\bigr),\\
    \rho_{\bm{q}} &= \sum_{\bm{k}\sigma} \,c^\dagger_{\bm{k}+\bm{q}\sigma}\Lambda(\bm{k},\bm{k}+\bm{q})c_{\bm{k}\sigma},\label{eq:H_Lambda_full}
\end{aligned}
\end{equation}
where $V_\bq$ is the Coulomb interaction, $\rho_{\bm{q}}$ is the band-projected density, $\xi_{\bm{q}}$ is the Fourier transform of a background charge distribution, and $\Lambda(\bm{k},\bm{k}')=\braket{u_{\bm{k}}|u_{\bm{k}'}}$ is the projection-induced form factor. In some physically-relevant situations (for instance, in twisted bilayer graphene \cite{Non-local-moments}) the momentum dependence of the form factor is concentrated around $\Gamma$ point: $\Lambda(\bm{k},\bm{k}+\bm{q})$ is nearly $\bm{k}$-independent when $\bm{k}$ and $\bm{k+q}$ are away from $\Gamma$, but varies strongly when one (or both) of these momenta lies within an $\mathcal{O}(s)$ neighborhood of $\Gamma$ (here the emergent small parameter $s$ is naturally tied to the size of the Berry curvature distribution). Moving to the coordinate space, this structure of interactions leads to both local Hubbard $U$ and weak extended correlated hopping terms, thus falling into the general class of models defined in Eq.~\eqref{eq:model_general} (see Sec.~\ref{sec:correlated_hop} for examples). Even in models which do not naturally contain such small a parameter, e.g. Hubbard model with only nearest neighbor hopping, 
it is plausible that computing corrections order-by-order in $s^2$ and setting $s = 1$ at the end may capture relevant physics, in a spirit similar to $1/N$ expansions in field theory.

We also emphasize that despite the asymptotically small size of the momentum region with concentrated dispersion or interaction form-factors, its contribution to physical observables can remain substantial. For example, in the concentrated single-particle dispersion case, the group velocity near the $\Gamma$ point scales as $1/s$, so the contribution of this $\sim s^2$ region to the current-current correlator remains finite. Our goal, therefore, is to develop a systematic expansion for various correlation functions (such as the single-particle Green's function, current-current correlators, and others) organized in powers of $s^2 \ll 1$. In order for this expansion to be meaningful and non-singular, we will also keep the ratio $k/s$ fixed for observables that depend on external momentum $k$ (e.g., the momentum-resolved spectral function, current vertex, etc.). This will allow us to probe long-distance physics associated with correlations on scales larger than $1/s \gg 1$, which makes our expansion distinct from models with all-to-all hopping which lack spatial resolution. 

More broadly, we will encounter the fact that the limit of $s \rightarrow 0$ does not commute with other limits. For instance, the limit $s^2\to 0$ and $T\to 0$ do not commute because the fluctuating moments should eventually freeze at zero temperature. This means that, in general, we will take $s^2$ to be small but not necessarily small compared to $T$. Indeed, we will later compute corrections $\propto s^2/T$, corresponding to exchange interactions $J\sim s^2$, that can be resummed into a Weiss temperature $\propto s^2$ in the magnetic susceptibility. 
For most of our manuscript, we will consider temperatures such that $T/s^2$ is large such that  our starting point is always a finite temperature state (but the ratio between temperature, bandwidth, and Hubbard $U$ could be arbitrary). In the section on spin susceptibility, we show how appropriate resummation can greatly extend the regime of analytic control.
While full perturbative control in all regimes is likely impossible, as it would imply an analytic treatment of critical fluctuations in $d=2$, our expansion provides a useful organizing principle for selection of important diagrams and sets the stage for renormalization group methods to take over.

In order to expose the main elements of the small‑$s$ perturbation theory, we adopt several complementary points of view. First, we use a version of the strong coupling cumulant expansion (originally developed for the expansion around the atomic large-$U$ limit \cite{Metzner1991_strongU}) which provides an efficient derivation of the diagrammatic rules. We also present a second derivation of the diagrammatic rules using the dual fermion formalism \cite{pairault2000strong}, in which the original fermionic degrees of freedom are integrated out and generate local multi‑point connected vertices for a new set of dual fermions. The small $s^2$ expansion then becomes a loop expansion for dual fermion correlators, which in turn map straightforwardly onto correlators of the physical fermions. In the Appendix~\ref{sec:Hubbard_expansion_app}, we also present an alternative formulation based on perturbation theory expressed directly in terms of the original fermionic degrees of freedom, with local interactions decoupled via a Hubbard–Stratonovich field. This construction is better suited for clarifying the combinatorics of the small‑$s$ expansion. Where appropriate, we comment on the correspondence between all approaches.

\section{Small-$s^2$ expansion: systematics}
\label{sec:systematics}

Similar to the conventional strong coupling expansion  \cite{Metzner1991_strongU,pairault2000strong}, it is useful to formulate our expansion around the local problem as a linked-cluster expansion in the non-local hopping. In the present case, however, this construction acquires an additional feature: in the small-$s$ limit, some arbitrarily long hopping sequences remain of order unity, while others are parametrically suppressed. This is the origin of the nontrivial resummation underlying the small-$s$ expansion.

From here until the end of this section we restrict, for notational simplicity, to the two-site term in Eq.~\eqref{eq:model_general}; the extension to terms coupling three or more sites is straightforward. We write the Euclidean action as
\begin{equation}
\begin{aligned}
S[c] &= S_{\rm loc}[c] + S_{\rm hop}[c],
\quad
S_{\rm hop}
=
\int_0^\beta d\tau \sum_{i j}
\bar{\bm{\gamma}}_i\,
\bm{\mathcal{T}}_{ij}\,
\bm{\gamma}_j,\\
&\bm{\mathcal{T}}_{ij} \equiv s^2 t(s|\bm r_i{-}\bm r_j|)\delta_{i\neq j}\bm M,\quad S_{\rm loc} = \sum_i S_{{\rm loc},i}
\end{aligned}
\end{equation}
Here $\bm{\gamma}_i$ is Grassmann-valued field (originating from the corresponding operator in Eq.~\eqref{eq:model_general}), which is either a fundamental fermion $c_i$ or a composite operator built from multiple $c$'s. In the Hubbard case, $a=\{\sigma=\uparrow/\downarrow \}$ labels spin, and $\gamma_{ia}=c_{i\sigma}$. In the more general correlated hopping case, $\gamma_{ia}$ may include composites such as $n_{i\bar{\sigma}} c_{i \sigma}$, etc., in which case $a$ also labels different composites. We
assume that the $\gamma$'s are already normal-ordered.

\subsection{Linked cluster Feynman rules}

We now review the linked-cluster Feynman rules (see Refs. \cite{Metzner1991_strongU,pairault2000strong} for more details) which correspond to a perturbative expansion in $\T$. We will then resum diagrams appropriately to obtain our desired expansion in $s$. We use the Green's function for $\gamma$ operators,
\begin{equation}
\begin{aligned}
\mathcal{G}_{\tau-\tau_0}^{ab}(\bm{r}_i{-}\bm{r}_j)
&=-\frac{1}{Z} \int \mathcal{D}\bar{c}\mathcal{D}c \;\gamma_{ia}(\tau)\bar{\gamma}_{jb}(\tau_0) e^{-S[c]}\\
&=-\langle\gamma_{ia}(\tau)\bar{\gamma}_{jb}(\tau_0) e^{-S_{\rm hop}}\rangle_0/\langle e^{-S_{\rm hop}}\rangle_0,\label{eq:G_interaction}
\end{aligned}
\end{equation}
as a representative example.
Here $Z$ is the full partition function of the theory, and  $\langle ... \rangle_0$ is the  expectation value with respect to the local part of the action $S_{\rm loc}$.

We now expand both factors in Eq.~\eqref{eq:G_interaction} in powers of $t$. The $n$'th-order contribution to the numerator of Eq.~\eqref{eq:G_interaction} is a sum of terms of the form
\begin{equation}
\begin{aligned}
&\frac{(-1)^n}{n!}
\int\limits_{\{\tau_l\}}
\sum_{\{i_\ell,j_\ell\}}
\langle
\bar{\gamma}_{jb}(\tau_0)\gamma_{ia}(\tau) \prod_{\ell=1}^n 
\bar{\gamma}_{i_\ell a_\ell}(\tau_\ell) \mathcal{T}_{i_\ell j_\ell}^{a_\ell b _\ell}\gamma_{j_\ell b_\ell}(\tau_\ell)
\rangle_0.
\end{aligned}
\label{eq:nth_term_raw}
\end{equation}
These terms can be decomposed into sums of products of connected correlators (cumulants). For example
\begin{equation}
\begin{aligned}
    -\langle& \gamma_{a}(\tau)\gamma_{c}(\tau_1) \bar{\gamma}_{b}(\tau_2) \bar{\gamma}_{d}(\tau_0)\rangle_{0} \\
    & = \Gamma^{(4),abcd}_{\tau,\tau_2,\tau_1,\tau_0} +g^{ab}_{\tau,\tau_2}g^{cd}_{\tau_1,\tau_0} - g^{cb}_{\tau_1,\tau_2}g^{ad}_{\tau,\tau_0},\label{eq:Gamma_def_main}
    \end{aligned}
\end{equation}
where the three disconnected contributions\footnote{The expectation value of a single $\gamma$ vanishes, $\langle \gamma_a\rangle_0 = 0$, due to $U(1)$ symmetry. We therefore do not need to include contributions involving one-point functions} are written in terms of the two-point functions
\begin{equation}
    g^{ab}_{\tau,\tau'} = \Gamma^{(2),ab}_{\tau \tau'} =  -\langle \gamma_a(\tau)\bar{\gamma}_b(\tau')\rangle_0\;,
\end{equation}
and the connected part $\Gamma^{(4),abcd}_{\tau,\tau_2,\tau_1,\tau_0}$ is the remainder.
More generally, the $n$'th order connected correlators $\Gamma^{(n)}$, for $n=2,4,6,\ldots$, can be obtained by differentiating the generating functional $\ln \langle e^{-\sum_{a}\int_0^\beta (\overline{\eta}_{a}(\tau) \gamma_{a}(\tau) + \ov{\gamma}_{a}(\tau)\eta_{a}(\tau)}\rangle_0 $ with respect to the source fields $\eta_{a}$~\cite{Metzner1991_strongU}.
Above we have dropped site labels for concision; expectation values with respect to $e^{-S_{\rm loc}}$ factorize trivially across different sites. 

We now discuss the Feynman rules~\cite{Metzner1991_strongU}. We will regard $\Gamma^{(n)}$ as an $n$-valent vertex\footnote{We note that the origin of vertices and lines has switched here. Indeed, typical Feynman rules, which arise from expanding around free theories, are based on Wick's theorem. Then $\Gamma^{(n>2)} = 0$ and one has two-point ``propagators'' $g=\Gamma^{(2)}$ which connect through higher-body interaction vertices at $n$'th order in perturbation theory. Here, averages with respect to $S_{\rm loc}$ generate higher-point connected correlators, whereas $\mathcal{T}$ is two-body. In the dual fermion approach in Sec. \ref{subsec:dualfermion}, however, the $\Gamma^{(n)}$ appear as genuine interaction vertices in the dual fermion action. To avoid confusion, we will thus refer to $\Gamma^{(n)}$ as vertices throughout.}
, drawn as an $n$-sided polygon (we use a circle for $n=2$). The hopping $\mathcal{T}$ is drawn as an arrow that connects two vertices of the polygonal $\Gamma^{(n)}$. At $n$'th order in perturbation theory one draws all diagrams with $n$ lines, and the appropriate operator insertions, that begin and end on vertices of $\Gamma$ polygons (see Fig. \ref{fig:fig1} for various examples to be discussed in detail later). As usual, the expansion of the denominator in \eqref{eq:G_interaction} cancels the disconnected contributions in the numerator such that we can represent $\mathcal{G}$ as the sum of connected diagrams.

\subsection{Small $s^2$ diagrammatics}

We now discuss the scaling of diagrams with $s$ and the associated resummation that enables us to expand in $s^2 \ll 1$ with $t \sim 1$. First, note that each hopping line contributes a factor of $s^2$ due to $\T_{ij} \propto s^2$. Conversely, each internal two point vertex $\bm{g}$ is associated with an additional lattice site to be summed over. The additional summation has a phase space volume $\propto s^{-2}$, which cancels out with the prefactor associated with the hopping line. As a result, arbitrary long strings of hopping lines and two-point vertices, $\prod (\bm{\mathcal{T}} \bm{g})$, contribute at the same order in $s^2$.  This is especially clear upon Fourier transforming to momentum space; each hopping line contributes with the factor $t(k,s)\bm{M}$, which is $O(1)$ near $k=0$. 

Contributions to $\mathcal{G}$ involving higher-point vertices (e.g. $\Gamma^{(4)}$) require the hopping lines to form a loop, and each hopping loop leads to a $s^2$ suppression. Indeed, a higher point vertex in $\mathcal{G}$ requires a particle to revisit a site (and thus form a loop); this constraint to return to a past site removes a spatial sum. Diagrammatically such diagrams necessarily involve one fewer internal vertex (free position summation) for each loop relative to the number of hopping lines. In general, including higher point vertices requires forming more hopping loops, and each hopping loop suppresses the diagram by a factor of $s^2$ relative to the tree level contribution. This counting can also be understood in momentum space. Each hopping loop involves a summation over the momentum $k$ and at least two factors of the dispersion $t(k,s)$. Since the dispersion $t(k,s)$ only has support for $k\sim s$, each loop causes a suppression by this small $\sim s^2$ phase space. 

We pause to comment on the form of the dispersion $t(k,s)$ for $s\ll 1$. Using the Poisson summation formula, we find 
\begin{equation}
\begin{aligned}
   t(k,s)&\equiv s^2\sum_{r_j\neq 0} t(s|r_j|)\, e^{ikr_j}\approx   \xi(k/s)+t(*,s)\;, \\
\xi(k)&=\int_{\bm{r}}t(r)\, e^{ir\cdot k} ,\quad t(*,s)\approx -s^2\int_{\bm{k}}\xi(k)+\mathcal{O}(s^4)\;.
 \end{aligned}  \label{eq:t_approx_small_s_main}
\end{equation}
Here the star $*$ schematically indicates momenta away from the Gamma point, $\int_{\bm{k}}=\int d^2k/(2\pi)^2$ is the unrestricted momentum integral, and $\xi(k)$ is $s-$independent function decaying at large $k$ (we assume that $t(r)$ decays faster than $1/r^2$; for models with topology-enforced $1/r^2$ tails, logarithmic corrections appear). In the scaling limit, $s \to 0$ with $k/s$ fixed, $t(k,s)$ simply becomes $\xi(k/s)$. For instance, the exponentially decaying hopping $t(r)=e^{-r}/2\pi$ leads to $\xi(k)=1/(1+k^2)^{3/2}$. We discuss similar identities for replacing $t(k,s)$ with $\xi(k/s)$ in Appendix \ref{app:small_s_integrals}.

\subsection{Green's function and density}

The above considerations lead to a recipe to compute correlation functions order by order in $s^2$, which we now illustrate explicitly through the Green's function of $\gamma$ operators \eqref{eq:G_interaction}. First, contributions to $\mathcal{G}$ involving arbitrarily long strings $\prod (\mathcal{T} \bm{g})$ should be resummed. 
This geometric sum (depicted diagrammatically in Fig.~\ref{fig:fig1}(b)) yields the zeroth-order Green's function:
\begin{equation}
\mathcal{G}^{(0),ab}_\varepsilon(k)
=\left(\frac{1}{\bm{g}_\varepsilon^{-1}-\bm{\mathcal{T}}(k)}\right)_{ab}
\approx \left(\frac{1}{\bm{g}_\varepsilon^{-1}-\xi(k/s) \bm{M}}\right)_{ab} ,
\label{eq:G0_chain}
\end{equation}
where $\ep$ is the fermionic Matsubara frequency, and we also used Eq.~\eqref{eq:t_approx_small_s_main} which implies $\bm{\mathcal{T}}(k)\approx \xi(k/s)\bm{M}$. Thus the zeroth-order term of the small-$s$ expansion is obtained by resumming all connected path diagrams with no repeated site visits. For the modified Hubbard model this coincides with the Hubbard-I Green's function. Notably, even at the order $s^0$ the Green's function acquires dispersion $\sim t$. 

The full Green's function has the form
\begin{equation}
\bm{\mathcal{G}}_\varepsilon( k)
=
\Big[
 \bm{g}_\varepsilon^{-1}
-\xi(k/s)\bm{M}
-\bm{\Sigma}_\varepsilon(k/s)
\Big]^{-1}.
\label{eq:G_dressed_matrix_cond}
\end{equation}
where the self energy $\bm{\Sigma}_\varepsilon(k/s)$ is the sum of 1PI contributions to $\bm{\mathcal{G}}$, according to the above Feynman rules, and can be expanded as
\begin{equation}
\bm{\Sigma}_\varepsilon(k/s)
=
\sum_{n\ge 1}\bm{\Sigma}_\varepsilon^{(n)}(k/s),
\qquad
\bm{\Sigma}_\varepsilon^{(n)} \sim O(s^{2n}) .\label{eq:Sigma_main_series}
\end{equation}
We note that our  self-energy $\bm{\Sigma}_\ep$ is distinct from the standard self-energy defined relative to the free fermion propagator; the leading order local self energy relative to the free fermion propagator is absorbed into $\bm{g}_{\ep}$, and in general $\bm{\gamma}$ includes composite operators.

The leading self energy correction is depicted in Fig. \ref{fig:fig1}c. It has the form 
\begin{equation}
\begin{aligned}
    \bm{\Sigma}_{\ep}^{(1),ab} &\approx s^{2} T\sum\limits_{\ep'}\int_{\bm{k}} \xi^2(k)\sum_{\{a_{1,2},b_{1,2}\}} [\bm{M}\bm{\mathcal{G}}^{(0)}_{\ep'}(sk)   \bm{M}]_{a_2b_2}
    \\
    &\times(g_{\ep}^{-1})_{aa_1}   \; \Gamma^{(4)a_1 a_2b_2 b_1}_{\ep,\ep';\omega=0} (g_{\ep}^{-1})_{b_1b}\label{eq:Sigma_general_k_k},
    \end{aligned}
\end{equation}
where we used that the hopping loop strings 
$\xi(k)\bm{M}\bm{g}_{\ep'} \cdots  \bm{g}_{\ep'} \xi(k) \bm{M}$ resum to $\xi(k) \bm{M} \bm{\mathcal{G}}^{(0)}_{\ep'}(sk)\bm{M} \xi(k)$. 
The two factors of $g_{\ep}^{-1}$ correspond to the amputation of the external legs associated with self energy diagrams. 

We will see that the nonzero bandwidth $\sim |t_0|$ associated with $ \bm{\mathcal{G}}^{(0)}$ generically leads to a nonzero quasiparticle lifetime $-\mathrm{Im}\, \bm{\Sigma} \propto s^2 |t_0|$. 
This follows from the Fermi's golden rule integral $s^2\int_\bk (\xi(k))^2 \rm{Im}\,\bm{\mathcal{G}}_\ep$, which originates from the elastic part of $ \Gamma^{(4)}$ \footnote{Additional inelastic contributions to the imaginary part can also arise, for example, from thermally activated charge fluctuations encoded in $\Gamma^{(4)}$.}. In this case, the $|t_0|$ dependence arises from a $t_0^2$ prefactor, associated with the minimum number of hops required for the hopping loop, together with density of states factor $\propto |t_0|^{-1}$ from the delta function arising from the imaginary part of $\bm{\mathcal{G}}_\ep$. The latter factor requires the resummation associated with $s^2 \ll 1$ and cannot be captured through perturbation theory in $\abs{t_0}$.
Thus, while the small $t_0$ expansion is generically uncontrolled, and the formally leading-order Hubbard-I approximation generically fails, the reorganization associated with $s^2 \ll 1$ enables us to capture such non-analytic broadening effects systematically.

Further corrections to the self energy can arise from $\Gamma^{(6)}$ hexagons, connecting multiple $\Gamma^{(4)}$ boxes, three-site terms, and beyond. We note that while $\Sigma^{(1)}$ above is momentum independent, the contribution to $\Sigma^{(2)}$ involving two $\Gamma^{(4)}$ boxes acquires momentum dependence. This diagram is depicted in Fig. \ref{fig:fig1}d. We will later see that it is associated with spin ordering and scales as $T^{-1}$ (see Sec. \ref{sec:spin_Hubbard_U_inf}). Such diagrams could be resummed to capture low temperature effects beyond DMFT (where the self energy is momentum-independent). If three-site terms are included, as in band-projected models with concentrated topology, then $\Sigma^{(1)}$ is already momentum dependent (see Sec. \ref{sec:concenttratedtopology}).

So far, we have worked at fixed chemical potential $\mu$. Importantly, the same $\mu$ enters both the non-local Green's function, and the local vertices $\Gamma^{(2n)}$. The physical electron density $n_{\rm phys}$ is then determined in the usual way from the physical 
$c$-fermion Green’s function $\mathcal{G}_\ep(k) =\mathcal{G}^{00}_\ep(k) $ as $n_{\rm phys}=2 T\sum_\ep \sum_{k\in \rm{BZ}} e^{i\ep 0^+}\mathcal{G}_\ep(k)$, where the factor of two comes from spin degeneracy. Using the first line of Eq.~\eqref{eq:approx_1_k-integral}), we find
\begin{equation}
\begin{aligned}
    n_{\rm phys} &= 2s^2T\sum_\ep e^{i\ep 0^+}\int_{\bm{k}} \left[\mathcal{G}_\ep(sk)-\mathcal{G}_\ep(k\rightarrow *) \right] \\&+2T\sum_\ep e^{i\ep 0^+}\mathcal{G}_\ep(k\rightarrow *)\;.\label{eq:n_phys_main}
    \end{aligned}
\end{equation}
This relation can be inverted perturbatively in $s^2$, as discussed in Sec. \ref{sec:Hubbard} and Appendix \ref{sec:appendix_chemical_fixed}. 
Physically, the second term in Eq.~\eqref{eq:n_phys_main} may be viewed as the contribution of the “bath” of heavy states away from the concentrated dispersion region (as implied by the notation $k\rightarrow *$, which denotes $k$ taken to be far from $\Gamma$); in particular, it reduces to the single-site density $n_0(\mu)= 2T\sum_\ep e^{i\ep 0^+ }g_\ep$ in the limit $s\to 0$. By contrast, the first term is associated with the small dispersive region near the concentrated momentum patch and therefore naturally scales as $s^2$. Nevertheless, despite the small fraction of the total density occupying this region, its contribution to transport can remain of order $\mathcal O(1)$, as we show next.

\subsection{Conductivity }\label{sec:conductivity_general}

The counting scheme outlined above allows for straightforward generalization to other observables. For instance, the Matsubara conductivity is given by
\begin{equation}
    \sigma_\omega= \frac{1}{\omega}\Big[\chi_{\omega}^{J_xJ_x}-K_{xx} \Big]\;, 
\end{equation}
where $\chi^{J_xJ_x}_{\omega}=\frac{1}{V} \int_0^\beta e^{i\omega \tau} \langle T J_x(\tau)J_x\rangle$ is the current-current correlator, $V$ is the system volume, and $K_{xx}$ is the diamagnetic term. The Ward identity implies that $ \chi^{J_xJ_x}_{\omega=0}=K_{xx}$.

The current operators are obtained from the fact that the components of $\bm{\gamma}_i$ carry conserved $U(1)$ charge, which allows the nonlocal term to be minimally coupled to an external vector potential through the usual Peierls substitution. For simplicity, we take all components of $\bm{\gamma}_i$ to have the same unit charge (the extension to more general charge assignments is straightforward). This leads to the following expressions
\begin{equation}
\begin{aligned}J_\alpha(\tau)
&=
-i\sum_{ij}(r_i-r_j)_\alpha\,
\bar\gamma_{ia}(\tau)\,\mathcal{T}^{ab}_{ij}\,\gamma_{jb}(\tau),\\
    K_{\alpha\beta}(\tau)
&=
-\sum_{ij}(r_i-r_j)_\alpha(r_i-r_j)_\beta\,
\bar\gamma_{ia}(\tau)\,\mathcal{T}^{ab}_{ij}\,\gamma_{jb}(\tau),
\end{aligned}
\label{eq:J_general}
\end{equation}
and $\alpha=x,y$. In momentum space, the current and diamagnetic vertices are given by derivatives of the hopping kernel,
$\bm{v}_\alpha( k,s)
=
\partial_{k_\alpha}t(k,s)\bm{M}$, and $\bm{w}_{\alpha\beta}(k,s)
=
\partial_{k_\alpha}\partial_{k_\beta}\!t(k,s)\bm{M}$. At the leading order in the small-$s$ expansion, one has to resum all self-avoiding paths, exactly as for the single-particle Green's function (the associated diagrams are depicted in Fig.~\ref{fig:fig1}(e)). This procedure yields
\begin{equation}
\chi^{J_x J_x}_\omega
\approx
T\sum_{\varepsilon}\int_{\bm k}(\xi'(k))^2
{\rm Tr}\Big[
\bm{M}\,
 \bm{\mathcal{G}}^{(0)}_{\varepsilon+\omega}(sk)\,
\bm{M}\,
\bm{\mathcal{G}}^{(0)}_{\varepsilon}( sk),
\Big],\label{eq:Pi0_matrix_general}
\end{equation}
up to $\mathcal{O}(s^2)$ corrections. We used identities derived in Appendix \ref{app:small_s_integrals} to pass to the ``continuum limit'' $t(k,s) \to \xi(k/s)$. Similarly, the leading order diamagnetic contribution is 
\begin{equation}
K^{}_{xx}
\approx
-\int_{\bm k} \left[\xi''(k)+\frac{\xi'(k)}{k}\right]
{\rm Tr}\Big[
\bm{M}
\bm{\mathcal{G}}^{(0)}_{\tau=0^-}(sk)
\Big],
\label{eq:K0_matrix_general}
\end{equation}
where we also assumed that $\xi(k)$ depends only on $|\bm{k}|$. Crucially, these expressions are not suppressed by any small factors of $s$.  This can be understood as
follows: near the concentrated patch around the Gamma point, $|k|\sim s$, the current vertex $v(k,s)\sim 1/s$ (since
$\xi(k/s)$ varies on scale $s$), while the momentum-space volume of the patch is $\sim s^2$, so the loop integral scales
as $s^2 \times 1/s^2 \sim \mathcal{O}(1)$. In this sense, transport is sensitive to the properties of the concentrated dispersion region even when $s$ is arbitrary small.

However, since the $0$'th order Green's functions typically do not include any spectral broadening effects, the real frequency conductivity obtained from Eq.~\eqref{eq:Jx} after analytic continuation remains finite only at finite frequency. In order to study DC conductivity, we need to include the finite quasiparticle lifetimes associated with the imaginary self energy. This is incorporated through self-energy insertions built from the local four-point vertex $\Gamma^{(4)}$ (i.e. repeated-site detours), as depicted in Fig.~\ref{fig:fig1}(f). Resumming such effects amounts to replacing the Green's functions $\bm{\mathcal{G}}^{(0)}$ with Eq.~\eqref{eq:G_dressed_matrix_cond} where the self-energy is given in Eq.~\eqref{eq:Sigma_general_k_k}. The resulting DC conductivity is then simply obtained as
\begin{equation}
\sigma
=-\pi\int d\Omega\,
\partial_\Omega n_F(\Omega) 
\int_{\bm{k}} (\xi'(k))^2 \operatorname{Tr}\left[(\bm{M}\bm{A}_\Omega(sk))^2\right]
\label{eq:kubo_bubble_main}
\end{equation}
where $n_F(\Omega)=1/(1+e^{\beta \Omega})$ is the Fermi-Dirac distribution, $\bm{A}_\Omega(k)$ is the spectral function $\bm{A}_\Omega(k)=- \operatorname{Im} \bm{\mathcal{G}}^R_\Omega(k)/\pi$, and $\bm{\mathcal{G}}^R_\Omega(k)$ is the retarded Green’s function
obtained from Eq.~\eqref{eq:G_dressed_matrix_cond} by analytic continuation  $i\ep\rightarrow \Omega +i0^+ $. It is important to mention that the frequency dependence of the self-energy in Eq.~\eqref{eq:Sigma_general_k_k} is explicit, and thus the analytic continuation to real frequencies is completely straightforward.

Since the life-time induced by $\Sigma^{(1)}$ is $\sim s^2$, then Eq.~\eqref{eq:kubo_bubble_main} naturally yields a conductivity of the order of $1/s^2$. This scaling could also be modified if the density is also $s$-dependent (see Appendix for details). In terms of the spectral function for physical $c$-electrons with fixed spin projection, $A^{(e)}_\Omega$, the total density is fixed using the relation
\begin{equation}
\begin{aligned}
     n_{\rm phys} &= 2s^2 \int d\Omega\;  n_F(\Omega)\int_{\bm{k}} \left[A^{(e)}_\Omega(sk)-A^{(e)}_\Omega(k\rightarrow *) \right] \\
     &+2\int d\Omega  \;n_F(\Omega)\;A^{(e)}_\Omega(k\rightarrow *)\;.\label{eq:n_fixed_A_electron}
     \end{aligned}
\end{equation}
Therefore, Eq.~\eqref{eq:Sigma_general_k_k} and Eq.~\eqref{eq:kubo_bubble_main}, combined with Eq.~\eqref{eq:n_fixed_A_electron}, form our main perturbative results for calculating DC transport. When the physical density by itself is $s$-dependent, then both terms in Eq.~\eqref{eq:n_fixed_A_electron} become important, allowing for a variety of transport regimes, as we illustrate in Sec.~\ref{sec:Hubbard}.

In general, vertex corrections contribute to conductivity at the same order as the self-energy corrections they lead to. Here, however, the lowest-order vertex correction associated with $\Gamma^{(4)}$ vanishes (as does the entire ladder built out of such vertices). The corresponding diagram is shown in Fig.~\ref{fig:fig1}(g). To see this, note that the vertex correction takes the schematic form $\sum_{k'} \Gamma^{(4)}\; \partial_{k_x}t(k',s)\; \mathcal{G}(k')^2$. Since $\Gamma^{(4)}$ is purely local (and thus momentum-independent) while the integrand is odd under $k'_x \to -k'_x$ (assuming inversion symmetry), the integral vanishes by parity. This is analogous to the cancellation of impurity-induced ladder corrections in disordered metals~\cite{Lee1985}. Crucially, this cancellation only holds at the lowest order: already at the order $s^4$ there are vertex corrections that contribute to conductivity.

Furthermore, if nonlocal hopping terms coupling three or more sites are present, then the current operators become more complicated than Eq.~\eqref{eq:J_general}. This does not affect the underlying power counting, and the small-$s$ rules can be derived in the same manner, with expressions such as Eq.~\eqref{eq:Pi0_matrix_general} appropriately generalized to account for multi-particle transport. A notable difference, however, is that in such models the $O(s^2)$ self-energy correction $\Sigma^{(1)}$ is momentum dependent. In that case, vertex corrections generally enter at the same order and should be retained consistently.

Finally, we note that other correlation functions can be obtained in a similar way (see Appendix A and \ref{sec:spin_appendix} for details). For instance, for the spin susceptibility one has to resum a ladder of diagrams consisting of $\Gamma^{(4)}$ vertices.

\subsection{Dual fermion perspective}\label{subsec:dualfermion}
The dual-fermion representation provides an alternative viewpoint in which the small-$s$ counting becomes especially transparent: after rescaling momenta to the concentrated momentum patch, the quadratic dual fermion action remains $\mathcal{O}(1)$, while each local $2n$-point vertex carries a factor $s^{2(n-1)}$.

To demonstrate this, we follow Ref.~\cite{pairault2000strong,RUBTSOV20121320}, and decouple the non‑local terms in Eq.~\eqref{eq:model_general} via a fermionic Hubbard–Stratonovich transformation, based on the Grassmann identity
\begin{equation}
   e^{\bar{\bm{\gamma}} A\bm{\gamma}} = \int \frac{d\bar{f} df}{\det A^{-1} }  \exp \left\{-\bar{\bm{f}} A^{-1}\bm{f} + \bar{\bm{f}} \bm{\gamma} +\bar{\bm{\gamma}} \bm{f} \right\}\;,
\end{equation}
which holds for an arbitrary matrix $A$. We apply this identity for each time point with
$A_{ij}^{\alpha \beta}=-\mathcal{T}_{ij}^{\alpha \beta}= s^2 \delta_{i\neq j}t(s|\bm{r}_i-\bm{r}_j|)M^{ab}$. This procedure yields the action
\begin{equation}
\begin{aligned}
    S[f,c] &= \int_0^\beta d\tau \Big\{ -\sum\limits_{ij a} \bar{f}_{i a}(\mathcal{T}^{-1})^{ab}_{ij}f_{j b}\\
    &-\sum\limits_{ia}(\bar{f}_{ia} \gamma_{ia} +\bar{\gamma}_{ia} f_{ia})\Big\}+  S_{\rm loc}[c]\;,\label{eq:dual_action_1}
    \end{aligned}
\end{equation}
where $ S_{\rm loc}[c]$ is the local part of the action for original $c$-fermions, and $f_{a}$ are the dual fermion fields.

Integrating out the $c$‑fermions via the cumulant expansion, the second line of Eq.~\eqref{eq:dual_action_1} generates an infinite series of spatially local, dynamical interaction vertices for the dual fermions, $S_{\rm int}[f]=\sum_{n=1}^{\infty} S_{2n}[f]$, where $2n$ denotes the number of fields involved. In particular, the quadratic term takes the form
\begin{equation}
    S_2[\zeta]= -\int_0^\beta   d\tau_{1,2} \sum\limits_i\bar{f}_{ia,\tau_1}\langle \gamma_{a,\tau_1} \bar{\gamma}_{b,\tau_2} \rangle  f_{ib,\tau_2}.
\end{equation}

Here all expectation values are evaluated with respect to the single‑site problem. Higher‑order terms involve multi‑point connected correlators of the schematic form $f_{\tau_1} \bar{f}_{\tau_2} f_{\tau_3}  \bar{f}_{\tau_4} \langle \bar{\gamma}_{\tau_1} \gamma_{\tau_2} \bar{\gamma}_{\tau_3}\gamma_{\tau_4}\rangle_c$, etc.

The Green’s function for the original composite $\gamma$ operators can be expressed in terms of the dual propagator.
Differentiating the generating functional with respect to appropriate sources and integrating
by parts
\cite{pairault2000strong}. The result is
\begin{equation}
    \bm{\mathcal{G}}_\ep(k)= -\bm{\mathcal{T}}^{-1}(k)-\bm{\mathcal{T}}^{-1}(k)  \bm{\mathcal{D}}_\ep(k)\bm{\mathcal{T}}^{-1}(k)\;,\label{eq:G_as_D}
\end{equation}
where $\bm{\mathcal{D}}_\ep(k)$ is the propagator for the dual fermions $f$.

After integrating out the original $c$-fermions, we arrive at the following action for dual fermions \cite{pairault2000strong}
\begin{equation}
\begin{aligned}
     &S[f] = \int\limits_{\tau_{1,2}}\sum_{\bm{k}\in \rm{BZ}} \bar{f}_{ka,\tau_1}\left[-\frac{\delta_{\tau_1,\tau_2}}{t(k,s)} M^{-1}_{ab}- \langle \gamma_{a,\tau_1} \bar{\gamma}_{b,\tau_2} \rangle\right]f_{kb,\tau_2}\\
   &+\sum\limits_{n\geq 2}\int\limits_{\{\tau_{j}\}}\hspace{-0.4em}\sum_{\{\bm{k}_j\}\in \rm{BZ}}' \bar{\bm{f}}_{k_1,\tau_1} ...\bar{\bm{f}}_{k_{n},\tau_{n}}  \frac{\bm{\Gamma}^{(2n)}_{\{\tau_j\}}}{(n!)^2}\bm{f}_{k_{n+1},\tau_{n+1}} ...\bm{f}_{k_{2n},\tau_{2n}} \label{eq:S_f_new}
   \end{aligned}
\end{equation}
where $\sum_{\bm{k}\in \rm{BZ}}$ denotes the normalized integration over the full BZ, and $\sum_{\{\bm{k}_j\}\in \rm{BZ}}'$ in the second line assumes total momentum conservation.

So far our manipulations have been exact. We now exploit the presence of the small parameter $s^2$. To this end, we approximate the dispersion in Eq.~\eqref{eq:S_f_new} using Eqs.~\eqref{eq:t_approx_small_s_main}, and perform the following rescaling of fields and momentum variables
\begin{equation}
k\rightarrow s k\;,\quad     s f_{sk,\tau} \rightarrow  \mathsf{f}_{k,\tau}\;,\label{eq:rescaling}
\end{equation}
while also extending all momentum integrals over the BZ to continuous unrestricted integrals.

This rescaling transfers all dependence on $s$ from the kinetic term for $\mathsf{f}$ to the interaction vertices. As a result, the quadratic part of the action for the $\mathsf{f}$ fermions (the first line of Eq.~\eqref{eq:S_f_new}) becomes manifestly $s$-independent, and all higher‑order interaction terms acquire explicit small prefactors such that they can be expanded in perturbatively.
\begin{equation}
\begin{aligned}
    & S[\mathsf{f}] = \int_{\bm{k}} \sum\limits_{\ep}\bar{\bm{\mathsf{f}}}_{k \ep}\left[-\frac{\bm{M}^{-1}}{\xi(k)}+\bm{g}_\ep \right]\bm{\mathsf{f}}_{k\ep},\\
    &+\sum\limits_{n\geq 2}\frac{s^{2(n-1)}}{(n!)^2}\hspace{-0.5em}\int'\limits_{\{\bm{k}_j\,\ep_j\}} \bar{\bm{\mathsf{f}}}_{k_1,\ep_1} ...\bar{\bm{\mathsf{f}}}_{k_{n},\ep_{n}} \bm{\Gamma}^{(2n)}_{\{\ep_j\}} \bm{\mathsf{f}}_{k_{n+1},\ep_{n+1}} ...\bm{\mathsf{f}}_{k_{2n},\ep_{2n}}
    \end{aligned}
\end{equation}
where we also switched to the Matsubara frequency representation. The primed integral should be understood as including a Matsubara sum with frequency conservation and unrestricted momentum integrals with total momentum conservation. 

Physically, the rescaling zooms into the concentrated momentum region, while the dynamical vertices capture correlations at large momenta.

We note that the dual fermion self-energy at one-loop order (consisting of a single $\Gamma^{(4)}$ vertex
with one internal dual propagator loop) reproduces the self-energy $\Sigma^{(1)}$ derived in the cumulant
approach (Eq.~\eqref{eq:Sigma_general_k_k}). This provides a concrete check of the equivalence between the two formulations
and confirms that the $s$-counting is consistent. The general correspondence at higher orders follows
from the standard relation between cumulant and dual fermion diagrams established in Ref.~\cite{pairault2000strong}.

A useful consequence of the dual-fermion representation is that it provides
a natural route to self-consistent approximations \cite{RUBTSOV20121320,Stepanov2016_dual}. One may truncate the dual
interaction to $\Gamma^{(4)}$ and define a Luttinger--Ward functional
$\Phi_{\mathrm{LW}}[\mathcal{D}]$ from skeleton diagrams built with the
dressed dual propagator $\mathcal{D}$. The dual self-energy
$\Sigma^{(\mathcal{D})} = \delta\Phi_{\mathrm{LW}}/\delta\mathcal{D}$ then satisfies
a self-consistent Dyson equation whose simplest realization replaces the bare
propagator in $\Sigma^{(1)}$ (Eq.~\eqref{eq:Sigma_general_k_k}) with the fully dressed one, resumming
nested chains of $\Gamma^{(4)}$ insertions to all orders. Functional
derivatives of $\Phi_{\mathrm{LW}}$ with respect to external fields further
generate the ladder resummation of Fig.~\ref{fig:fig1}(f) in a manifestly consistent way.
At the dual level this construction is $\Phi$-derivable and thus conserving. The translation to
physical response functions via Eq.~\eqref{eq:G_as_D} is controlled in the small-$s$ approximation since the
truncation to $\Gamma^{(4)}$ is parametrically justified (higher vertices
carry extra powers of $s^2$). This self-consistent scheme might become useful when temperature is pushed
below the emergent $\sim s^2$ scales, where individual higher-order
corrections are enhanced relative to their naive scaling. For the present analysis, however, we work in the regime
where the perturbative expansion is well-controlled.

\begin{figure*}[t!]
    \centering
\includegraphics[width=0.99\linewidth]{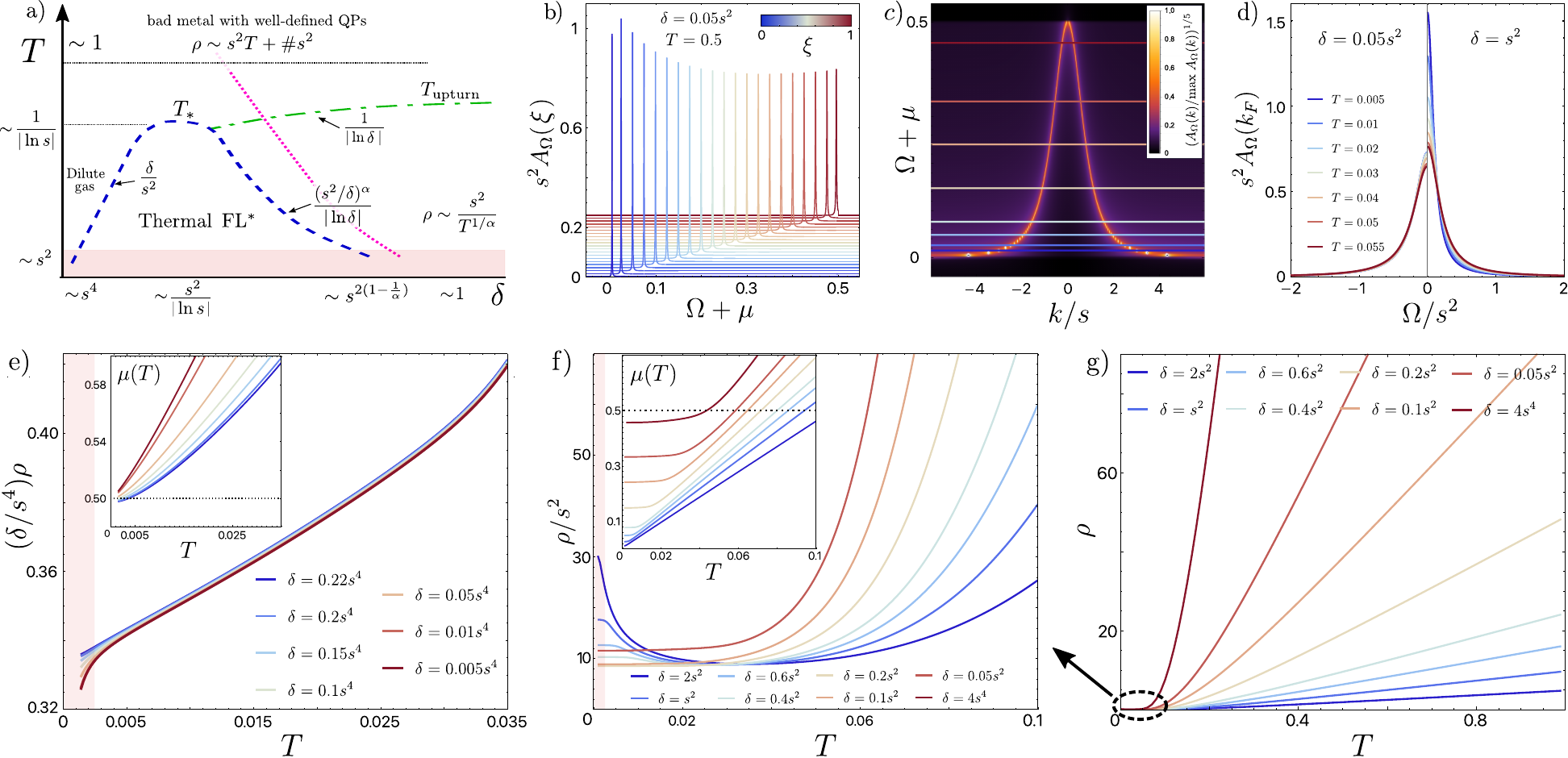}
    \caption{Summary of results for transport in the modified Hubbard model with dispersion $\xi=1/(1+k^2/s^2)^{\alpha}$, at large $U$ and for hole doping $\delta=1-n_{\rm phys}$. a) Crossover diagram for different transport regimes. "QPs" stands for "quasiparticles". FL stands for "Fermi liquid". The region above the pink dashed line is characterized by an approximate relation $|\mu|\sim T$. b) Normalized spectral function for $T=0.5$ and $\delta=0.05s^2$, and for different values of momenta parametrized via the dispersion $\xi(k)$. Each curve is shifted vertically by $0.25\xi$ relative to zero. c) Normalized and rescaled spectral function at $T=s^2$ and $\delta=0.05 s^2$. Horizontal lines depict the position of the chemical potential at this temperature for different densities (represented by the same color as in panels (f) and (g)). d) Normalized quasiparticle peaks at the Fermi momentum $k_F$ in the spectral function for two different doping levels $\delta=0.05s^2$ (left) and $\delta=s^2$ (right) and different temperatures. e) Temperature dependence of the DC resistivity at smallest doping levels $\delta\sim \mathcal{O}(s^4)$. Inset: temperature dependence of the chemical potential for the same range of densities. f) Resistivity in the $\sim s^2$ density regime. g) Same as in (f) but for an extended  temperature range showing high-$T$ asymptotics. For all numerical plots (c-g), parameters $s=0.05$ and $\alpha=3/2$ (exponential hopping) were used. }
    \label{fig:hubbard_hole_mega_fig}
\end{figure*}

\section{Results for the Hubbard model with extended hopping}\label{sec:Hubbard}

In this section, we provide details of how the general formalism discussed in Sec.~\ref{sec:systematics} is applied to a specific case of the modified Hubbard model in Eq.~\eqref{eq:Hubbard_model}.

For concreteness, we will consider hopping profile $t(r)$ such that
\begin{equation}\label{eq:dispersion_3/2}
    \xi(k)\equiv \int_{\bm{r}}t(r)e^{ikr} =\frac{1}{(1+k^2)^{\alpha}}\;,
\end{equation}
where $\alpha$ is a tunable parameter, and the bandwidth $t_0\equiv 1$ is set to unity. The special case $\alpha=3/2$ corresponds to $t(r)= e^{-r}/(2\pi)$ which decays exponentially at long distances, while $\alpha=1$ arises from $t(r)= K_0(r)/(2\pi)$, where $K_0(r)$ is the modified Bessel function. All energy scales will be measured in units of the bandwidth. 

The modified Hubbard case corresponds to a simple choice of operators $\gamma_{ia}\equiv c_{i\sigma}$, and $\bm{M}$ is just an identity matrix in the spin space $\bm{M}=\operatorname{diag}\{1,1\}$. The local Green's function $g_\ep$ acquires a well-known form
\begin{equation}
g^{-1}_{\varepsilon}=i\ep+\mu-\frac{n_0U}{2}-\frac{n_0(2-n_0)U^2/4}{i\ep+\mu+U(n_0/2-1)}\;,\label{eq:G_0_loc_main}
\end{equation}
where $n_0= 2(e^{\mu\beta}{+}e^{(2\mu-U)\beta})/(1{+}2e^{\mu\beta}{+}e^{(2\mu-U)\beta})$ is the averaged single-site density, and we suppressed the spin index due to spin invariance. Correspondingly, the $0$th order Green's function $\mathcal{G}^{(0)}_{\ep'}(k)$ can be obtained from Eq.~\eqref{eq:G0_chain}, and it formally coincides with the result of the Hubbard-I approximation. We emphasize that this result is controlled only at finite temperature exceeding any emergent $\mathcal{O}(s^2)$ scale. The four-point connected vertex $\Gamma^{(4)}$ is also well known (e.g. see \cite{pairault2000strong}), and for certain cases is given in the Appendix \ref{sec:Hubbard_U-loop_full}. 

Using these ingredients we can evaluate the $O(s^2)$ self-energy given in Eq.~\eqref{eq:Sigma_general_k_k}, which in the present case simplifies to \footnote{We assume a paramagnetic state and suppress a spin index for the Green's function and the self-energy.}
\begin{equation}
 \Sigma_{\ep}^{(1)} \approx s^{2}g_\ep^{-2}T\sum\limits_{\ep'} \sum\limits_{\sigma'}\Gamma^{(4) \sigma  \sigma ' \sigma '\sigma }_{\ep,\ep';\omega=0}\int_{\bm{k}} \xi^2(k)
\mathcal{G}^{(0)}_{\ep'}(sk)  \;.\label{eq:Sigma_Hubbard_main}
\end{equation}
This self-energy is depicted diagrammatically in Fig.~\ref{fig:fig1}(c), and the factor $g_\ep^{-2}$ corresponds to the amputation of the external legs in $\Gamma^{(4)}$.
The explicit result is given in the Appendix. However, as discussed in the introduction, our main regime of interest is the one in which charge fluctuations are effectively frozen, while local moments remain fluctuating. This setting is well captured by the $U\to\infty$ limit, and we therefore focus primarily on that case below. Another reason for discussing this limit is its structural simplicity: all resulting expressions are compact and have transparent physical meaning. We discuss separately the hole-doped regime, $n_{\rm phys}<1$, and the electron-doped regime, $1<n_{\rm phys}<2$.

\subsection{Hole-doped regime at large $U$}

In the hole-doped case $n_{\rm phys}<1$, the limit $U \rightarrow \infty$ corresponds to the Gutzwiller-projected Hamiltonian:
\begin{equation}
\begin{aligned}
\label{eq:Uinf_holes_main}
    \mathcal{H}
    =s^2 \sum\limits_{i\neq j,\sigma} t(s|\bm{r}_i{-}\bm{r}_j|) (1{-}n_{i\bar{\sigma}})c_{i\sigma}^\dagger c_{j\sigma }  (1{-}n_{j\bar{\sigma}}){-}\mu \sum\limits_i n_i 
    \end{aligned}
\end{equation}
where $\bar\sigma$ denotes the spin opposite to $\sigma$. We note that Eq.~\eqref{eq:Uinf_holes_main} is itself of the form of Eq.~\eqref{eq:model_general} with $\gamma_{i\sigma}=(1-n_{i\bar{\sigma}}) c_{i\sigma }$ and $\bm{M}=\operatorname{diag}\{1,1\}$, so the small$-s$ expansion applies directly. We have also confirmed that the results we obtain agree with the $U\to \infty$ limit of the corresponding finite $U$ expressions.

Before discussing the detailed calculations, it is instructive to overview the various parametric regimes of density and temperature and the physics expected in each regime, as summarized in Fig.~\ref{fig:hubbard_hole_mega_fig}a. First, we note that as discussed earlier, we are always working in the regime $T_{\rm min} \gtrsim s^2$. Unless stated otherwise, when we say ``low temperature'' we mean $T$ smaller than other scales but still $T>T_{\rm min}$. In this initial summary we will also drop logarithmic factors.

For hole doping $\delta \sim s^2$ and low temperature, below a crossover scale $T_*$, the holes are doped into the dispersive part of the band and coexist with fluctuating local moments. The chemical potential $\mu = \mu_F$ saturates, and the holes form an associated Fermi sea with volume $\delta$. They occasionally scatter off of fluctuating moments, leading to a nonzero resistivity.
We dub this regime ``thermal FL$^*$ for reasons that we will explain later.

The $T\gg T_*$ regime above the thermal FL$^*$ depends on the doping. For $\delta \gtrsim s^2$, the scale $T_*$ corresponds to when the flat portion of the band is activated such that the active carriers are no longer restricted to the vicinity of the Fermi surface. $T_*$ decreases with increasing hole doping $\delta$ as $\mu_F$ moves towards the flat part of the band. 
There is another crossover scale $T_{\rm upturn}>T_*$ associated with activating the dispersive part of the band. 

In transport, $T_{\rm upturn}$ corresponds to an upturn in resistivity; as the temperature is lowered, the fast carriers become less available for transport (Fig.~\ref{fig:hubbard_hole_mega_fig}f). 
Conversely, at very small density $\delta \ll s^2$, the scale $T_*\sim \delta/s^2$ corresponds to the (small) Fermi energy. Here, for $T>T_*$, 
the holes form a dilute (non-degenerate) classical gas moving in a background of fluctuating local moments. Interestingly, this phase has $T$-linear resistivity albeit with a large intercept. The bad metal phase has $T$-linear resistivity and long-lived quasiparticles. It is realized for $T \gtrsim 1$, though numerically it seems to persist to $T \approx 0.1 - 0.2$ (Fig.~\ref{fig:hubbard_hole_mega_fig}g).

We now begin deriving concrete expressions corresponding to the physics described above. The starting point is single-site Green's function,
\begin{equation}
    g_\ep=\frac{1-n_0/2}{i\ep+\mu}\;, \quad n_0=\frac{2e^{\beta \mu}}{1+2e^{\beta \mu}}\;,\label{eq:g_LHB}
\end{equation}
where $n_0$ is the density in a single-site problem. The full non-local Green's function acquires the following form
\begin{equation}
    \mathcal{G}_\ep(k)=\frac{1-n_0/2}{i\ep+\mu -(1{-}n_0/2)t(k,s) -(1-n_0/2)\Sigma^{(1)}_\ep},\label{eq:G_LHB_main}
\end{equation}
where from Eq.~\eqref{eq:t_approx_small_s_main} we have $t(k,s)=\xi(k/s)-s^2\int_{\bm{k}}\xi(k)$, and the self-energy is given by
\begin{equation}
     \begin{aligned}
   \Sigma^{(1)}_\ep &=\frac{s^2}{1-2/n_0}\Big\{ (i\ep+\mu )\Upsilon_1+\Upsilon_2\\
   &+\int_{\bm{k}}\frac{\left(n_0/2-2\right)\xi^2(k)}{i\ep+\mu-(1-n_0/2)\xi(k)}\Big\}\;,
   \label{eq:Upsilon_large_U_1_main}
  \end{aligned}
\end{equation}
where the various terms arise from the $\Gamma^{(4)}$ expression in the \eqref{eq:Gamma_U_inft}
The first two terms contain smooth negative functions $\Upsilon_{1,2}$ that depend only on $\mu$ and $T$, and their explicit form can be found in the Appendix (see Eq.~\eqref{eq:Upsilon_hole_doped_12}). The terms with $\Upsilon_{1,2}$ do not contribute to the imaginary part of the self-energy after the analytic continuation. Their main role is to provide a weak temperature and doping dependent renormalization of the  quasiparticle residue and chemical potential. At fixed $\mu$ and high temperature $\beta\ll 1$, functions $\Upsilon_{1,2}$  decay as $-(\beta^2/12) \int_{\bm{k}}\xi^2(k)$ and $-(\beta^2/2) \int_{\bm{k}}\xi^2(k)$ respectively. On the other hand, at $T\rightarrow 0 $ and $1/2>\mu>0$ they approach finite values $\Upsilon_{1}=-4\int_{\bm{k}}\theta(\xi(k)/2-\mu)$ and $\Upsilon_{2}=-2\int_{\bm{k}}\xi(k)\theta(\xi(k)/2-\mu)$. For $\mu>1/2$ we find $\Upsilon_{1}\approx-2 e^{-(\mu-1/2)\beta}[\alpha+2(1+\alpha)/\beta+...]/(\pi \alpha^2 \beta) +\mathcal{O}(e^{-\mu \beta})$.

The remaining term in Eq.~\eqref{eq:Upsilon_large_U_1_main} formally resembles the elastic self-energy produced by impurities in disordered metals \cite{Lee1985}, and can be associated with scattering off the fluctuating local moments with scattering rate $\sim |t_0|s^2$ (see Fig.~\ref{fig:fig0} for a schematic depiction of this scattering mechanism). The momentum integral can be evaluated in closed form for our choice of dispersion; the result is given in Eq.~\eqref{eq:int_Sigma_full}. After performing analytic continuation and taking the imaginary part, we obtain
\begin{equation}
      \operatorname{Im}\Sigma^R_\Omega =- \frac{\pi s^2 n_0(4-n_0)}{(2-n_0)^2}\Gamma\left(\frac{\Omega+\mu}{1-n_0/2}\right),\label{eq:Sigma_--e}
\end{equation}
where we introduced the broadening profile 
\begin{equation}
    \Gamma(x)=\int_{\bm{k}}\xi^2(k)\delta(x-\xi(k))\;,\label{eq:broadening_prof}
\end{equation}
which for our dispersion takes the form $\Gamma(x)=\frac{1}{4\pi \alpha} x^{1-1/\alpha} \theta(x)\theta(1-x)$. Therefore the on-shell broadening is parametrically of order $s^2n_0$ throughout the lower Hubbard band.  This scaling has a
simple Fermi's golden rule interpretation. The decay rate is obtained
by integrating the density of final dispersive states weighted by the
square of the hopping matrix element. The
dispersive part of the band occupies only an $O(s^2)$ patch near
$k=0$, while the hopping matrix element in this patch is $O(1)$. This
gives the overall $s^2$ factor in Eq.~\eqref{eq:Sigma_--e}. States
outside this patch occupy an $O(1)$ phase volume, but the 
hopping matrix elements  decay at large $k$ and are suppressed there as $O(s^2)$. The frequency dependence in Eq.~\eqref{eq:Sigma_--e} is controlled entirely by the band-edge exponent $1-1/\alpha$. In particular, for $\alpha=3/2$ it vanishes near the flat part of the band as $\sim |\Omega+\mu|^{1/3}$. Similarly, the real part of the self-energy is given by
\begin{equation}
\begin{aligned}
     \operatorname{Re}\Sigma^R_\Omega &= \frac{s^2}{1-2/n_0}\Big\{ \Upsilon_1(\Omega+\mu)+\Upsilon_2 \\
     &- \frac{(4-n_0)}{(2-n_0)} \int dx \;\mathcal{P} \frac{\Gamma(x)}{\frac{\Omega+\mu}{1-n_0/2} - x} \Bigg\} \;.
     \end{aligned}
\end{equation}
We also note that the standard self-energy defined with respect to the free fermion propagator is $\Sigma_\ep +(i\ep+\mu)/(1-2/n_0)$.

The spectral function for the lower Mott band is shown in Fig. \ref{fig:hubbard_hole_mega_fig}c. There is a clear separation of scales between the quasiparticle bandwidth, $\sim t_0$, and the lifetime $\sim \abs{t_0}s^2$. The velocity $v\sim t_0/s$, near $k=0$, is further enhanced. Linecuts shown in Figs. \ref{fig:hubbard_hole_mega_fig}b,d reveal quasiparticle peaks that depend on energy and temperature respectively but always remain parametrically sharp. We therefore see that hole-like quasiparticles can propagate quickly and mostly freely, while occasionally scattering elastically off of thermally fluctuating moments, which behave similarly to random disorder.

We now discuss the relation between the density and the chemical potential based on Eq.~\eqref{eq:n_phys_main}. To this end, we first assume that the chemical potential is fixed and  positive, i.e. $\mu>0$. In this case, the total density can be obtained from Eq.~\eqref{eq:n_phys_main}, which in this particular case becomes
\begin{equation}
   n_{\rm phys}=2T\sum\limits_{\ep}e^{i\ep 0^+} \mathcal{G}_\ep(k{\rightarrow} *) \Big[1+s^2\int_{\bm{k}}\xi(k) \mathcal{G}_\ep(sk)\Big].\label{eq:n_LHB_2}
\end{equation}
In addition, if $\mu$ is not within an asymptotically small $s$-dependent window near the flat part of the band, then we can further expand Eq.~\eqref{eq:n_LHB_2} in $s^2$. As we show below, at fixed density this difficult regime of $\abs{\mu}\propto s^2$ is reached only when $T\sim s^2$, which already lies beyond the regime of control of our expansion (without further resummation). We thus find 
\begin{equation}
    n_{\rm phys} \approx  n_0 +s^2 n_1\;,\label{eq:density_approx_LHB}
\end{equation}
where $n_0$ is given in Eq.~\eqref{eq:g_LHB}, and $n_1$ is a smooth function of $T$ and $\mu$ defined as
\begin{equation}
\begin{aligned}
n_1 &= \frac{n_0}{2}(n_0-2)\,n_F'(-\mu)\,\Upsilon_2 -2 n_F'(-\mu)\int_{\bm p}\xi(p)-\frac{n_0^2}{2}\,\Upsilon_1
 \\
&\quad + \frac{4}{2-n_0}\int_{\bm k}
\left[
n_F\!\left(\frac{2-n_0}{2}\,\xi(k)-\mu\right)-n_F(-\mu)
\right],\end{aligned}
\label{eq:n_for_fixed_mu_LHB_main}
\end{equation}
where $\int_{\bm{k}}\xi(k ) = \frac{1}{4\pi (\alpha-1)}$ for Eq.~\eqref{eq:dispersion_3/2}. The first line here originates from $s^2$-corrections to the ``local'' part of the density in the flat portion of the band, while the second line reflects the contribution from the dispersive region.

When the temperature is approaching zero, the local contribution to the density $n_0 \to 1$, reflecting full occupation of the local moment states that have most of their weight in flat region of the band. The first two terms of Eq.~\eqref{eq:n_for_fixed_mu_LHB_main} vanish at low temperatures. The remaining two terms (associated with the residue renormalization in the flat part of the band and a small density of holes occupying the itinerant states in the dispersive part of the band) contribute as $s^2$ corrections leading to
\begin{equation}
    n_{\rm phys}(T\rightarrow 0) \approx 1-2 s^2 \int_{\bm{k}} \theta(\xi(k)/2 - \mu)\;,\label{eq:nn11}
\end{equation}
such that $s^2 n_1$ reduces to the density associated with preferentially emptying the dispersive part of the band, as opposed to the local moment density $n_0$ which has equal weight at all $\bk$.

On the other hand, in the high temperature limit, $n_1$ goes to zero, while $n_0$ remains finite and approaches $2/3$ at $T\gg 1$. This implies an estimate for the crossover scale $T_{n\; \text{cross}}$ at which the local  density $n_0$ and the density of itinerant holes $|n_1|$ become comparable at fixed $\mu$. This leads to a crossover scale $\sim \mu/|\ln[s^2 \int_{\bm{k}}\theta(\xi(k)/2-\mu)]|\ll \mu$. One can also easily verify that additional finite temperature effects in the second term in Eq.~\eqref{eq:n_for_fixed_mu_LHB_main} contribute a relative correction of the order of $\sim (T/|\mu|)^2$ at low temperatures, which remains much smaller than one at the crossover. Finally, if $\mu$ is within an asymptotically small $\mathcal{O}(s^{2})$ window near the flat part of the band, then the approximate expression in Eq.~\eqref{eq:density_approx_LHB} breaks down and one should use Eq.~\eqref{eq:n_phys_main} instead, combined with appropriate resummations to control the expansion at the associated temperatures $T\sim s^2$.

In general, we can invert Eq.~\eqref{eq:density_approx_LHB} and obtain the chemical potential as a function of the density $n_{\rm phys}$. We can obtain crossover temperatures by comparing the low and high temperature limits of the resulting expression. For our choice of the dispersion, the low-temperature value of the chemical potential becomes $\mu_F(\delta)=1/[2(1+2\pi \delta/s^2)^{\alpha}]$, where $\delta=1-n_{\rm phys}$ is the density of doped holes. At higher temperatures where the flat part of the band is activated we can use the single site relation $\mu = T\ln \frac{1-\delta}{2\delta }$. For $\delta \gtrsim s^2$ the crossover scale is then 
\begin{equation}
    T_* \sim \frac{1}{2(1+2\pi \delta/s^2)^{\alpha}\ln \frac{1-\delta}{2\delta}}\;.
\end{equation}
Below this scale, the chemical potential is already saturated to $\mu_F(\delta)$. From Eq.~\eqref{eq:n_for_fixed_mu_LHB_main}, we see that for $T\ll \mu_F(\delta)$ we effectively have $n_0 = 1$ and $n_1 = - \delta$. 
We will soon discuss the low density limit $\delta \ll s^2$, where the small Fermi energy sets $T_*$ instead.

Let us consider different levels of hole doping. If $\delta\sim s^0$ (i.e. if we dope a $\mathcal{O}(1)$ amount of holes), then $T_*\sim s^{2\alpha}\lesssim s^2$, which is in the low temperature regime where we do not have analytic control. If instead the doped density is small as $\delta\sim s^2$, then $T_*\sim 1/|\ln s|$. More generally, for $\delta\sim s^{\varkappa}$ with $0<\varkappa<2$,
\begin{equation}
    T_*\sim \frac{s^{\alpha(2-\varkappa)}}{|\ln s|}\;,
    \qquad
    \mu(T\ll T_*)\sim s^{\alpha(2-\varkappa)}\;.\label{eq:T_*_scale}
\end{equation}
Thus, in the dilute-hole regime $\delta\sim s^2/|\ln s|$, there exists a parametrically broad temperature window
\begin{equation}
    s^2 \ll T \ll \frac{1}{|\ln s|}\ll 1 \;,
\end{equation}
in which our leading-order expressions remain controlled while the chemical potential has already saturated to $\mu_F(\delta)$. As a result, the compressibility is approximately temperature-independent throughout this regime. A schematic illustration of the corresponding crossover lines is shown in Fig.~\ref{fig:hubbard_hole_mega_fig}(a). The onset of the crossover at $T_*$ in the chemical potential dependence is shown in the inset of Fig.~\ref{fig:hubbard_hole_mega_fig}(f).

We refer to the regime below $T_*$ as a ``thermal Fermi liquid$^*$'' because the holes form a small Fermi surface that violates the naive Luttinger count\cite{luttingerFermiSurfaceSimple1960}. This is possible since we are at finite temperature, $T\gg s^2$; the fate of this regime at lower temperatures cannot be resolved within the present leading-order treatment. One way to intuitively interpret our small Luttinger volume is through comparing and contrasting with zero temperature FL$^*$ states\cite{senthilFractionalizedFermiLiquids2003,senthilWeakMagnetismNonFermi2004,bonettiFractionalizedFermiLiquids2026}. Such states are also characterized by a small Fermi surface of holes doped on top of a spin liquid. The topological order (or spinon Fermi surface volume) of the spin liquid shifts the required Luttinger volume to be that of the small Fermi surface\cite{oshikawaTopologicalApproachLuttingers2000,paramekantiExtendingLuttingersTheorem2004,elseNonFermiLiquidsErsatz2021b}. Here, instead, our holes are doped on top of a thermal state of local moments, which can be interpreted as an infinite temperature state of the spinons.

From Eq.~\eqref{eq:T_*_scale} we also see that the thermal FL$^*$ regime also becomes unresolved (i.e. pushed below the $\sim s^2$ temperature window) at the doping of the order $s^{2(1-1/\alpha)}$ (up to logarithmic factors), which for the exponential hopping case $\alpha=3/2$ gives $\sim s^{2/3}$. As we will show below, the thermal FL$^*$ regime also exhibits interesting transport signatures throughout this low-temperature window.

Finally, if $\delta$ is parametrically smaller than $s^2$ (e.g. $\sim s^4$), then at low temperatures the chemical potential is mostly above the band edge and varies with temperature strongly, i.e. $\mu(T)>1/2$ (see inset of Fig.~\ref{fig:hubbard_hole_mega_fig}(e)), whereas $n_0$ is already saturated to $1$. In this case, we can approximate the momentum integral in Eq.~\eqref{eq:n_for_fixed_mu_LHB_main} as $\int_{\bm k}[...]\approx-e^{-(\mu-1/2)\beta}[\alpha+2(1+\alpha)/\beta+...]/(2\pi \alpha^2 \beta) +\mathcal{O}(e^{-\mu \beta})$. Combining this with our previous estimates for $\Upsilon_1$, we find 
\begin{equation}
    1-n_{\rm phys}\approx \frac{s^2}{\pi \alpha^2 \beta} e^{-(\mu-1/2)\beta}[\alpha+2(1+\alpha)/\beta]+...\label{eq:low_density_mu}
\end{equation}
Inverting this expression at $T\ll 1/2$, we find the chemical potential at fixed density of doped holes $\mu\approx 1/2+ T\ln [s^2 T/\delta]$. In order for our calculation to be self-consistent, we have to require $\mu(T)-1/2\gg T$, and thus $T\gg \delta/s^2$. For even smaller temperatures, the chemical potential eventually saturates to a finite value below the band edge $\mu\approx 1/2- \alpha \pi \delta/s^2$.

\subsubsection{Transport}
We now turn to the DC resistivity. As explained in Sec.~\ref{sec:systematics}, the $O(s^2)$ vertex correction vanishes because the vertex $\Gamma^{(4)}$ is spatially local, whereas the current operator is odd under inversion $k\to -k$ (see Fig.~\ref{fig:fig1}(g)). Therefore, up to $O(s^4)$ corrections, the DC conductivity is obtained from the bubble diagram shown in Fig.~\ref{fig:fig1}(e-f) and written explicitly in Eq.~\eqref{eq:kubo_bubble_main}. At the same level of approximation, the real part of the self-energy can be neglected when extracting the leading behavior of the DC conductivity; its effect only appears at higher order (see the Appendix \ref{app:conductivity_small_s} for details). The numerical results shown in Fig.~\ref{fig:hubbard_hole_mega_fig} were obtained by evaluating Eq.~\eqref{eq:kubo_bubble_main} without the real part of the self-energy, together with the solution of the implicit equation \eqref{eq:density_approx_LHB}, which fixes the chemical potential at a given density.

To gain analytical insight into the transport regimes displayed in Fig.~\ref{fig:hubbard_hole_mega_fig}, we now isolate the leading $1/s^2$ contribution to the conductivity. Since the $O(s^2)$ self-energy is momentum-independent, it is convenient to introduce the velocity-resolved density of states 
\begin{equation}
\Phi(x)\equiv \int_{\bm{k}} (\xi'(k))^2\,\delta(x-\xi(k)),
\label{eq:transport_function_main}
\end{equation}
which for $\xi(k)=1/(1+k^2)^\alpha$ evaluates to $\Phi(x)=\frac{\alpha}{\pi} x (1-x^{1/\alpha})\theta(x)\theta(1-x)$. Thus, using Eq.~\eqref{eq:transport_function_main}, we can re-write the momentum integral in Eq.~\eqref{eq:kubo_bubble_main} as $\int_{\bm{k}}(\xi'(k))^2 f(\xi(k))=\int d\xi\,\Phi(\xi)f(\xi)$. The integral over $\xi$ then can be carried out in the small$-s$ limit as, leading to a simple expression
\begin{equation}
\begin{aligned}
\sigma_{}
&\approx
\frac{ (2-n_0)^3}{16\pi s^2 T n_0(4-n_0)} \int_0^1\frac{dx \; \Phi(x)/\Gamma(x)}{\cosh^2\left(\frac{ (2-n_0)x}{4T} -\frac{\mu}{2T}\right) }
\end{aligned}
\end{equation}
After substituting explicit expressions for $\Phi(x)$ and $\Gamma(x)$, we find
\begin{equation}
\sigma_{}
\approx
\frac{\alpha^2 (2-n_0)^3}{4\pi T s^2 n_0 (4- n_0)}\int_{0}^{1}dx \;\frac{ x^{1/\alpha }(1-x^{1/\alpha}) }{\cosh^2\left(\frac{ (2-n_0)x}{4T}-\frac{\mu}{2T}\right)},
\label{eq:sigma_leading_fixed_mu__23_main}
\end{equation}

As a first application of this formula, we analyze the high-temperature limit when $T\gg 1$. In this case, the correction to the density in Eq.~\eqref{eq:density_approx_LHB} is suppressed, and the chemical potential can be determined from the single-site relation $n_{\rm phys}=n_0$. Thus, the resistivity $\rho=1/\sigma$ can be obtained by expanding the integral in Eq.~\eqref{eq:sigma_leading_fixed_mu__23_main} at large $T$, leading to the following result
\begin{equation}
\begin{aligned}\label{eq:T-linear_high}
    \rho &\approx \frac{\pi s^2T  (1+\alpha)(2+\alpha)  (4-n_{\rm phys})}{2\alpha^3 (2-n_{\rm phys}) (1-n_{\rm phys})} \\
    &-  \frac{\pi s^2  (4-n_{\rm phys}) (3n_{\rm phys}-2) (2+\alpha)^2(1+\alpha)}{8\alpha^3 (2-n_{\rm phys}) (1-n_{\rm phys}) (1+2\alpha)},
    \end{aligned}
\end{equation}
with the subleading terms of the order of $s^2/T$. This high-temperature behavior is demonstrated in Fig.~\ref{fig:hubbard_hole_mega_fig}(g). The $T-$linear resistivity scaling in Eq.~\eqref{eq:T-linear_high} has sometimes been associated to the disappearance of quasiparticles. Here, however, we have quasiparticles with parametrically long lifetimes $\propto s^{-2}$, as can be clearly seen from Eq.~\eqref{eq:Sigma_--e} and  Figs.~\ref{fig:hubbard_hole_mega_fig}(b-d). The temperature dependence instead comes from the compressibility (the chemical potential in this regime scales linearly with temperature, see e.g. Fig.~\ref{fig:hubbard_hole_mega_fig}(f) for large doping levels). This regime, with $T$-linear resistivity and long lived quasiparticles, has been argued to exist quantitatively in DMFT studies\cite{Deng2013_resistivity}; the small $s^2$ parameter here makes such characteristics qualitatively well-defined.

Next, we consider low temperature regimes $T\ll 1$. If the chemical potential is fixed in the range $0<\mu<1/2$, then $n_0$ approaches $1$ at $T\ll \mu$. The remaining integral over $x$ is then dominated by a vicinity of the point $x=2\mu$, and produces a factor $8T  \times (2\mu)^{1/\alpha}(1-(2\mu)^{1/\alpha})$. Thus, we obtain
\begin{equation}
\sigma(T\rightarrow 0)
\approx
\frac{2\alpha^2 }{3\pi  s^2 }  \times (2\mu)^{1/\alpha}(1-(2\mu)^{1/\alpha})\;.
\end{equation}
If instead a small density of holes $\delta \ll 1$ is fixed, then  the chemical potential eventually saturates to a finite value determined from Eq.~\eqref{eq:nn11}, $\mu_F(\delta)=1/[2(1+2\pi \delta /s^2)^{\alpha}]$, and thus we then obtain
\begin{equation}
\sigma(T\rightarrow 0)
\approx
\frac{4\alpha^2 \delta }{3  s^4 \left(1+\frac{2\pi \delta}{s^2}\right)^2}  \;.\label{eq:sigma_saturation}
\end{equation}
For a $\mathcal{O}(s^2)$ doped density this saturation is demonstrated in Fig.~\ref{fig:hubbard_hole_mega_fig}(f). The finite DC conductivity we obtain as $T\to 0$ (but not smaller than $s^2$) originates from hole-like quasiparticles elastically scattering off of fluctuating moments. The fluctuating moments behave like impurities at each lattice site that holes can lose momentum to.

Now let us suppose that the chemical potential exceeds the top of the band, $\mu>1/2$. In this case, at low temperatures $T\ll \mu$ one can replace $n_0\to1$ up to exponential corrections of the order $e^{-\beta \mu}$. As long as $\mu(T)-1/2 \gg T$, the integral in Eq.~\eqref{eq:sigma_leading_fixed_mu__23_main} is dominated by the upper edge. Expanding the numerator of the integrand near $x=1$ we find the activational behavior of the conductivity
\begin{equation}
  \sigma\approx  \frac{4 }{3\pi s^2\beta} e^{-(\mu-1/2)\beta}\left(\alpha +2(\alpha-3)/\beta \right)\;,\label{eq:sigma_activational}
\end{equation}
where corrections involve terms of the order of $e^{-(\mu-1/2)\beta}/\beta^3$ and $e^{-\beta \mu}$. In order to fix the density of holes, we can use Eq.~\eqref{eq:low_density_mu}. Substituting the resulting $\mu(T)$ into Eq.~\eqref{eq:sigma_activational} eliminates the exponential factor and results in $ \sigma\approx  4 \alpha^2 (\delta/s^4) \left(1-8 /(\alpha \beta)+\mathcal{O}(1/\beta^2)\right)/3$. The resistivity then has the following low-temperature expansion
\begin{equation}
  \rho\approx  \frac{3 s^4}{4 \alpha^2 \delta } \left(1+\frac{8 T}{\alpha}+\mathcal{O}(T^2)\right).\label{eq:T-linear_holes}
\end{equation}
We also note that the apparent residual resistivity obtained this way matches the expression in Eq.~\eqref{eq:sigma_saturation}. The $T$-linear behavior in Eq.~\eqref{eq:T-linear_holes} persists above the temperatures of the order of $\delta/s^2$. Thus, for the doping levels $\delta \sim \mathcal{O}(s^4)$ we find $T$-linear resistivity that persists all the way down to the lowest accessible temperatures $T\sim s^2$ (see Fig.~\ref{fig:hubbard_hole_mega_fig}(e)). This regime is best thought of as a dilute classical (non-degenerate) gas of holes moving in a background of well-formed but dynamically disordered local moments, as schematically emphasized in Fig.~\ref{fig:hubbard_hole_mega_fig}(a). However, it should be emphasized that the $T$-linear term here is a small correction relative to the constant term, similar to the Drude weight calculation for a finite number of holes in Ref. \cite{liu2026transporttemperature1exact} and the Yukawa-SYK theory in Ref. \cite{patelUniversalTheoryStrange2023} (though the latter theory also has marginal Fermi liquid physics). The physics here is likely distinct from the strange metallicity in samples where the linear part vastly exceeds the residual resistivity, e.g. \cite{rullier-albenqueUniversalTcDepression2000a}.

Finally, let us consider the situation with $\mathcal{O}(1)$ doped density.
In this case, the relevant part of the transport spectral function in Eq.~\eqref{eq:sigma_leading_fixed_mu__23_main} vanishes as a power-law $x^{1/\alpha}$. After expanding around $x=0$ and performing the integral, we find
\begin{equation}
   \sigma(T\rightarrow 0)
\approx  \frac{ 2^{1+1/\alpha} \alpha \Gamma(1/\alpha) \operatorname{Li}_{1/\alpha}\left(-e^{\mu/T}\right) }{\pi n_{0}(n_{0}-4) (2-n_{0})^{1/\alpha-2} }  \times \frac{T^{1/\alpha}}{s^2 }\;.\label{eq:sigma_low_T}
\end{equation}
where $\operatorname{Li}$ is the poly-logarithm. This result corresponds to the upturn in resistivity at low temperatures. The scale $T_{\rm cross}$ at which the upturn sets it can be estimated by noticing that the integral in Eq.~\eqref{eq:sigma_leading_fixed_mu__23_main} is large when the derivative of the Fermi-Dirac distribution is centered at the maximum of the spectral weight in the numerator. This implies a condition $2-n_0(T_{\rm cross})\sim \mu(T_{\rm cross})$. Assuming a single-site relation for $\mu$, we find that the crossover scale decays as $1/|\ln \delta|$ at small density of doped holes. The appearance of this upturn can be seen in Fig.~\ref{fig:hubbard_hole_mega_fig}(f) at doping levels $\delta \gtrsim s^2$.

\subsubsection{Spin susceptibility}\label{subsec:maintext_spinsusceptibility}

\begin{figure}[t!]
    \centering
\includegraphics[width=0.95\linewidth]{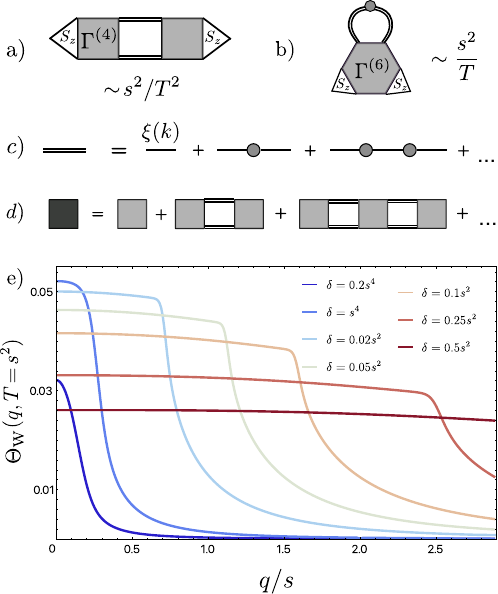}
   \caption{
a,b) Leading perturbative $\mathcal O(s^2)$ corrections to the static spin susceptibility $\chi^{S_zS_z}$. The non-local contribution in panel (a) contains two local $\Gamma^{(4)}$ cumulants placed on different sites and connected by arbitrarily long lattice paths.  The local contribution in panel (b) involves a $\Gamma^{(6)}$ cumulant and a single non-local lattice detour.  Our diagrammatic notation is the same as in Fig.~\ref{fig:fig1}, and the white triangles represent the contraction of the external legs with the spin operator $S_z$. 
c) Hopping amplitude dressed by an infinite series of lattice detours (dual-fermion propagator). 
d) Four-point non-local vertex dressed by the infinite ladder of local $\Gamma^{(4)}$ cumulants. 
e) Momentum dependence of a single rung of the ladder in panel (d), shown for $T=s^2$ and several hole densities between $\delta\sim s^4$ and $\delta\sim s^2$. This rung enters the spin susceptibility through $\chi^{S_zS_z}\propto [\,T-s^2\Theta_{\rm W}(q)\,]^{-1}$. The cusps are analogous to conventional Kohn's $2k_F$ singularities at the diameter of the hole Fermi surface.
}
\label{fig:spin_main}
\end{figure}

Our next goal is to discuss the structure of the static spin susceptibility $\chi^{S_zS_z}_{\omega=0}(q)=\frac{1}{V} \int_0^\beta d\tau e^{i \tau 0^+} \langle T S_z(\tau,q)S_z(-q)\rangle$, where $S_z(q)=\frac{1}{2}\sum_{\bm r\sigma} e^{i\bm{r}\bm{q}} \sigma n_{i\sigma}$ is the electron spin operator. At zeroth order this is simply the Curie law $\chi_0^{S_z S_z} = \beta \langle S_z^2\rangle = \beta n_0/4.$ In addition to assessing the ordering tendencies of the model, we will also carry out a resummation of diagrams that enable us to extend the regime of validity of our expansion to lower temperatures $\sim s^2$.

Let us begin by discussing qualitatively the expected spin ordering in the model. It is helpful to restore finite $U$ for this discussion. At half-filling for large but finite $U$, the relevant physics is captured by the effective spin Hamiltonian, in which the  superexchange interaction between local moments takes the standard form $ \mathcal{H}_{t/U} = \sum_{\bm q} J(\bm q)\,\bm S_{\bm q}\cdot \bm S_{-\bm q}$, where 
\begin{equation}\label{eq:H_spins_half-filling}
    J(\bm q) = \frac{s^4}{U}\sum_{\bm r_i\neq 0} t^2(s|\bm r_i|)\,e^{i\bm q \bm r}\;.
\end{equation}
The preferred low-temperature ordering is determined by the momentum at which $J(\bm q)$ is minimized. Using Eqs.~\eqref{eq:t_approx_small_s_main} and \eqref{eq:dispersion_3/2} for $\alpha=3/2$, one finds $J(\bm q)=\frac{s^2}{8\pi U\left(1+(q/2s)^2\right)^{3/2}}$, which is positive and monotonically decreasing as a function of $q$. This suggests that the dominant magnetic instability occurs at large momenta, $q\gg s$, i.e. near the boundary of the Brillouin zone. Determining the actual ordering vector, however, requires going beyond the continuum approximation of Eqs.~\eqref{eq:t_approx_small_s_main},\eqref{eq:dispersion_3/2} and retaining the sub-leading corrections that encode short-distance details of the dispersion. These corrections generate a negative tail of $J(\bm q)$ at large $q$, of order $s^4/U$. For example, for the hopping $t(r)=(e^{-r}-\delta_{r,0})/2\pi$ on the square lattice, the ordering vector is $\bm q_*=(\pi,\pi)$. The corresponding exchange scale satisfies $J(\bm q_*)\sim -s^4/U$, which implies $  T_{\rm AFM}\sim s^4/U$ in units where the bandwidth is set to unity. This is parametrically small, indicating that antiferromagnetic order at half-filling is very fragile within our expansion.

Upon doping holes relative to half filling, and when $U$ is very large, we anticipate Nagaoka ferromagnetism. In this scenario, doped holes gain kinetic energy of order $\delta$ in a ferromagnetic background, whereas the antiferromagnetic state gains only $\sim s^4/U$. This suggests that a hole density as small as $\delta \sim s^4/U$
is already sufficient to polarize the spins. We indeed find that doped holes contribute to ferromagnetism, though for $\delta \gg s^2$ spirals with wavenumber $q \sim s\sqrt{\delta}$ are also quite competitive. This is likely because spirals with $q \ll k_F$ have limited effect on the Fermi sea relative to $q=0$.

We now begin the calculation, focusing on $U\to \infty$ and the ferromagnetic tendencies induced by doping for concreteness. The computation of $\chi$ is similar in spirit to the current-current correlator, but with an important difference: the spin vertex is local. As a result, both self-energy and vertex corrections contribute on equal footing already at the leading order in $s^2$. We will also discuss diagrammatic contributions $\propto s^2/T$, their resummation, and the regime of validity for the (resummed) expansion.

Fig.~\ref{fig:spin_main}(a) depicts one of the $s^2$ corrections originating from two arbitrarily long hopping sequences connecting spin operators on distant sites. This correction involves a product of two local cumulants with a single spin insertion,  $\sum_\sigma\langle S_z(\tau) c_\sigma(\tau_2')c^\dagger_\sigma(\tau_1')\rangle_{0,c}\langle c^\dagger_\sigma(\tau_2)c_\sigma(\tau_1)S_z\rangle_{0,c}$, and two dressed hopping amplitudes shown in Fig.~\ref{fig:spin_main}(c). In momentum space, these amplitudes contribute the following frequency-dependent loop 
\begin{equation}
   \Pi_{\ep}(q)=\int_{\bm{k}}\frac{\xi(k+q/s)\xi(k)}{(1-g_{\ep}\xi(k+q/s))(1-g_{\ep}\xi(k))}.
\end{equation}
The diagram in Fig.~\ref{fig:spin_main}(a) can then be expressed via the two four-point vertices in the spin channel $\Gamma^{(4,s)}_{\ep,\ep'} \equiv\sum_{\sigma'} \sigma' \Gamma^{(4) \sigma \sigma' \sigma' \sigma }_{\ep,\ep'}$ (its full expression is given in the Appendix) connected by $\Pi_{\ep}(q)$. Crucially, each vertex contains a term $\frac{n_0 \beta \delta_{\omega}}{2(i\ep+\mu)(i\ep'+\mu)}$, which is of the order of $\propto n_0 \beta$. These terms dominate at low temperature $T \ll 1$, and we will soon discuss their role at temperatures $T \propto s^2$.
At such low temperatures, we thus find the leading order correction to be of the form $s^2\beta^2 \Theta_{\rm W}(q)/4$ where $\Theta(q)\equiv -\frac{n_{0} }{2}T\sum_{\ep}\Pi_{\ep}(q)/(i\ep+\mu)^{2}$ evaluates to
\begin{equation}
\begin{aligned}
    \Theta_{\rm W} &=\frac{n_0}{2{-}n_0} \int_{k}
\frac{\xi(k{+}q/s)\,\xi(k)}
{\xi(k{+}q/s)-\xi( k)}\Big[n_F\!\left(\left(1{-}\frac{n_0}{2}\right)\xi(k)-\mu\right)
\\
&-
n_F\!\left(\left(1-\frac{n_0}{2}\right)\xi(k+q/s)-\mu\right)
\Big].\label{eq:Theta_main}
\end{aligned}
\end{equation}
The momentum dependence of this function at low temperature $T\sim s^2$ and different densities is shown in \ref{fig:spin_main}(e). A sub-leading term in this diagram is of the order $\beta$ and is given explicitly in Eq.~\eqref{eq:chi_2_T=0_q}. Its role is to provide a $q$-dependent renormalization of the Curie constant in the single-site susceptibility $\chi^{S_zS_z}_0(T\rightarrow 0)= \beta \langle S_z^2\rangle_0=\beta n_0/4$.

Another $s^2$ correction to the Curie constant originates from the diagram in Fig.~\ref{fig:spin_main}(b). It involves an infinite lattice detour that begins and ends on the same site with two $S_z$ insertions. This vertex involves a cumulant $\sum_\sigma \int_\tau \langle S_z(\tau) \bar{c}_\sigma(\tau_2) c_\sigma(\tau_1) S_z \rangle_{0,c} $ which is then convolved with $\int_k \xi^2(k)\mathcal{G}_{\tau_2-\tau_1}^{(0)}(sk)$. This diagram could also be expressed via the six-point vertex $\Gamma^{(6)}$, and its explicit expression is given in the Appendix (see Eq.~\eqref{eq:Chi_s2_01}). Crucially, 
this diagram contains a $\sim \beta^2 n_0(1-n_0)$ term which at finite $n_0<1$ would be of the same order as the diagram in Fig.~\ref{fig:spin_main}(a). However, when the chemical potential is already saturated to its low-temperature value for $T\ll T_*$, then $n_0\approx 1$, and this term vanishes exponentially. The remaining correction contributes a factor $-s^2\beta \int_k \theta(\xi(k)/2-\mu)$ and is therefore parametrically subleading, for $T\ll 1$, compared to the $\beta^2$ enhancement generated in the diagram (a).

We now discuss low temperature, $T\to \mathcal{O}(s^2)$. At such temperatures, the diagram in Fig. \ref{fig:spin_main}a, and its corresponding ladder series Fig. \ref{fig:spin_main}d, are of the same order as the bare susceptibility $\propto T^{-1}$. Such diagrams can be resummed\footnote{The resummation is straightforward due to the Matsubara-space factorization of the vertex $\Gamma^{(4,s)}_{\ep,\ep'}$ (see SI).} to produce the Curie-Weiss form
\begin{equation}
    \chi^{S_zS_z}(q) \approx 
    \frac{C_0}{T- s^2\Theta_{\rm W}(q)}\;,\label{eq:Chi_zzz}
\end{equation}
where the zeroth order Curie constant $C_0=\frac{1}{4} n_0 = \frac{1}{4}$ at low $T$, . The expression \eqref{eq:Chi_zzz} can be formally regarded as the contribution from diagrams with fixed $\beta s^2 = \mathcal{O}(1)$ but otherwise leading order in $s^2$. 

Physically, however, the expression \eqref{eq:Chi_zzz} must break down when $T = s^2 \Theta_{\rm W}(q)$. Indeed, \eqref{eq:Chi_zzz} corresponds to mean-field-like order onset, which is only valid in dimensions $d \geq 4$. Eq. \eqref{eq:Chi_zzz} can remain valid remarkably close to this point, however. Indeed, temperature scale for the breakdown can be estimated by noting that an $O(s^4)$ correction to the denominator of \eqref{eq:Chi_zzz} is unimportant as long as $\frac{T-s^2 \Theta_{\rm W}}{s^2} \gtrsim s^2$. On the other hand, temperatures $T\lesssim s^2\Theta_{\rm W}(q)$ require us to deal with non-analyticities associated with fluctuating orders in low dimensions. This is typically done through renormalization group analyses of Landau-Ginzburg theories. The role of small $s^2$ could then be to systematically calculate the (bare) terms in such theories.

We pause to comment that the perturbative correction in Fig. \ref{fig:spin_main}b, and a subleading term in Fig \ref{fig:spin_main}a, can be interpreted as corrections to the Curie constant
\begin{equation}
    C=\frac{1}{4}+\frac{s^2}{2}\hspace{-0.3em}\int_{\bm k}\hspace{-0.2em}
\frac{
\xi(k)\,\theta\!\left(\xi(k){-}2\mu\right)
{-}
\xi(k{+}q/s)\,\theta\!\left(\xi(k{+}q/s){-}2\mu\right)
}{
\xi(k+q/s)-\xi(k)
}
\end{equation}
Let us focus on the case when the chemical potential is positive $\mu>0$. For $T\ll T_*$, $n_0\rightarrow 1$ and the chemical potential saturates to a finite value $\mu<1/2$. In this case, the corrected Curie constant is given by
\begin{equation}
    C(q)=\frac{1}{4}+s^2\int_{|\bm k|<k_F}\mathcal{P}
\frac{
\xi(k)
}{
\xi(k+q/s)-\xi(k)
}+\mathcal{O}(s^4)
\end{equation}
where the symbol $\mathcal{P}$ denotes the principal value, and $k_F$ is the $1/s$-rescaled Fermi momentum determined through the relation $\xi(k_F)=2\mu$.

We note the absence of an incipient Kondo effect at this order, which would appear as a logarithmic correction to the Curie constant. This is not a surprise, as the Kondo logarithm first enters the susceptibility at order $J_K^2$, which should be $O(s^4)$ here at least. The associated Kondo temperature would be exponentially small in $s$. We therefore focus on the incipient orders, as for sufficiently small $s$ we expect them to set in before any Kondo-like effects.

We now discuss the ordering tendencies, associated with the $q$-dependence of $\Theta_W(q)$. After replacing the distribution functions in Eq.~\eqref{eq:Theta_main} with their $T=0$ limit, we similarly find $\Theta_{\rm W}(q)\approx 2\int_{|k|>k_F}\mathcal{P}\xi(k{+}q/s)\xi(k)[\xi(k{+}q/s)-\xi(k)]^{-1}$. At $q=0$, this integral reduces to $\Theta_{\rm W}(q{=}0)=\Gamma(2\mu)$, where $\Gamma(x)=\int_{k}\xi^2(k) \delta(x-\xi(k))$ was defined in Eq.~\eqref{eq:broadening_prof}. At finite $q\ll s$, and in case of the dispersion relation $\xi(k)=1/(1+k^2)^{3/2}$, we find 
\begin{equation}
\begin{aligned}
    \int_{|\bm{k}|>k_F}\mathcal{P}\frac{\xi(k)\xi(k{+}q/s)}{\xi(k{+}q/s)-\xi(k)} &\approx \frac{1}{12\pi \sqrt{1+k_F^2}}\\
    &- \frac{(q/s)^2}{144\pi (1+k_F^2)^{5/2}}+... 
    \end{aligned}
\end{equation}

Crucially, $\Theta_{\rm W}$ is maximal at $q=0$, indicating that the leading long-wavelength tendency on the hole-doped side is ferromagnetic with a characteristic temperature scale $T_{\rm CW}\sim  s^2 \Theta_{\rm W}(q=0,T\rightarrow 0)$. At the same time, because the momentum dependence enters through $q/s$, spiral states with $q\sim s$ remain close competitors. The emergence of the scale $T_{\rm CW}\sim s^2$, at which collective spin fluctuations become important, also provides the natural low-temperature cutoff for the leading-order perturbative results for DC transport discussed in the previous subsection. Through resumming the spin ladder contributions to the single particle self energy and current vertex, transport results could be extended to temperatures down to an $\mathcal{O}(s^4)$ fluctuation window above $s^2\Theta_W(0)$; we leave this extension to future work.

Another important point is that $\Theta_{\rm W}(q)$ does not change  sign as $\delta\to0$, so the ferromagnetic tendency survives even in the dilute limit. This is naturally consistent with a Nagaoka-like picture in which a small density of holes gains kinetic energy in a more uniformly polarized spin background. In particular, in the single-hole limit, the Hamiltonian in Eq.~\eqref{eq:Uinf_holes_main} satisfies the assumptions of the Nagaoka theorem. However, the temperature scale $T_{\rm CW}$ is further suppressed for very small densities of holes $\delta\ll s^2$ such that $T_{\rm CW}\lesssim T_*\sim \delta/s^2$. In this case, we enter the dilute gas regime (see Fig.~\ref{fig:hubbard_hole_mega_fig}(a)) where the Fermi surface is not formed yet, and the chemical potential is above the band edge. In this regime $\Theta_{\rm W}\sim e^{-(\mu(T)-1/2)\beta}$. Using Eq.~\eqref{eq:low_density_mu}, $\mu\approx 1/2+ T\ln [s^2 T/\delta]$ and thus $\Theta_{\rm W}\sim \delta /(s^2T)$. This indicates that the susceptibility only becomes large at the temperature scale $T_{\rm CW}\sim \sqrt{\delta}\ll s^2$, which is parametrically reduced compared to the $s^2$ estimate.

\subsection{Electron-doped case at large $U$}

\begin{figure*}[t!]
    \centering
\includegraphics[width=0.99\linewidth]{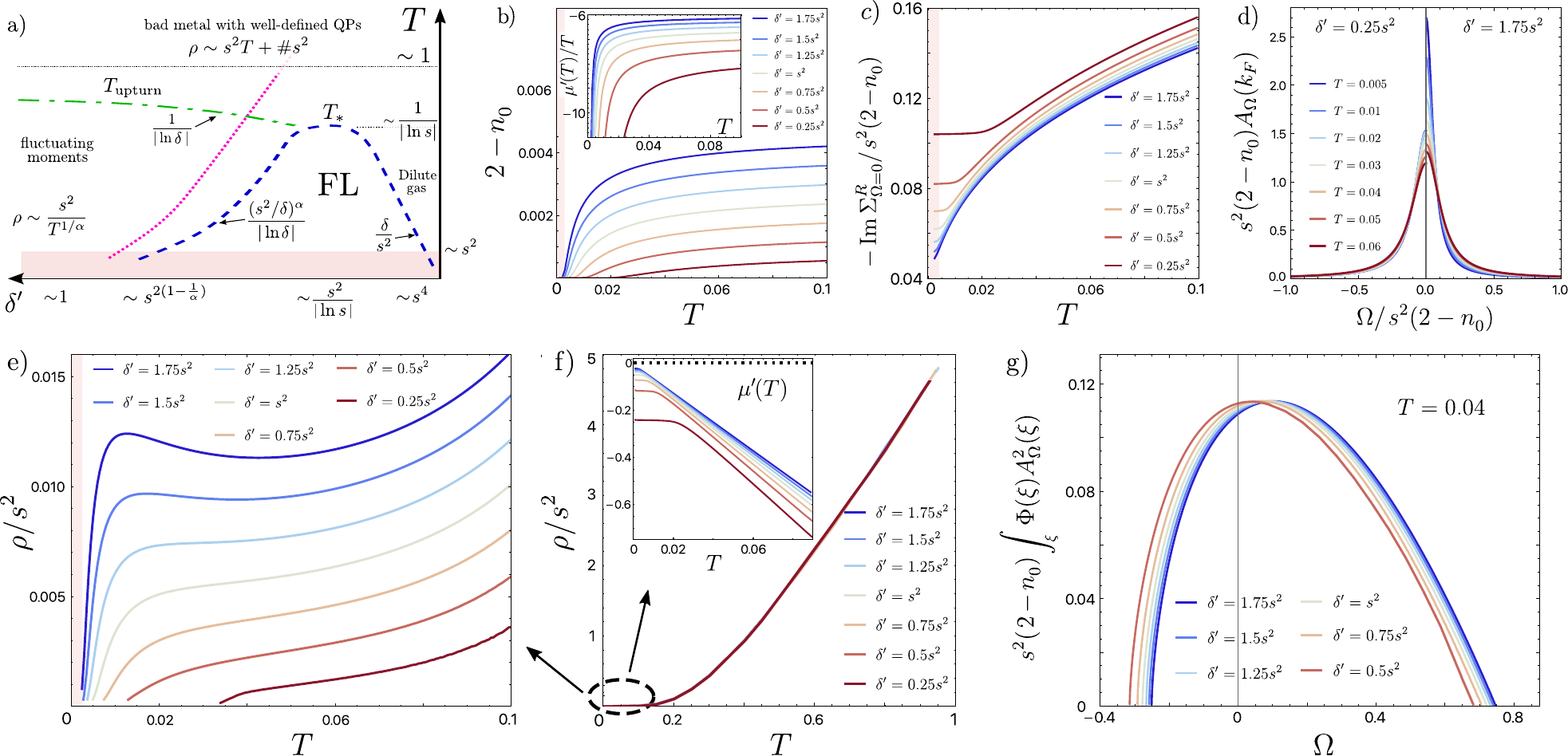}
    \caption{
    Transport on the electron-doped side of the modified Hubbard model with concentrated dispersion $\xi(k)=1/(1+k^2/s^2)^\alpha$ in the large-$U$ limit, with filling parameterized as $n_{\mathrm{phys}}=2-\delta'$. 
a) Schematic crossover diagram showing the transport regimes as a function of temperature and electron doping away from full filling $n_{\mathrm{phys}}=2$. FL stands for Fermi liquid. The region above the pink dashed line is characterized by an approximate relation $|\mu'|\sim T$.
b) Density of thermally active local moments as a function of temperature for several values of $\delta'$. 
c) Imaginary part of the retarded self-energy at $\Omega=0$, normalized by the active-moment density. 
d) Quasiparticle peaks in the spectral function at the Fermi momentum $k_F$ defined as via $\operatorname{Re}\mathcal{G}^{-1}_{\Omega=0}(k_F)=0$ for $\delta'=0.25\,s^2$ (left) and $\delta'=s^2$ (right), for several temperatures. 
e): DC resistivity $\rho(T)$ for $\delta' \sim O(s^2)$: depending on filling, the resistivity either rises monotonically or develops a shallow low-$T$ minimum before crossing over to the high-temperature bad-metal regime shown in (f); the inset in (f) shows the corresponding chemical potential $\mu(T)$. 
(g) Normalized velocity-weighted spectral density $\int_\xi \Phi(\xi)A^2_\Omega(\xi)^2$ at fixed temperature. In the Kubo formula, $\sigma$ is obtained by integrating this quantity against $-\partial_\Omega n_F(\Omega)$. 
In all numerical panels, $s=0.05$ and $\alpha=3/2$ (exponentially decaying hopping).
    }
    \label{fig:hubbard_electron_mega_fig}
\end{figure*}

The results on the electron side are similar in several regards to the hole side (see Fig.~\ref{fig:hubbard_electron_mega_fig}). On the hole side, we focused on doping the dispersive part of the band as all transport regimes could be understood from this perspective. This corresponds to doping holes relative to the Mott insulator of fluctuating moments. On the electron side, the dispersive part of the band is close to full filling, such that we should consider doping holes relative to full filling. 

The main difference associated with doping the band insulator, as opposed to the Mott insulator, is the lack of fluctuating moments. While activation of the flat part of the upper Mott band creates fluctuating moments, at temperatures below this activation scale the the resistivity vanishes due to the lack of momentum-relaxation. Relatedly, the Fermi surface state is an ordinary FL rather than a thermal FL$^*$. Otherwise, identification of regimes and crossover scales is very similar to that of the hole side.

The analysis on the electron-doped side mostly parallels the hole-doped side. To this end, we shift the chemical potential as 
$\mu=U-\mu'$, where $\mu'$ is fixed, and take the large $U$ limit.
We note that $\mu'<0$ corresponds to doping away from the flat part of the band. The Green's function in this regime can be easily obtained from the results of the previous subsection using the exact relation
\begin{equation}
     \mathcal{G}_{\ep}^{}(k) = -   \mathcal{G}_{-\ep}^{(\rm hole)}(k)|_{\substack{\xi(k)\rightarrow -\xi(k)\\\mu\rightarrow \mu'}}\;,\label{eq:G_h_e_corr}
\end{equation}
where $\mathcal{G}_{\ep}^{(\rm hole)}$ is given in Eq.~\eqref{eq:G_LHB_main}.
This correspondence can be easily understood from the standard particle-hole transformation performed on the projected Hamiltonian $\mathcal{H}'=\lim\limits_{U\rightarrow \infty}\mathcal{H}_{\mu=U-\mu'}$ given by
\begin{equation}
    \mathcal{H}'=\mu'\sum\limits_{i}n_i +s^2\sum\limits_{i\neq j\sigma} t(s|\bm{r}_i-\bm{r}_j|)n_{i\bar{\sigma}}c_{i\sigma}^\dagger c_{j\sigma} n_{j\bar{\sigma}}\;.
\end{equation}
Specifically, this transformation maps $\mathcal{H}'$ to the Hamiltonian of holes in Eq.~\eqref{eq:Uinf_holes_main} with the chemical potential $\mu'$ and the opposite sign of hopping. For the Green's function this transformation leads to $\mathcal{G}_{\tau}(k)  = -\langle T  \gamma_{\tau,k\sigma} \gamma^\dagger_{k\sigma} \rangle_{\mathcal{H}'} = \langle T \tilde{\gamma}_{-\tau,k\sigma} \tilde{\gamma}^\dagger_{k\sigma} \rangle_{\mathcal{H}|_{\xi(k)\rightarrow -\xi(k)}} =- \mathcal{G}_{-\tau}^{(h)}(k)$
where $\gamma_{i\sigma}=n_{i\bar{\sigma}}c_{i\sigma}$, and $\tilde{\gamma}_{i\sigma}=(1-n_{i\bar{\sigma}})c_{i\sigma}$. This relation in turn implies Eq.~\eqref{eq:G_h_e_corr}.

This results in the following Green's function 
\begin{equation}
    \mathcal{G}_\ep(k)=\frac{n_0/2}{i\ep-\mu' -(n_0/2)t(k,s) -(n_0/2)\Sigma'^{(1)}_{\ep}},\label{eq:G_UHB_main}
\end{equation}
where $n_0= 2- \frac{2e^{\beta \mu'}}{1+2e^{\beta \mu'}}$ is the density in the single-site problem, and the self-energy reads as
\begin{equation}
     \begin{aligned}
   \Sigma'^{(1)}_\ep &=\frac{s^2(2-n_0)}{n_0}\Big\{ -(i\ep-\mu' )\Upsilon_1'+\Upsilon_2'\\
&+\int_{\bm{k}}\frac{\left(1+n_0/2\right)\xi^2(k)}{i\ep-\mu'-(n_0/2)\xi(k)}\Big\}\;.
   \label{eq:Upsilon_electron_large_U_1_main}
  \end{aligned}
\end{equation}
Here we also defined $\Upsilon_{1,2}'\equiv \Upsilon_{1,2}|_{\substack{\xi\rightarrow -\xi\\\mu\rightarrow \mu'}}$, where $\Upsilon_{1,2}$ are given in Eq.~\eqref{eq:Upsilon_hole_doped_12}.

The physical density can be obtained by noticing that $    \mathcal{G}_{\tau=0^-} = \langle   \gamma^\dagger_{i\sigma} \gamma_{i\sigma}\rangle_{\mathcal{H}'}=\langle  n_{i\uparrow}  n_{i\downarrow}\rangle_{\mathcal{H}'}$ which is the density of doublons. Thus, the physical density is $n_{\rm phys}= 1+ \sum_{\bm{k}}\mathcal{G}_{\tau=0^-}(k)$, which leads to an equation similar to Eq.~\eqref{eq:n_LHB_2} with an overall factor of two removed and a background offset $1$ added. As before, the physical density can be further expanded up to the $s^2$ order as $n_{\rm phys}=n_0+s^2 n_1$, where $n_1$ is given by
\begin{equation}
\begin{aligned}
&n_1 = \frac{n_0(2{-}n_0)}{4}\,n_F'(\mu')\,\Upsilon_2'-(1{-}\frac{n_0}{2})(n_0{-}1)\Upsilon_1'
 \\
&\quad - n_F'(\mu')\int_{\bm p}\xi(p)  + \frac{2}{n_0'}\int_{\bm k}
\left[
n_F\!\left(\frac{n_0}{2}\xi(k){+}\mu'\right)-n_F(\mu')
\right]\end{aligned}
\label{eq:n_for_fixed_mu_UHB_main}
\end{equation}
At low temperatures and fixed $\mu'<0$, the single-site contribution approaches
$n_0\to 2$, while the first three terms in Eq.~\eqref{eq:n_for_fixed_mu_UHB_main}
vanish. The remaining momentum integral then gives
\begin{equation}
    n_{\rm phys}(T\to 0)\approx 2-2s^2\int_{\bm k}\theta\!\left(\xi(k)+\mu'\right).
    \label{eq:n_UHB_T0}
\end{equation}
It is therefore natural to define the density of holes relative to full filling, $\delta' = 2-n_{\rm phys}$. For the dispersion $\xi(k)=1/(1+k^2)^\alpha$, Eq.~\eqref{eq:n_UHB_T0} implies
\begin{equation}
   \delta'
    \approx
    \frac{s^2}{2\pi}\left[(-\mu')^{-1/\alpha}-1\right],
    \quad
    \mu_F'(\delta')=
    -\frac{1}{\left(1+2\pi \delta'/s^2\right)^\alpha}.
    \label{eq:muF_UHB}
\end{equation}
Thus, the low-energy degrees of freedom on the electron-doped side are holes
forming a conventional small Fermi pocket around the $\Gamma$ point, whose volume is set by $\delta'$.
At high temperature, by contrast, the $s^2$ correction to the density is small,
$n_1\to 0$, and the chemical potential is determined by the single-site relation $ \mu' \approx T\ln\frac{2-n_{\rm phys}}{2(n_{\rm phys}-1)}$. This provides the crossover between the incoherent high-$T$ bad metal regime and the low-$T$ hole Fermi liquid in the upper Hubbard band. 

\subsubsection{Transport}

The conductivity on the electron-doped side follows from the same bubble
expression as on the hole-doped side. Using Eq.~\eqref{eq:G_UHB_main},
ignoring the real part of the self-energy at leading order, and introducing the
same velocity-resolved density of states $\Phi(x)$ as in
Eq.~\eqref{eq:transport_function_main}, one finds
\begin{equation}
\begin{aligned}
\sigma
&\approx
\frac{n_0^3}{16\pi s^2 T\left(4-n_0^2\right)}
\int_0^1
dx\;
\frac{\Phi(x)/\Gamma(x)}
{\cosh^2\!\left(\frac{n_0 x}{4T}+\frac{\mu'}{2T}\right)}.
\end{aligned}
\label{eq:sigma_UHB_general1}
\end{equation}
up to $O(s^0)$ corrections. This is the electron-doped counterpart of
Eq.~\eqref{eq:sigma_leading_fixed_mu__23_main}. 

In the high-temperature limit $T\gg 1$, the correction to the density $n_1$ is negligible, so that $ \mu' \approx T\ln\frac{2-n_{\rm phys}}{2(n_{\rm phys}-1)}$ can be used inside
Eq.~\eqref{eq:sigma_UHB_general1}. Expanding the integral in powers of $1/T$
yields
\begin{equation}
\begin{aligned}
\rho \equiv \sigma^{-1}
&\approx
\frac{\pi s^2 T(1+\alpha)(2+\alpha)(2+n_{\rm phys})}
{2\alpha^3 n_{\rm phys}(n_{\rm phys}-1)}
\\[0.3em]
&\quad
+
\frac{\pi s^2(2+n_{\rm phys})(4-3n_{\rm phys})(2+\alpha)^2(1+\alpha)}
{8\alpha^3 n_{\rm phys}(n_{\rm phys}-1)(1+2\alpha)}.
\end{aligned}
\label{eq:rho_UHB_highT}
\end{equation}
Thus the electron-doped side also exhibits a broad $T$-linear bad-metal regime,
with a density-dependent constant offset. The constant term changes sign at
$n_{\rm phys}=4/3$. As on the hole-doped side, this $T$-linear behavior does
not signal the absence of quasiparticles; rather, it originates from the
temperature dependence of the compressibility and the thermal occupation of the
dispersive $\Gamma$ pocket. 

We now turn to the low-temperature behavior and focus on the physically most
interesting regime: close to full filling, $\delta'=2-n_{\rm phys}\ll 1$. The identification and dependence of the crossover scales, $T_*$ and $T_{\rm upturn}$, mirror that of the hole side and are depicted in Fig. \ref{fig:hubbard_electron_mega_fig}a. The main difference occurs for $T\ll T_*$, where we now have $n_0 \to 2$. This corresponds to a background band insulator rather than a background of fluctuating local moments. As a result, for $T \ll T_*$ the local moment density $2-n_0 \to 0$ and the self energy \eqref{eq:Upsilon_electron_large_U_1_main} vanishes. The state becomes an ordinary FL with zero resistivity\footnote{Higher order in $s^2$ corrections should generate a self energy $\propto T^2$ according to standard Fermi liquid arguments. We do not expect this to lead to nonzero resistivity for small $\delta'$, however. The state has a small Fermi surface of holes surrounding the $\Gamma$ point, such that momentum-relaxing Umklapp processes are kinematically forbidden at low energy.}.

A useful way to think about the electron-doped side is therefore the following.
When $\mu_F'(\delta')$ lies deep inside the $\Gamma$ pocket, transport is
dominated by the high-velocity states already at the lowest temperatures and
the resistivity rises monotonically with $T$. When $\mu_F'(\delta')$ moves
closer to the flat part of the band, low-$T$ transport is more strongly influenced by
thermally activated but relatively immobile states near the band edge, which
can produce a shallow minimum in $\rho(T)$ before the system eventually crosses
over to the high-$T$ bad-metal regime described by Eq.~\eqref{eq:rho_UHB_highT}.
This is the same qualitative behavior as the hole side except for the fact that we obtain an ordinary FL below $T_*$ rather than a thermal FL$^*$.

\section{Results for the model with correlated hopping}\label{sec:correlated_hop}
\begin{figure}[t!]
    \centering
\includegraphics[width=0.95\linewidth]{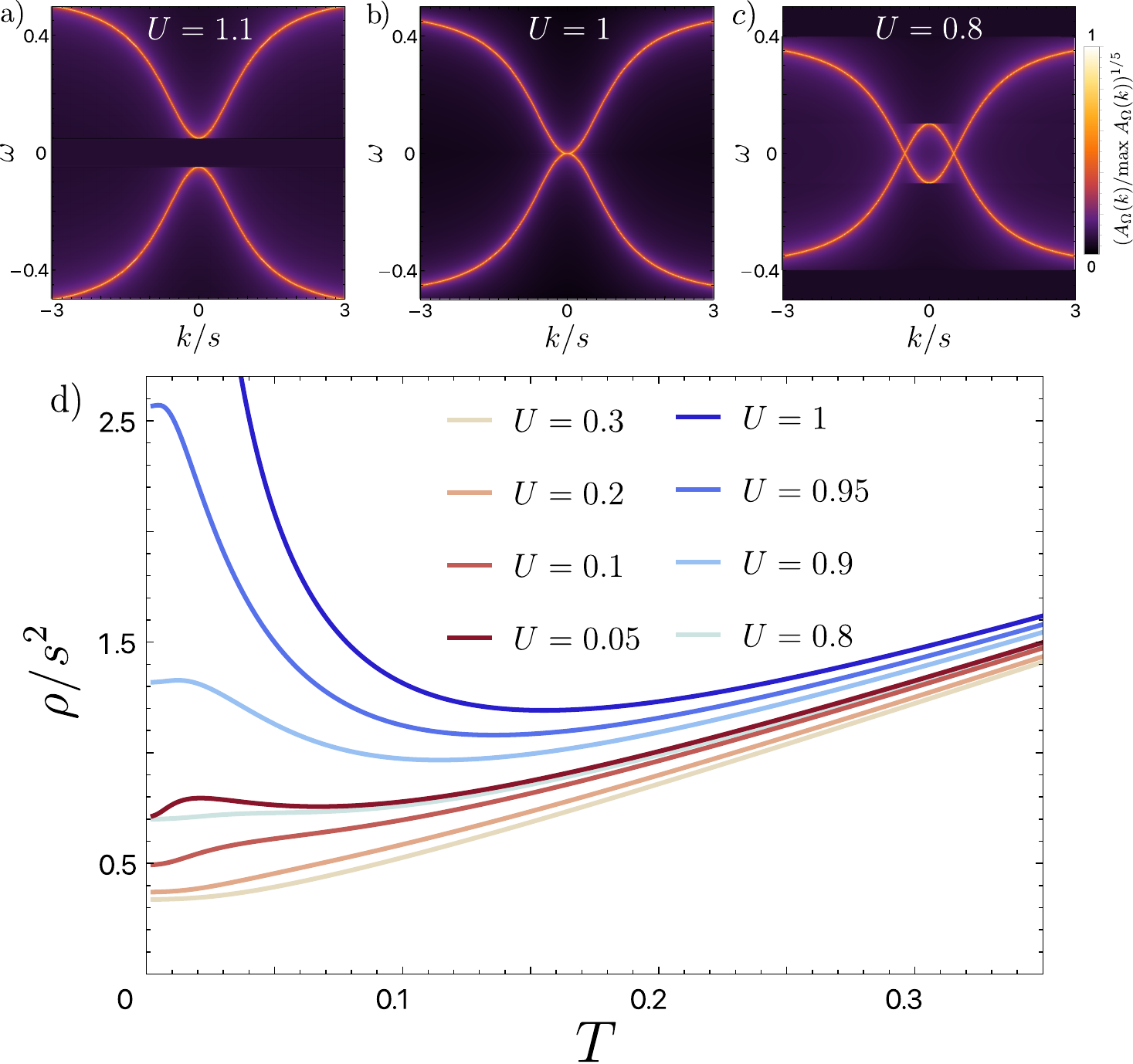}
    \caption{Spectral and transport properties of the correlated-hopping model in Eq.~\eqref{eq:Hubbard_model_corr_hops} at half filling for $t(r)=0$ and $V(r)=e^{-r}/(2\pi)$. 
(a-c) Single-particle spectral function. 
Panel (a) shows the case $U=1.1$, where the system is in the insulating Mott state. Panel (b) shows the critical point $U=1$ at which the system undergoes a metal-insulator transition with a quadratic band touching, and panel (c) shows $U=0.8$, corresponding to a compensated metallic state. 
(d) DC resistivity as a function of temperature for several values of $U$. In all plots, $s=0.05$ was used. }
\label{fig:corr_hop_1}
\end{figure}

Moving beyond the simplest setup of the modified Hubbard model, we now discuss other models that fall into the general class described by Eq.~\eqref{eq:model_general}. One simple extension is a correlated hopping term, which we focus on in this section. It is useful to write it using the local trion operator
\begin{equation}
    F_{i\sigma}
    = \{n_i{-}1, c_{i \sigma} \} = 
    \left(2n_{i\bar\sigma}{-1}\right)c_{i\sigma},\quad \{  F_{i\sigma},   F_{i\sigma}^\dagger\}=1,
\end{equation}
where we recall $\bar{\sigma}$ denotes the spin opposite to $\sigma$.
The electron-trion anti-commutator is non-canonical, $\{c_\sigma, F^\dagger_{\sigma}\} = (2n_{\bar{\sigma}}-1)$, but its expectation value vanishes in the half filling paramagnetic state. 
In terms of the notation of Eq.~\eqref{eq:model_general}, we have $\gamma_{ia}=(c_{i\sigma}, \;  F_{i\sigma})$. 
In this convention the correlated-hopping Hamiltonian is
\begin{equation}
\begin{aligned}
\label{eq:Hubbard_model_corr_hops}
    \mathcal{H}
    &=
    U \sum_i n_{i\uparrow}n_{i\downarrow}
    -\mu\sum_i n_i
    +s^2\sum_{i\neq j,\sigma}
     t(s|\bm r_i-\bm r_j|)
    c^\dagger_{i\sigma}c_{j\sigma}
    \\
    &\quad
    -\frac{s^2}{2}
    \sum_{i\neq j,\sigma}
    V(s|\bm r_i-\bm r_j|)
    \left(
     F^\dagger_{i\sigma}c_{j\sigma}
    +
    c^\dagger_{i\sigma}F_{j\sigma}
    \right).
\end{aligned}
\end{equation}
The first line here is the Hubbard model that we studied in the previous section, whereas the second like could be interpreted as an additional electron-trion hopping process. Similar correlated hopping terms have been previously studied in conventional strongly correlated systems \cite{Wojtkiewicz2001_correlated_hopping,Westerhout2022_correlated,Samajdar2024_correlated,gleis2026pseudogapped}. Such models can arise by projecting which has both kinetic and potential energy to a band with concentrated quantum geometry such that the Bloch states are trivial for most of the BZ. Such bands has Wannier function with a central peak containing most of the weight and a shallow tail containing weight $\sim s^2$ but extending to distance $\sim 1/s$. The overlaps of such Wannier functions controls both the single particle hopping and correlated hopping, giving a Hamiltonian of the form (\ref{eq:Hubbard_model_corr_hops}) We will find that the model can lead to interesting electron-trion physics. From a technical standpoint, it will also serve as a warm up for the topological model in the next section. 
For simplicity, we will specialize to half filling: $\mu=U/2$ and $n_0=1$. We will also take $V(r)=e^{-r}/(2\pi)$, such that $v(k)=\frac{1}{(1+k^2)}^{3/2}$.

The local two-point function in the electron-trion basis is
\begin{equation}
\begin{aligned}
    g_\tau^{ab} & =
    -\left\langle
   [2n_{\bar{\sigma}}(\tau)-1]^a c_{\sigma}(\tau)
      [2n_{\bar{\sigma}}-1]^b \bar{c}_{\sigma}
    \right\rangle_0, \\
    \bm{g}_\varepsilon & = (\omega - U\tau_x/2)^{-1},\label{eq:g_ep_corr_hop}
    \end{aligned}
\end{equation}
where $a,b=0,1$ and $\tau_{x,y,z}$ are the Pauli matrices in the electron trion space. 
We see that the on-site Hubbard interaction acts as a tunneling term between electrons and trions and that, in this basis, the Mott insulator Green's function looks like that of a band insulator. The zeroth order in $s^2$ Green's function also has a non-interacting form in this basis:
\begin{equation}
\begin{aligned}
    \bm{\mathcal{G}}_\varepsilon^{(0)}(k) & = (\omega - U\tau_x/2 - \mathcal{T}(k,s))^{-1}\\
    \mathcal T(k,s)& =
   \frac{1}{2} t(k,s) (\tau^0+\tau^z)
    -
    \frac{1}{2}V(k,s)\tau^x.
    \label{eq:corrhopp_hubbardone}
    \end{aligned}
\end{equation}

Let us first discuss the electron-trion ``band structure'' associated with \eqref{eq:corrhopp_hubbardone}. For $t=0$ there is a unitary particle hole symmetry $c_{i\sigma} \to c^\dag_{i\sigma}$ under which $F_{i \sigma} \to -F^\dag_{i \sigma}$.The condition $t=0$ has further dynamical implications: processes that create doublon-holon pairs out of a singly occupied background are forbidden. Holons and doublons can propagate once present, but the kinetic term no longer creates them at leading order from the Mott background. At half filling this strongly constrains the charge dynamics and effectively decouples the lower and upper Hubbard sectors. Away from half filling, the same cancellation suppresses the usual virtual doublon-holon fluctuations that generate antiferromagnetic superexchange, thereby favoring kinetic, Nagaoka-like ferromagnetic tendencies \cite{Westerhout2022_correlated}.

The system can be in three phases depending on $U$
, as depicted in Fig.~\ref{fig:corr_hop_1}(a-c). For $U>v(k=0)=1$ the spectrum is gapped, for $U=1$ the bands touch at $k=0$, and for $U<1$ there is a nodal ring corresponding to the electron-like and hole-like Mott bands passing through each other. This nodal ring is protected by particle-hole symmetry, Breaking particle-hole through $t\neq 0$ gaps out the ring, producing a trionic band edge on one side and an electronic band edge on the other side.

We will now compute the self energy corrections and conductivity, focusing on the particle-hole symmetric semimetal, $t=0$ and $U \leq 1$, for concreteness.

The calculation of the self-energy in Eq.~\eqref{eq:Sigma_general_k_k} involves a four-point cumulant function
\begin{equation}  
\begin{aligned}
\Gamma_{\tau\tau_2\tau_1\tau_0}^{(4),as'sb} &=-\langle [2n_{\bar{\sigma}_a}(\tau){-}1]^ac_{\sigma_a} (\tau)  [2n_{\bar{\sigma}_{s}}(\tau_1){-}1]^{s}c_{\sigma_{s}}(\tau_1) \\
& \hspace{-2em}\times   [2n_{\bar{\sigma}_{s'}}(\tau_2){-}1]^{s'}\bar{c}_{\sigma_{s'}}(\tau_2) [2n_{\bar{\sigma}_b}(\tau_0){-1}]^b\bar{c}_{\sigma_b}(\tau_0) \rangle_{0,c}
\end{aligned}
\end{equation}
where $a,s,s',b=0,1$, and we suppressed independent spin indices $\sigma_{a,b,s,s'}=\uparrow/\downarrow$ on the l.h.s. For $a=s=s'=b=0$, this object reduces to the four-point function that appeared in the previous section for the pure Hubbard case. The remaining components involve additional inserted density operators. Importantly, as explained in Sec.~\ref{sec:systematics}, the cumulant that appears here is defined with respect to the disconnected propagators of composite operators $c_{i\sigma}$ and $F_{i\sigma}$, and so $\Gamma_{\tau\tau_2\tau_1\tau_0}^{(4),as'sb}= -\langle... \rangle_0 +\delta_{\sigma_a \sigma_b}\delta_{\sigma_s \sigma_{s'}}g_{\tau ,\tau_0}^{ab} g_{\tau_1 ,\tau_2}^{ss'}-\delta_{\sigma_a \sigma_{s'}}\delta_{\sigma_s \sigma_b}g_{\tau ,\tau_2}^{as'} g_{\tau_1,\tau_0}^{sb}$.

The resulting expression for the matrix elements of $\Sigma^{(1),ab}_\ep$ has the form
\begin{equation}
\begin{aligned}
    \Sigma_{\ep}^{(1),ab} &= s^{2} T\sum\limits_{\ep'} \sum_{\{a_i,b_i\}}  (g_{\ep}^{-1})_{aa_1}\hspace{-0.6em}\sum\limits_{\sigma_{a_2}=\sigma_{b_2}} \Gamma^{(4)a_1 a_2b_2 b_1}_{\ep,\ep';\omega=0} (g_{\ep}^{-1})_{b_1b} 
    \\
    & \times \int_{\bm{k}}[\mathcal{T}^{ }(sk) \mathcal{G}^{(0)}_{\ep'} (sk)\mathcal{T}^{}(sk)]_{a_2b_2} \;.\label{eq:self-energy_hopp+corr}
    \end{aligned}
\end{equation}
Here we also assume a paramagnetic state, so the external spin induces $\sigma_{a}=\sigma_{b}$ of $ \Gamma^{(4)a_1 a_2b_2 b_1}_{\ep,\ep';\omega=0} $ are coinciding, while the internal ones $\sigma_{a_2}=\sigma_{b_2}$ are summed over (note that $\mathcal{G}^{(0)}$ is diagonal in spin indices).  The explicit expressions for $\Gamma^{(4)a_1a_2 b_2 b_1}_{\varepsilon\varepsilon',0}$ are given in the Appendix~\ref{app:Im_correlated_hopping_model}.

The zeroth-order Green's function in this regime is especially simple:
\begin{equation}
    [\mathcal G_\varepsilon^{(0)}(k)]^{-1}
    =
    i\varepsilon \;\tau^0
    -
    \frac{1}{2}\left[U-v(k/s)\right]\tau^x,
\end{equation}
and its electron-electron component is given by
\begin{equation}
    \mathcal G_{\varepsilon}^{(0),00}(k)
    =
    \frac{1}{
    i\varepsilon
    -
    \frac{\left[U-v(k/s)\right]^2}{4i\varepsilon}
    }.
\label{eq:Mott_semimetal_2}
\end{equation}

From Eq.~\eqref{eq:Mott_semimetal_2}, we find that the gap closes when $U=v(k/s)$ for some momentum $k$. For example, if $V(r)=e^{-r}/(2\pi)$, then $ v(k)=\frac{1}{(1+k^2)^{3/2}}$, and the gap closes at the Gamma point when $U=1$, with a quadratic band touching. The associated spectral function is shown in Fig.~\ref{fig:corr_hop_1}(a-c) for different values of $U$.

Making use of the structure of the $0$th order Green's function at the particle-hole symmetric point, we obtain the following simplified expression for the self-energy in Eq.~\eqref{eq:self-energy_hopp+corr} 
\begin{equation}
    \Sigma_\varepsilon^{(1)}
    =
    \frac{3s^2}{4}\left(\mathcal{I}_\ep^+ \;\tau^0 + \mathcal{I}_\ep^- \;\tau^x\right).
\label{eq:Sigma_full_corr_trion}
\end{equation}
Here we defined the following combination
\begin{equation}
    \mathcal{I}^{\pm}_\varepsilon
    =
    \frac{1}{2}
    \int_k 
   \Big[ \frac{v^2(k)}
    {i\varepsilon-\frac{1}{2}\left[U-v(k)\right]} \pm \frac{v^2(k)}
    {i\varepsilon+\frac{1}{2}\left[U-v(k)\right]}\Big].
\label{eq:I_eps}
\end{equation}
The dressed Green's function is therefore diagonal in the $\tau^x$ basis, and can be written as
\begin{equation}
\begin{aligned}
    \mathcal G_\varepsilon(k)
    &=
    \frac{\frac{1}{2}(\tau^0+\tau^x)}
    {i\varepsilon-\frac{1}{2}\left[U-v(k/s)+3s^2I_\epsilon\right]}
    \\
    &\quad
    +
    \frac{\frac{1}{2}(\tau^0-\tau^x)}
    {i\varepsilon+\frac{1}{2}\left[U-v(k/s)+3s^2I_{-\varepsilon}\right]},
\label{eq:G_corr_hopp_1loop}
\end{aligned}
\end{equation}
where $ I_{\ep}= \frac{1}{2}(\mathcal{I}^+_{\ep}+\mathcal{I}^-_{\ep})$.

We can now calculate the conductivity using Eq.~\eqref{eq:kubo_bubble_main}. In the particle-hole symmetric correlated-hopping case, the current vertex is proportional to \(\partial_k v(k/s)\tau^x\). Since the Green's function is diagonal in the \(\tau^x\) eigenbasis, this gives
\begin{equation}
    \sigma =-\frac{1}{\pi} \int_\Omega \partial_\Omega n_F(\Omega)\int_x \Phi(x) \Big[\operatorname{Im}\frac{1}{\Omega-\frac{1}{2}(U-x+3s^2 I_\Omega^R)}\Big]^2\label{eq:cond_corr_hop}
\end{equation}
Here $ \Phi(x)
    =
    \int_k
    \left[v'(k)\right]^2
    \delta\left(x-v(k)\right)$
is the velocity-weighted density of states, and $I_\Omega^R$ is the analytic continuation of $I_{\ep}$. Fig.~\ref{fig:corr_hop_1}(d) demonstrates the behavior of the conductivity in Eq.~\eqref{eq:cond_corr_hop}for the exponential hopping case $V(r)=e^{-r}/(2\pi)$ as the system is tuned away from the metal-insulator transition at $U=1$ to the compensated metal regime.

In order to extract low-temperature asymptotics of the conductivity we repeat the steps outlined in Sec.~\ref{app:conductivity_small_s}. The resulting $1/s^2$ behavior is given  by
\begin{equation}
    \sigma
    \approx
    \frac{2}{3s^2}
    \int_\Omega
    \partial_\Omega n_F(\Omega)
    \frac{\Phi(U-2\Omega)}
    {\operatorname{Im}I_\Omega^R}
\end{equation}
up to $O(s^0)$ corrections. For a positive isotropic $v(k)$ that decreases monotonically with $k=|\bm k|$,
\begin{equation}
    \Phi(x)
    =
    \left.
    \frac{k\left[v'(k)\right]^2}{2\pi |v'(k)|}
    \right|_{k=k(x)}
    \theta(x)\theta(v(0)-x),
\end{equation}
where $k(x)$ is the positive root of $v(k(x))=x$. Similarly,
\begin{equation}
    \operatorname{Im}I_\Omega^R
    =
    {-}
    \left.
    \frac{k v^2(k)}{2|v'(k)|}
    \right|_{ k=k(U{-}2\Omega)}
  \hspace{-0.3em}   \theta\left(\frac{U}{2}{-}\Omega\right)
   \hspace{-0.1em} \theta\left(\Omega{-}\frac{U{-}v(0)}{2}\right)
\end{equation}

If $v(0)=U$, so that the band edge lies at $\Omega=0$, then at low temperature
\begin{equation}
\sigma
=
\frac{[v'(0)]^2}{3\pi s^2U^2}
+
\frac{8\ln2}{3\pi s^2U^2}
\left(
\frac{[v'(0)]^2}{U}
-
v''(0)
\right)T
+...
\end{equation}
For the exponential hopping case we have $v(k)=1/(1+k^2)^{3/2}$ and so $v'(0)=0$, $v''(0)<0$. As a result, we obtain insulating-like behavior at the band-touching point $\rho \sim 1/T$.

If $v(0)>U$, so that the system has a compensated Fermi surface, then for temperatures much smaller than $v(0)-U$ we obtain
\begin{equation}
\begin{aligned}
\sigma
\approx
\frac{2[v'(k_F)]^2}{3\pi s^2U^2}
\left\{
1
+
\frac{4\pi^2T^2}{3}
\left[
\frac{v'''}{v'^3}
-
\frac{4v''}{Uv'^2}
+
\frac{3}{U^2}
\right]_{k=k_F}
\right\},
\end{aligned}
\end{equation}
where $k_F\equiv k(U)$ is defined by $ v(k_F)=U$. This implies a Fermi-liquid-like $T^2$ scaling of the resistivity. Finally, if $v(k)<U$ for all $k$, then the system is gapped, and the conductivity is thermally activated.

\section{Results for a model with concentrated topology}\label{sec:concenttratedtopology}

We now discuss applying our expansion to band projected models with concentrated topology. This context, and specifically the Chern bands of twisted bilayer graphene, is where the small parameter $s$ was initially introduced~\cite{Non-local-moments} and used to compute the leading $\mathcal{O}(s^0)$ order contribution to the spectral function (the analog of the spectrally-sharp Hubbard-I Green’s function in Eq.~\eqref{eq:G0_chain}). This resulted in the identification of a ``Mott semimetal'' regime~\cite{Non-local-moments} where the Mott gap closes at $\Gamma$ due to band topology. The linear crossing of the Mott semimetal was later understood as a Dirac cone formed by the hybridization of electrons and trions~\cite{ledwithExoticCarriersConcentrated2025}.

While the methodology developed in the preceding sections allows for a systematic calculation of broadening, transport, and susceptibilities for the full model of twisted bilayer graphene defined in Ref.~\cite{Non-local-moments}, such a detailed analysis is beyond the scope of the present paper. Instead, we consider a simplified effective model that retains the essential structure of concentrated projection-induced form factors while omitting many details specific to twisted bilayer graphene, such as multiple flavors. At the same time, the model is sufficiently simple and physically rich to be of interest in its own right. Concretely, we study the following real-space Hamiltonian at half-filling
\begin{equation}
    \mathcal{H} = \frac{U}{2}\sum_i (\delta n_i)^2\label{eq:H_model_main}
\end{equation}
where the projected density operator is given by
\begin{equation} 
\delta n_i = \sum_{jk,\sigma} a(s|r_i-r_j|)a(s|r_i-r_k|)\,\Big[c^\dagger_{j\sigma}c_{k\sigma} -\frac{1}{2}\delta_{jk} \Big]\;. \label{eq:n_def_corr}\end{equation} 

The function $a(r)$ describes the real space "smearing" of orbitals, $c_{i\sigma}\to \sum_j a(s|r_i-r_j|) c_{ j\sigma}$, associated with projecting to a topological band. Indeed, a topological band is formed through the hybridization orbitals, $c_\bk = \lambda_\bk f_\bk + \ldots $. We can then obtain \eqref{eq:H_model_main} through projecting Hubbard interactions for $f_\bk$ onto to the band $c_\bk$ and identifying $a(sr)$ as the Fourier transform of $\lambda_\bk=\lambda(k/s)$. This is a simplification relative to the full density-density interaction \eqref{eq:H_Lambda_full} as it neglects interactions involving the other orbitals.
In the twisted bilayer graphene context, $f_\bk$ corresponds to the $f$-electrons of the Song-Bernevig model \cite{BernevigTHF}, and \eqref{eq:H_model_main} can be understood as a two-flavor version of the interacting Song-Bernevig model with only $f$-electron Hubbard interactions included. 

The orbital projection $\lambda$ must vanish for some $\bk$ if the band is topological; otherwise the projection can be chosen real and positive, which corresponds to a trivializing gauge. Motivated by the microscopic description of the Chern bands for TBG (Ref.~\cite{Non-local-moments}), we will focus on the following choice:
\begin{equation}
    \lambda(k)=\frac{|k|}{\sqrt{k^2+1}},\label{eq:lambda_defintion}
\end{equation}
where the non-analytic behavior as $k\to 0$ is associated with the fact that there is no smooth gauge in a topological band and we choose the singularity at $k=0$.

It is useful to split $a$ into its local and nonlocal parts, corresponding to the average and $k$-dependent parts of $\lambda$:
\begin{equation}
\begin{aligned}
    a(s|r_i-r_j|) & = a_0 \delta_{ij} + s^2 \beta(s|r_i-r_j|) \delta_{i\neq j} \\
    s^2\sum_{r_j \neq 0} \beta(s|r_j|)e^{ik r_j} &  = \lambda(k/s)-a_0 \approx \lambda(k/s)-1.
\end{aligned}
\label{eq:a_beta_def}
\end{equation}
The first term in the expression for $a(s|r_i{-}r_j|)$ is a local on-site contribution of the order of unity, $a_0\approx 1+\mathcal{O}(s^2)$, with $s$-dependent corrections. The second non-local term encodes overlap of Wannier state power-law tails. For the choice in Eq.~\eqref{eq:lambda_defintion}, one finds $\beta(r)=\frac{ 1 }{4\pi }\left[K_1(r/2)I_1(r/2)-K_0(r/2)I_0(r/2)\right]$, which decays as $1/r^3$ at long distances. There is also a universal $1/r^2$ part of Wannier tails in topological bands\cite{Li2024constraints}, but it has support on the $k=0$ wavefunction associated with the gauge singularity. Since we only include interactions for the $f$-orbitals, and the orbital projection of $f$ electrons satisfies $\lambda(k=0)=0$, the $1/r^2$ tail does not enter our expressions. Relatedly, the Hamiltonian obeys the exact property $[\mathcal{H},c_{k=0,\sigma}]=0$, and therefore all self-energy corrections to the electron Green's function vanish at $k=0$. 
As we will show, the resulting $0$'th order spectral function is equivalent to the one obtained in Ref.~\cite{Non-local-moments} and corresponds to the Mott semimetal state.

The Hamiltonian in Eq.~\eqref{eq:H_model_main} can be decomposed into local and nonlocal terms in a way that makes its connection to the general class of models in Eq.~\eqref{eq:model_general} manifest:
 \begin{equation}
     \mathcal{H}=\mathcal{H}_{\rm loc}+ \mathcal{H}_{\rm 2-site}+\mathcal{H}_{\rm 3-site}+...\label{eq:H_corr_top_23}
 \end{equation}
After normal ordering, the local term and the two-site and three-site correlated hopping processes take the form
\begin{equation} \begin{aligned} 
\mathcal{H}_{\rm loc} &=  \bar{U}\sum_i n_{i\uparrow}n_{i\downarrow} -\frac{\bar{U}}{2}\sum_i n_i  \\
\mathcal{H}_{\rm 2-site}&=
\frac{s^2}{2}\sum_{i\neq j,\sigma} X(s|r_i{-}r_j|)\,(F^\dagger_{i\sigma}c_{j\sigma}+\mathrm{h.c.})
\\ \mathcal{H}_{\rm 3-site}&=-\frac{s^4}{2}\hspace{-0.7em}\sum_{i\neq j\neq k,\sigma} M(s(r_i{-}r_j),s(r_i{-}r_k))\Big\{ c^\dagger_{i\sigma}c_{i\bar\sigma}^\dagger \, c_{j\sigma}c_{k\bar\sigma}  \\
&\quad+\mathrm{h.c.}+\sum_{\sigma'} [2c^\dagger_{i\sigma'}c_{i\sigma}-\delta_{\sigma\sigma'}]\, c^\dagger_{j\sigma}c_{k\sigma'} \Big\}\label{eq:H_2_3_site} \end{aligned}  \end{equation} 
Here the effective local repulsion $\bar{U}$ is given by 
\begin{equation}
   \bar{U} = \sum_{j} a(s|r_j|)^4\;,\label{eq:U_def_corr}
\end{equation}
while the hopping amplitudes are 
\begin{equation} \begin{aligned} 
\hspace{-2em}s^2X(s|r_i{-}r_j|) &= \sum_{m} a(s|r_m|)^3\,a(s|r_i{-}r_j {-}{r_m}|),
\\[0.3em]  s^4M(s(r_i{-}r_j),s(r_i{-}r_k))&= \sum_m a(s|r_m|)^2\\
&\hspace{-2em}\times a(s|r_i{-}r_j {-}{r_m}|)\,a(s|r_i{-}r_k {-}{r_m}|) \label{eq:def_x_M}\end{aligned} \end{equation} 
The two-site term describes electron--trion hopping, while the three-site term contains correlated pair hopping and density- or spin-assisted hopping processes. The dots in Eq.~\eqref{eq:H_corr_top_23} denote terms involving four or more distant sites, together with additional pair-hopping and exchange-like contributions. Their explicit form is given in Appendix. It turns out that for the $O(s^2)$ self-energy calculation considered below, only the terms displayed explicitly above contribute, and the remaining terms can be neglected.

\begin{figure*}[t!]
    \centering
\includegraphics[width=0.99\linewidth]{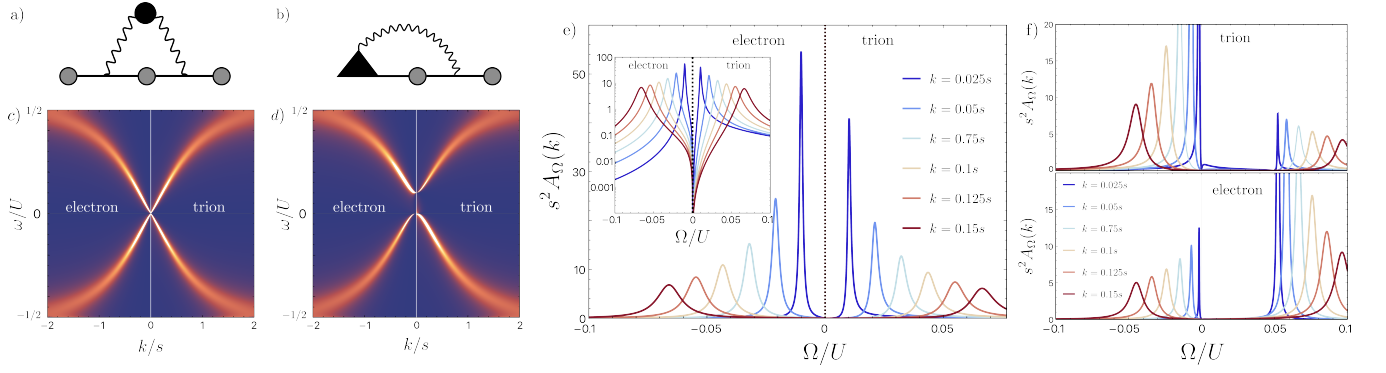}
    \caption{(a) Self-energy contribution generated at second order in the three-site correlated hopping amplitude $M$. 
The wavy line denotes the nonlocal leg of $M$ associated with an additional density or pair operator on a distant site. 
The black circle represents the corresponding local two-point cumulant of these density or pair operators. 
(b) Mixed self-energy contribution involving one three-site hopping vertex $M$ and one two-site hopping process. 
The black triangle represents a local three-point cumulant involving two $\gamma_a$ operators and one additional density or pair operator supplied by the nonlocal leg of $M$. 
In both diagrams, the internal two-point function can be dressed by an arbitrary sequence of two-site hopping processes. 
(c) Spectral function including the $O(s^2)$ self-energy in Eq.~\eqref{eq:self-energy_topological_model}. The electron component is shown for $k<0$, while the trion component is shown for $k>0$.
(d) Spectrum in the presence of a weak electron--electron hopping term
$\xi(k)=0.1/(1+(k/s)^2)$ that breaks particle-hole symmetry. (e) Spectral function cuts at fixed momenta as a function of frequency (the electron component is shown for $\Omega<0$, while the trion is depicted for $\Omega>0$). Inset shows the same plot on the logarithmic scale. (f) The trion and electron spectral functions at fixed $k$ and with $\xi(k)=0.05/(1+(k/s)^2)$. In panels (c-f), we use $s=0.5$.  }
\label{fig:topological_model}
\end{figure*}

Before discussing the technical details, let us summarize the main results given in Fig.~\ref{fig:topological_model}. It is instructive to express the Green's function as a matrix in the electron-trion space and plot the spectral function for both electron $00$ component and trion $11$ component. In panel c, we see the spectra of both electron and trion become gapless at $\bk = 0$ indicating the formation of a Mott semimetal\cite{Non-local-moments,ledwithExoticCarriersConcentrated2025,zhaoTopologicalMottLocalization2025,zhao2025mixedvalencemottinsulator}. The broadening is overall of order $s^2$, but unlike the Hubbard case, has noticeable momentum dependence; for electrons it vanishes at the $\Gamma$ point and increases as we move away from it as shown in panel e. The broadening for the trion displays a similar behavior but does not identically vanish at $\Gamma$, instead becoming parametrically small.

A further diagnosis for the Mott semimetal is its response to particle-hole breaking perturbations which lead to gapping out the spectral function with asymmetric distribution of the spectral weight between the upper and lower bands that is opposite for electrons and trions. This behavior is naturally understood by modeling the effect of particle-hole breaking perturbation as an electron-trion Dirac mass (similar to a sublattice potential in graphene) which makes one of the conduction/valence band edges electron-like and the other trion like. This is confirmed by the spectral function shown in panel d and the linecuts shown in panel f.

Our first step is the calculation of the $0$th order Green's function. At this order, only the electron-trion hopping processes contribute, and the hopping matrix is given by
\begin{equation}
    \mathcal{T}^{ab}(k)=\frac{1}{2}X(k,s)\tau^x \approx \frac{\bar{U}}{2}[\lambda(k/s)-1]\tau^x\;.
\end{equation}
The approximation above can be understood in the following way
\begin{equation}
\begin{aligned}
X(k,s)&=s^2\sum_{r_j\neq 0}X(s|r_j|)e^{ik r_j} \approx a_0^3s^2\sum_{r_j\neq 0} \beta(s|r_j|)e^{ik r_j}\\ 
&\approx  a_0^3 [\lambda(k/s)-a_0] \approx   \bar{U}[\lambda(k/s)-1], \label{eq:X_smooth}
\end{aligned}
\end{equation}
where we first singled out the $r_m=0$ term in Eq.~\eqref{eq:def_x_M}, and then used Eq.~\eqref{eq:a_beta_def} and the leading order expression for $\bar{U}$ in Eq.~\eqref{eq:U_def_corr}, i.e. $\bar{U}\approx 1+\mathcal{O}(s^2)$. After solving the resulting Dyson equation, Eq.~\eqref{eq:G0_chain}, we arrive at the inverse Green's function of the form
\begin{equation}
  [\mathcal{G}^{(0)}_\ep (k)]^{-1}=\begin{pmatrix}
        i\ep &      - \frac{\bar{U}}{2} \lambda(k/s)\\
     - \frac{\bar{U}}{2} \lambda(k/s)& i\ep
    \end{pmatrix}\label{eq:G_0th_order_main}
\end{equation}
Crucially, the off-diagonal terms here manifestly vanish at $k=0$, reflecting the exact property $[\mathcal{H},c_{k=0,\sigma}]=0$. The electron Green's function corresponds to the $a=b=0$ component, and it is given by
\begin{equation}
    \mathcal{G}^{(0),00}_\ep(k)= \frac{1}{i\ep- \frac{\bar{U}^2 \lambda^2(k/s)}{4i\ep}}\;.\label{eq:Mott_semimetal}
\end{equation}
After analytic continuation, $i\ep\to \Omega+i0^+$, the small-momentum form
$\lambda(k/s)\simeq |k|/s$ shows that the $0$th order Green's
function has two sharp poles at
$\Omega=\pm \bar U|k|/(2s)$. Thus, the leading-order spectrum consists of a
Dirac-like cone whose velocity is set by $\bar U/(2s)$, while the spectral
weight at $k=0$ remains protected by $\lambda(0)=0$. In the opposite regime $|k|\gg s$ we recover the standard Mott self-energy. 

The $O(s^2)$ self-energy receives several contributions, which we summarize here and evaluate in detail in Appendix~\ref{sec:topological_model_appendix}.

The first contribution is the already familiar self-energy in Eq.~\eqref{eq:Sigma_general_k_k}. It arises from a repeated site visit along a propagation path built from the two-site electron--trion hopping amplitude $X$. As was emphasized before, since this process only involves returning to the same site through two-site hopping detours, the resulting self-energy is momentum independent. In the electron--trion basis it contributes to all matrix elements of the self-energy.

The three-site assisted hopping terms in $\mathcal{H}_{\rm 3-site}$ generate two additional effects, shown diagrammatically in Fig.~\ref{fig:topological_model}(a,b). The process in Fig.~\ref{fig:topological_model}(a) involves two three-site vertices $M$, each of which transfers an electron between two sites, while also introducing an additional electron bilinear at a distant site. The non-trivial $s^2$ effect occurs when these bilinears from both amplitudes $M$ appear on the same site. That  site then contributes a local two-point cumulant of density or pair operators, for example $\sum_{\sigma'}
\left\langle
\bar c_{\sigma'}(\tau)c_{\uparrow}(\tau)\,
\bar c_{\uparrow}c_{\sigma'}
\right\rangle_{0,c}$ and $\left\langle
\Delta(\tau)\bar\Delta
\right\rangle_{0,c}$ (here $\Delta=c_{\downarrow}c_{\uparrow}$). In contrast to the local contribution in Eq.~\eqref{eq:Sigma_general_k_k}, this process produces an explicitly momentum-dependent self-energy. 
The momentum dependence follows from the leading form of a single three-site vertex 
\begin{equation}
\begin{aligned}
M(k,p,s)
&=
s^4
\sum_{r_j,r_k\neq 0}
M(s|r_j|,s|r_k|)
e^{i k r_j+i p r_k}
\\
&\approx
a_0^2 s^4
\sum_{r_j,r_k\neq 0}
\beta(s|r_j|)\beta(s|r_k|)
e^{i k r_j+i p r_k}
\\
&\approx
\bar{U}
\left[\lambda(k/s)-1\right]
\left[\lambda(p/s)-1\right],
\end{aligned}
\end{equation}
where we isolated the
$r_m=0$ term in Eq.~\eqref{eq:def_x_M}. Here one momentum is external, while the other is integrated over in the internal loop. 
The resulting contribution to the self-energy is given in Eq.~\eqref{eq:Sigma_MM}.  Because the extra nonlocal bilinear generated by the three-site vertex attaches to the electron component of the external electron--trion spinor, this contribution appears only in the electron--electron block, $a=b=0$.

The second new contribution, shown in Fig.~\ref{fig:topological_model}(b), is a mixed process involving one three-site vertex $M$ and one two-site hopping vertex $X$. It occurs when an additional electron bilinear introduced by $M$ is placed on a site that is also visited by an ordinary two-site hopping process. The corresponding local object is a three-point cumulant involving two $\gamma_a$ operators and one additional density or pair operator: $\langle
[2n_{\bar\sigma}(\tau)-1]^a c_\sigma(\tau)\,
[2n_{\sigma}(\tau')-1]^b c_{\bar\sigma}(\tau')\,
\bar\Delta
\rangle_{0,c}$ and $
\sum_{\sigma}
\langle
[2n_{\bar\sigma'}(\tau)-1]^a c_{\sigma'}(\tau)\,
[2n_{\bar\sigma}(\tau')-1]^b \bar c_{\sigma}(\tau')\,
\bar c_{\sigma'}c_{\sigma}
\rangle_{0,c}$. This contribution is also momentum dependent, and it is given explicitly in Eq.~\eqref{eq:MXSigma}. In the electron--trion basis it contributes to the $00$, $01$, and $10$ components of the self-energy, but not to the trion--trion component.

Combining the local contribution with the two momentum-dependent contributions generated by the three-site assisted hopping terms, and evaluating the local cumulants in the low-temperature regime $T\ll U$, we find that the self-energy takes the simple form
\begin{equation}
\Sigma_{\ep}(k)=
\frac{3}{4}s^2 \bar{U}^2\begin{pmatrix}\displaystyle
\lambda^2(k/s) \mathcal{I}_\ep^+ & \displaystyle \lambda(k/s) \mathcal{I}_\ep^-\\[10pt] \displaystyle
\lambda(k/s)\mathcal{I}_\ep^- & \displaystyle   \mathcal{I}_\ep^+
\end{pmatrix}\;,\label{eq:self-energy_topological_model}
\end{equation}
where the frequency-dependent integrals $\mathcal{I}^{\pm}_\varepsilon$ are defined as
\begin{equation}
    \mathcal{I}^{\pm}_\varepsilon
    =
    \frac{1}{2}
    \int_k 
  (\lambda(k){-}1)^2 \left[ \frac{1}
    {i\varepsilon-\frac{\bar{U}}{2} \lambda(k)} \pm \frac{1}
    {i\varepsilon+\frac{\bar{U}}{2} \lambda(k)}\right].
\label{eq:I_eps_222}
\end{equation}
We note that both the off-diagonal matrix elements, as well as the $00$ element vanish at $k=0$, as expected. For our choice of $\lambda$ in Eq.~\eqref{eq:lambda_defintion} the integrals in Eq.~\eqref{eq:I_eps_222} can be evaluated in the straightforward way, and we obtain
\begin{equation}
    \begin{aligned}
\mathcal{I}^{+}_{\varepsilon} = -\,\frac{i\varepsilon}{4\pi\!(\varepsilon^{2}+\tfrac{\bar{U}^{2}}{4})^{\!2}}\bigg[\,&
\Bigl(\varepsilon^{2}-\tfrac{\bar{U}^{2}}{4}\Bigr)\ln\!\frac{4\varepsilon^{2}}{\varepsilon^{2}{+}\frac{\bar{U}^{2}}{4}}
-\Bigl(\varepsilon^{2}{+}\frac{\bar{U}^{2}}{4}\Bigr)\\
&+\,2\bar{U}|\varepsilon|\arctan\!\frac{\bar{U}}{2|\varepsilon|}\,\bigg]
\end{aligned}
\end{equation}
and similarly
\begin{equation}
    \begin{aligned}
\mathcal{I}^{-}_{\varepsilon} = -\,\frac{\bar{U}}{8\pi(\varepsilon^{2}+\tfrac{\bar{U}^{2}}{4})^{\!2}}&\bigg[\,
\Bigl(\varepsilon^{2}+\tfrac{\bar{U}^{2}}{4}\Bigr)
-2\varepsilon^{2}\ln\!\frac{4\varepsilon^{2}}{\varepsilon^{2}{+}\frac{\bar{U}^{2}}{4}}\\
&+\,\frac{4|\varepsilon|}{\bar{U}}\Bigl(\varepsilon^{2}-\tfrac{\bar{U}^{2}}{4}\Bigr)\arctan\!\frac{\bar{U}}{2|\varepsilon|}\,\bigg]\;.
\end{aligned}
\end{equation}
After performing analytic continuation to real frequencies $\Omega$, and expanding at small $|\Omega|\ll \bar{U}$, we find $\mathcal I^{+,R}_\Omega
=
-\frac{2\Omega}{\pi \bar{U}^2}
\ln\frac{\bar{U}}{4\sqrt{e}|\Omega|}
-
i\frac{|\Omega|}{\bar{U}^2}
+\mathcal{O}(\Omega^2)$ and $\mathcal I^{-,R}_\Omega
=
-\frac{1}{2\pi \bar{U}}
-
i\frac{\Omega}{\bar{U}^2}+\mathcal{O}(\Omega^2 \ln|\Omega|)$.

Given the full $\mathcal{O}(s^2)$ self-energy, we are now in a position to analyze the spectrum of single electron and trion excitations. After performing analytic continuation to real frequencies, we find the following expression for the components of the Green's function
\begin{equation}
\begin{aligned}
\mathcal G^{aa,R}_{\Omega}(k)
&=
\frac{\Omega-\frac{3}{4}s^2\bar{U}^2\lambda^{2a}(k/s)\mathcal I^{+,R}_\Omega}
{D^R_\Omega(k/s)}\;,\\
\mathcal G^{01,R}_\Omega(k)
&=
\mathcal G^{10,R}_\Omega(k)
=
\frac{
\lambda(k/s)
[
\frac{\bar U}{2}
+
\frac{3}{4}s^2\bar U^2\mathcal I^{-,R}_\Omega
]
}{
D^R_\Omega(k/s)
}\;.
\end{aligned}\label{eq:G_full}
\end{equation}
where $a=0$ in the first line corresponds to the electron Green's function and $a=1$ is the trion Green's function. \color{black}We also defined
\begin{equation}
    \begin{aligned}
   D^R_\Omega(k)
&=
\left(\Omega-\frac{3}{4}s^2\bar{U}^2\lambda^2(k)\mathcal I^{+,R}_\Omega\right)
\left(\Omega-\frac{3}{4}s^2\bar{U}^2\mathcal I^{+,R}_\Omega\right)
\\
&-
\lambda^2(k)
\left(
\frac{\bar{U}}{2}+\frac{3}{4}s^2\bar{U}^2\mathcal I^{-,R}_\Omega
\right)^2 \;. \label{eq:D_def_main}    
    \end{aligned}
\end{equation}
The corresponding spectral functions $A^{aa}_\Omega(k)=-(1/\pi)\operatorname{Im}\mathcal G^{aa,R}_{\Omega}(k)$ are shown in Fig.~\ref{fig:topological_model}(c). The spectrum consists of two broadened Dirac-like branches. The electron component remains sharply constrained near $k=0$ because both $\Sigma_{00}$ and $\Sigma_{01}$ in Eq.~\eqref{eq:self-energy_topological_model} vanish as $\lambda(k)\to0$, while the trion component retains a non-zero frequency-dependent self-energy even at zero momentum. Both the electron and trion spectra diverge as $k,\omega \to 0$, however, since all imaginary self energies vanish at zero frequency here.

To gain further analytical insight, we expand the full Green's function near the Dirac point, in the regime $|\Omega|\ll \bar U$ and $|k|\ll s$. Using the low-frequency asymptotics of $\mathcal I^{\pm,R}_\Omega$ and the small-momentum form $\lambda(k/s)\simeq |k|/s$, we find
\begin{widetext}
\begin{equation}
\bm{\mathcal{G}}^{R,-1}_\Omega(k)
\simeq
\begin{pmatrix}
\Omega
+
\frac{3k^2}{2\pi}\left(
\Omega
\ln\frac{\bar U}{4\sqrt e|\Omega|}
+
\frac{i\pi}{2}|\Omega|
\right)
&
-\frac{\bar{U}|k|}{2s} +s|k|
\left(
\frac{3\bar U}{8\pi}
+\frac{3i}{4}\Omega
\right)
\\[0.8em]
-\frac{\bar{U}|k|}{2s} +s|k|
\left(
\frac{3\bar U}{8\pi}
+\frac{3i}{4}\Omega
\right)
&
\Omega
+
\frac{3s^2}{2\pi}\left(
\Omega
\ln\frac{\bar U}{4\sqrt e|\Omega|}
+
\frac{i\pi}{2}|\Omega|
\right)
\end{pmatrix},
\label{eq:G_matrix_small_Omega_k_main}
\end{equation}
\end{widetext}
up to higher powers in $|k|/s$ and subleading low-frequency terms $\Omega^2/\bar{U}^2$. For the electron-electron component, this leads to the generalization of the $0$th order result in Eq.~\eqref{eq:Mott_semimetal}
\begin{equation}
\mathcal G^{00,R}_{\Omega}(k)
\simeq
\frac{1}{
\Omega
-
\frac{\bar U^2 k^2}{4s^2\Omega}
\mathcal{F}(\Omega, s)
},\label{eq:corrected_Mott_semimetal}
\end{equation}
where the function $\mathcal{F}(\Omega, s)$
behaves as
\begin{equation}
 \mathcal F(\Omega,s)\approx    1
-\frac{3s^2}{2\pi}\Big[
\ln\frac{\sqrt e\,\bar U}{4|\Omega|}
+\frac{i\pi}{2}\operatorname{sgn}\Omega
+\frac{2i\pi \Omega}{\bar U}\Big]
\end{equation}
in the intermediate frequency range $\bar{U}e^{-2\pi/3s^2}\ll |\Omega|\ll \bar U$, while 
\begin{equation}
 \mathcal F(\Omega,s)\approx  \frac{2\pi}{3s^2 \ln\frac{\bar U}{4\sqrt e|\Omega|}}
\Bigg[
1-\frac{i\pi \operatorname{sgn}\Omega}{2\ln\frac{\bar U }{4\sqrt e|\Omega|}}
\Bigg],
\end{equation}
for the frequencies below an exponentially small scale $|\Omega|\ll \bar{U}e^{-2\pi/3s^2}$.

These simplified expressions allow us to easily extract the structure of the poles. First we focus on the regime $|k|\lesssim s$ but $|k|\gg s e^{-2\pi/3 s^2}$. In this case, using Eq.~\eqref{eq:G_matrix_small_Omega_k_main} we find \color{black} 
\begin{equation}
    \Omega_{\pm}(k)\approx  \pm \frac{\bar{U}|k|}{2s}\left(1- \frac{3s^2}{4\pi} \ln \frac{\sqrt{e} s}{2|k|}\right) -\frac{3is\bar{U} |k|}{16},
\end{equation}
with the corrections to to both real and imaginary parts of the order $\mathcal{O}(k^2)$, and the condition $|k|\gg s e^{-2\pi/3 s^2}$ guarantees that $s^2\ln(s/|k|)\ll1$. The corresponding residues for the electron and the trion Green's functions are given by
\begin{equation}
\begin{aligned}
\mathcal{Z}_{\pm,c}(k)&\approx \frac{1}{2}\!\left(1+\frac{3s^{2}}{4\pi}\right)\;\mp\;i\,\frac{3s|k|}{8}\;, \\
\mathcal{Z}_{\pm,F}(k)&\approx \frac{1}{2}\!\left(1-\frac{3s^{2}}{2\pi}\ln\frac{s}{2e|k|}\right)\;\mp\;i\,\frac{3s^{2}}{8}\;.
\end{aligned}
\end{equation}
We also note that the full Green's function obeys the sum rule $\int_\Omega A^R_\Omega(k) = \tau^0$ which is expected from the anticommutation relations
$\{c_\sigma,c_\sigma^\dagger\}=1$,
$\{F_\sigma,F_\sigma^\dagger\}=1$, and
$\langle\{c_\sigma,F_\sigma^\dagger\}\rangle
=\langle 2n_{\bar\sigma}-1\rangle=0$
in the paramagnetic half-filled state. This can be seen directly from the fact that $\mathcal{G}^R_\Omega(k)$ is analytic in the upper complex frequency half-plane, and decays as $\mathcal{G}^R_\Omega(k)=\tau^0/\Omega+O(\Omega^{-2})$ at large complex frequency 
since $\Sigma^R_\Omega(k)=O(\Omega^{-1})$.

The logarithmic renormalizations in the above expressions become large for smallest momenta such that $|k|\ll s e^{-2\pi /3s^2}$. For such $k$, the real part of the dispersion changes its leading behavior to $\Omega_\pm(k)\approx \pm  \sqrt{2\pi/3}\bar{U}|k|/[2s^2\ln^{1/2}(s/|k|)]$, while the broadening is logarithmically weaker $\sim |k|/\ln^{3/2}(s/|k|)$. The residues then behave as $\mathcal{Z}_{\pm,c}(k)\approx \frac{1}{2}\!\left[1+\frac{1}{2\ln(s/|k|)}\right]\;\mp\;\frac{i\pi}{8\ln^2(s/|k|)}$ and $\mathcal{Z}_{\pm,F}(k)\approx \frac{\pi}{3s^2\ln(s/|k|)}\mp \frac{i\pi^{2}}{6s^{2}\ln^{2}(s/|k|)}$. This result is consistent with the expectation that the electron excitations decouple at $k=0$, because $[H,c_{k=0,\sigma}]=0$, while the trion at $k=0$ has a logarithmically vanishing residue.

Finally, we also analyze the role of an additional weak hopping in the electron-electron sector that breaks the particle-hole symmetry. For concreteness, we choose the full hopping matrix to be $ \mathcal{T}^{ab}(k)=\frac{\bar{U}}{2}[\lambda(k/s)-1]\tau^x + \frac{\kappa}{2}(1-\lambda^2(k/s))(\tau^0+\tau^z)$, and assume that $\kappa$ is small so that we can neglect its effect on the $\mathcal{O}(s^2)$ self-energy. The resulting spectral functions for electrons and trions are shown in Fig.~\ref{fig:topological_model}(d). The weak electron hopping splits the crossing by breaking particle-hole symmetry, and it acts as a mass term for the electron-trion Dirac cone. As a result, the lower (upper) band edge becomes purely trionic (electronic) for $k\to0$. This is especially clear in Fig. \ref{fig:topological_model}f, where the electrons in the lower Mott band have vanishing spectral weight as $k\to 0$. Note that while the electron peak height is still visible at small $k$, this is a result of the $y$-axis scale and the fact that the broadening vanishes as $k\to0$ in the model studied here.

\section{Discussion and conclusions}\label{sec:Discussion}

In this work, we introduced a controlled expansion for systems in which the nontrivial momentum dependence of either the single-particle dispersion or the band-projected interaction form factors is concentrated in a small region of size $s\ll 1$. In real space, this leads to single-particle or correlated hopping amplitudes of order $s^2$ and decay range $1/s$. The smallness of $s$ suppresses real-space paths with repeated site visits, while self-avoiding paths of arbitrary length remain $O(1)$. This yields a systematic perturbation theory around the local problem that retains nonlocal propagation and 
and can systematically access transport. Our expansion does not assume a parametric separation of interaction and bandwidth scales; we do not require the interaction scale to be large compared to the bandwidth. 
The resulting spectral broadening, transport, and magnetic response for the modified Hubbard model are presented in Secs.~\ref{sec:Hubbard}, while the applications to correlated-hopping models and band projected models with concentrated topology are discussed in Sec.~\ref{sec:correlated_hop} and Sec.~\ref{sec:concenttratedtopology} respectively.

For the modified Hubbard model, our most notable findings are (i) a bad metal regime for temperatures $T \gtrsim 1$, which persists to numerically small temperatures $\sim 0.2$, that coexists with sharply defined quasiparticles with parametrically small broadening $\sim s^2$, (ii) an intermediate temperature phase $s^2 \ll T \ll 1$ at filling $1 - \delta$ with a hole pocket with volume $\delta$ whose transport is dominated by scattering off thermally fluctuating local moments. For the band projected model with concentrated topology, we show we can describe the broadening in the electron and trion spectral functions of a Mott semimetal.

A further extension is to include weak but long-range density-density interactions scaled in the same way as the hopping. For density interactions alone, this reduces to the Kac limit of classical statistical mechanics \cite{Kac1959,Lebowitz1966}, and its quantum extensions \cite{Lieb1966}, where interactions of strength $\sim s^2$ and range $1/s$ provide a controlled route to mean-field-like behavior. The present construction may be viewed as a generalization of this idea to coherent nonlocal hopping processes, thereby retaining intrinsically quantum dynamics. If such density interactions are included as well, then both charge transfer and nonlocal interactions are governed by the same combinatorics of rare return paths and large effective coordination volume. This may provide a useful setting for extended Hubbard-type models \cite{Hartnoll2018} and for studying the competition between Mott localization, charge order, and collective soft modes. The requirement to start with a local reference problem can also be relaxed; any spatially coupled subsystem can be used provided its cumulants can be computed and its correlation length is small compared to $1/s$. This extension could be useful for incorporating stronger antiferromagnetic correlations.

While we have mainly motivated our construction by its analytic tractability, we note that the kind of hoppings we introduced can be realized in projected models with bands with strongly concentrated quantum geometry. This is most developed for the Chern bands in twisted bilayer graphene where the concentration of the Berry curvature at $\Gamma$ leads to Wannier functions whose main weight is concentrated in a narrow peak within the unit cell but with a power law tail containing $\sim s^2$ weight\cite{Non-local-moments}. When projecting an interaction onto such band, the form factors induce correlated hopping terms of the type we discussed here, captured by the toy model of Sec.~\ref{sec:concenttratedtopology}.

Other models with hopping and correlated hopping having the form discussed here can be realized in a similar way. For instance, a band which looks trivial for most of the BZ, but which has a small region of concentrated quantum metric
(within area $s^2$), will give rise to Wannier functions with a main narrow peak in the unit cell (coming from the trivial BZ region), with a tail with weight and spread $s^2$. If the Berry curvature vanishes, or integrates to zero, the tail will decay exponentially rather than as a power law due to the vanishing Chern number. This will realize the exponentially decaying version of the correlated hopping model in Sec. \ref{sec:correlated_hop}. We note that when projecting a Hamiltonian with both kinetic energy and interaction onto some band, the resulting Hamiltonian contains both hopping and correlated hopping terms in addition to local Hubbard term. The ratio between hopping and correlated hopping can be tuned by tuning the ratio of the bandwidth to interaction. When considering such setup for a band with concentrated quantum geometry, we can achieve the Hubbard model with extended hopping if $U \gg t_0 s^2 \gg U s^2$ which justifies dropping the correlated hopping terms and focusing on the single particle hopping (Note however that this is not compatible with the $U \rightarrow \infty$ limit that we mostly focused on). We leave a more detailed discussion for specific realizations of these models for future work.

A core challenge for future work will be addressing low temperature physics, where $T$ is at or below the emergent $O(s^2)$ scales. For such temperatures, we have seen that selected classes of diagrams, especially in the spin channel, must be resummed. In the modified Hubbard model, the resulting static spin susceptibility already identifies the relevant ordering tendencies, including weak antiferromagnetism at half filling and an itinerant ferromagnetic tendency upon doping holes. At and below this exchange scale, a natural next step is to derive an effective long-wavelength spin action. In this framework, familiar from large-$N$ theories, small $s^2$ can used to systematically calculate the coefficients of terms in the action. The non-analyticities associated with fluctuating order parameters can be subsequently treated within the derived continuum theory.
The onset of magnetic correlations at low temperature may be related to the question of pseudogap physics \cite{Kyung2006,RUBTSOV20121320,Simkovic2024}. As moments become correlated, spin fluctuations modify the fermionic self-energy and can generate strong momentum dependence beyond the leading local broadening (a representative diagram is shown in Fig.~\ref{fig:fig1}(d)). The present formalism provides a natural starting point for addressing the role of these effects in a controlled setting. Similar considerations apply in other channels. In particular, superconducting instabilities can be analyzed through the particle-particle vertex \cite{Stepanov2026}, and in correlated-hopping models the pair-hopping terms already present in Eq.~\eqref{eq:H_2_3_site} may provide a direct pairing mechanism.

The small $s^2$ expansion shares some features with the large-$d$ limit underlying DMFT: in
particular, certain classes of real-space paths are parametrically suppressed, enabling controlled resummation of diagrams. 
One difference is that, in our case, genuinely nonlocal correlations remain present and can be calculated systematically. 
As a result, transport coefficients receive $\mathcal{O}(1)$ contributions already at leading order, and the self-energy acquires momentum dependence at subleading orders in $s$ (and in some cases at leading order). 
Furthermore, as we always take $1/s$ to be much smaller than the system size, the small-$s$ limit describes local Hamiltonians that satisfy all constraints of a genuine 2D system. This poses certain challenges with low temperature ordering, as discussed above, whereas the exactness of mean field in $d\to \infty$ enables DMFT to be applied at any temperature. At the same time, the fact that the small $s^2$ expansion can ``see its own breakdown'' in certain parameter regimes can also be viewed as an advantage; it signals when to incorporate renormalization group techniques that are equipped to handle non-analyticities arising from long-wavelength order parameter fluctuations.

A particularly striking feature of the small $s^2$ expansion is the ability to systematically compute finite DC transport order by order in $s^2\ll1$. This has already enabled us to identify several interesting transport regimes including a high temperature bad metal regime with $T$-linear resistivity that persisted to numerically small but parametrically $O(1)$ temperature, and a thermal FL$^*$ regime at lower temperature. It is tantalizing to speculate that this expansion could ultimately shed light on strange metal resistivity, where $\rho = \rho_0 + \alpha T$\cite{phillipsStrangerMetals2022,hartnollColloquiumPlanckianDissipation2022,varmaColloquiumLinearTemperature2020,proustRemarkableUnderlyingGround2019,legrosUniversalTlinearResistivity2019,bruinSimilarityScatteringRates2013}. 
We note that $\rho_0$ can be extremely small in sufficiently clean samples\cite{rullier-albenqueUniversalTcDepression2000a}. This motivates the question: Can we obtain $\rho \propto T$, in a clean model with small residual resistivity, down to the lowest temperatures for which we have analytic control? This did not occur for the $T\gg 1$ bad metal resistivity; while it persisted to rather small temperatures $T \approx 0.1$, ultimately $\rho$ approaches zero (for the FL), a constant (for the thermal FL$^*$), or has an upturn (for $O(1)$ doping) as $T\to O(s^2)$. Furthermore, while the ``dilute hole gas'' regime $\delta \sim s^4$ has a $T$-linear resistivity down to temperatures $T\sim s^2$, the residual resistivity is large.   

Indeed, it would have been surprising to obtain strange metal phenomenology through only the leading order bubble diagram in the modified Hubbard model, here corresponding to long lived quasiparticles elastically scattering off of fluctuating moments that act as disorder.
If strange metal phenomenology can be accessed through small $s^2$ it is likely in a regime where vertex corrections are important. 
In the Hubbard model studied here this occurs for $T\sim s^2$, through, e.g., the necessity of resumming the spin ladder in Sec. \ref{subsec:maintext_spinsusceptibility}. This should apply fluctuating self-consistent exchange fields to the local moments, which may enable them to mimic random two-level systems with a broadened distribution of energies\cite{Eliashberg_two_level,Noga2024_strange_metals,Noga2026_SC}.
It is interesting to note that vertex corrections enter the conductivity at leading order in models of concentrated topology, and strange metallicity has been observed in twisted bilayer graphene\cite{caoStrangeMetalMagicAngle2020,jaouiQuantumCriticalBehaviour2022a}, where $s^2 \ll 1$ is quantitatively reasonable\cite{Non-local-moments}.

One promising direction for future work is to study a transition regime between a low temperature FL and thermal FL$^*$. In addition to its intrinsic interest, such a transition may shed light on strange metallicity. For example, the change in Fermi volume associated with heavy fermion criticality is the origin point of the strange metal fan in heavy fermion systems \cite{gegenwartQuantumCriticalityHeavyfermion2008}. Studying such a transition and its impact on transport would be one motivation for introducing $s^2 \ll 1$ into a Kondo lattice model. Alternatively, the dispersion in the Hubbard model could be modified to be concentrated at two points in the BZ, with one region hole-like and one region electron-like. For hole doping $\delta$ near half filling, and low temperatures $T \lesssim T_* \sim s^2/\delta$ (here $\alpha = 1$), one obtains the thermal FL$^*$ as above. Closer to empty filling and $T \lesssim s^2/n$ the state is a FL. Resolving the transition would thus require studying temperatures $T \sim s^2$, which we discuss below. One potential complication is that this charge scale coincides with the spin exchange scale for the local moments on the thermal FL$^*$ side. These scales could be separated by introducing another parameter, such as large $N$, which could also aid with the resummations required for $T \sim s^2$. Alternatively, one could study $s^2 \ll 1$ deformations of models with no local moments but two different Fermi liquids, one doped relative to the vacuum and another doped relative to a strong-coupling rung-singlet insulator, with a Fermi volume jump\cite{rungsinglet}. 

More broadly, the small-$s^2$ expansion provides a controlled deformation of two-dimensional strongly correlated lattice models that keeps local Mott physics and itinerant motion on the same footing. Systems in this category host many of the most interesting and most puzzling phenomena in condensed matter physics. Given this context, we look forward to the insights that the small-$s^2$ expansion will generate.

\emph{Note added}.--- During the final stages of preparing the manuscript, three related drafts appeared on the arXiv \cite{vituri2026controlled, wei2026lifetime, hu2026twisted} which computed broadening within the topological heavy fermion model of TBG. The calculations can be understood within the framework of the small $s$ expansion applied to the specific problem of quasiparticle lifetime in TBG. Our results for the broadening in the Mott semimetal broadly agree with these works.

\begin{acknowledgements}
We acknowledge Subir Sachdev, Sri Raghu, Steven Kivelson, Tom Devereaux, Bert Halperin, Carl Zelle, Zhaoyu Han, Zhengyan Darius Shi, and Felix Desrochers for fruitful discussions. We are grateful to Erez Berg and Yaar Vituri for informing us about their manuscripts~\cite{vituri2026controlled} and for fruitful discussions as well as Nemin Wei, Felix von Oppen, and Leonid Glazman \cite{wei2026lifetime} for informing us about their manuscript. E.~K. is supported by NSF CAREER grant DMR award No. 2441781.
\end{acknowledgements}

\appendix
\begin{widetext}

\section{Small-$s^2$ perturbation theory in the Hubbard-Stratonovich representation}\label{sec:Hubbard_expansion_app}

In this Appendix we present a derivation of the small-$s^2$ expansion formulated directly in the original fermionic variables, using a Hubbard-Stratonovich (HS) decoupling of the local interaction. This formulation is complementary to the cumulant expansion and the dual-fermion approaches discussed in the main text. The main advantage of the HS formulation is that it simplifies the count of diagrams by pushing the interacting nature of the problem to the final steps of the calculation: one first counts and resums paths for a formally free particle propagating in a fixed realization of the fluctuating HS field, and only then performs the HS average for the relevant classes of diagrams contributing to a given order in $s^2$.

The derivation proceeds as follows. First, we decouple the on-site interaction with a bosonic HS field $\varphi$, so that for each fixed field configuration the fermions are free, and the full Green’s function is written as the normalized functional average of the field-dependent partition function $\mathcal{Z}^{[\varphi]}$ and the Green's function $\mathcal{G}^{[\varphi]}$ (precisely defined below). We then expand both $\mathcal{Z}^{[\varphi]}$
and $\mathcal{G}^{[\varphi]}$ in powers of the nonlocal hopping, evaluate the resulting terms with the local free but field-dependent propagator $g^{[\varphi]}$, and organize them as sums over real-space paths: for paths with no repeated sites the HS average factorizes site by site and, after restoring translation invariance, these contributions resum to the $0$th order Green’s function. Similarly, the paths with repeated site visits produce non-factorizable HS averages, i.e. connected cumulants  such as $\Gamma^{(4)}$, which generate the $\mathcal{O}(s^2)$ self-energy and vertex corrections.

For concreteness, we illustrate the procedure for the modified Hubbard model \eqref{eq:Hubbard_model}, but the same logic extends straightforwardly to models with correlated hopping terms.

\subsubsection{Green's function}

We begin with the single-particle Green's function. In order to deal with local interactions, we introduce a HS field $\varphi$ with an associated action $S_\varphi = \int_0^{\beta} d\tau \frac{\varphi^2}{2U}$. The full non-local Green's function in Euclidean time can be expressed as follows
\begin{equation} \mathcal{G}_{\tau-\tau_0}(\bm{r},\bm{r}_0)=\frac{1}{Z} \int \mathcal{D}\varphi e^{-S_\varphi}\;\mathcal{Z}^{[\varphi]} \mathcal{G}^{[\varphi]}_{\tau,\tau_0}(\bm{r},\bm{r}_0), \end{equation}
where $\mathcal{G}^{[\varphi]}_{\tau,\tau_0}(\bm{r},\bm{r}_0)$ and $\mathcal{Z}^{[\varphi]}$ are the Green's function and the partition function evaluated at $U=0$ but in the presence of a fixed configuration of the field $\varphi(\tau)$, and $Z= \int \mathcal{D}\varphi  e^{-S_\varphi}\mathcal{Z}^{[\varphi]}$ is the full partition function of the model.

In order to deduce the leading order terms in the small-$s$ expansion, we first need to consider the series in powers of the hopping amplitude for both $\mathcal{Z}^{[\varphi]}$ and $\mathcal{G}^{[\varphi]}_{\tau,\tau_0}(\bm{r},\bm{r}_0)$. Importantly, the Wick contractions are performed using an on-site propagator $g^{[\varphi]}_{\tau,\tau'}(\bm{r})$ evaluated in a free single-site theory, in the presence of a fixed background $\varphi$ \cite{Kamenev1996,salasnich2025coherent}. Its explicit expression is given by 
\begin{equation}
    g^{[\varphi]}_{\tau,\tau'} =   \exp\left\{-i\int_{\tau'}^{\tau}d\bar{\tau} \left[\varphi_{}(\bar{\tau}){-}T\tilde{\varphi}_{0}\right]\right\} g_{\tau-\tau'}^{[0]}|_{\mu\rightarrow \mu+U/2-iT\tilde{\varphi}_{0}},\label{eq:G_fluctuating}
\end{equation}
where $\tilde{\varphi}_{\omega}=\int_0^\beta d\tau \;e^{i\omega \tau}\varphi_{\bm{r}}(\tau)$ is the Fourier transform of $\varphi$, and $g^{[0]}$ is the standard free fermion Green's function
\begin{equation}\label{eq:g}
    g^{[0]}_{\tau} = T\sum\limits_{\ep} \frac{e^{-i\varepsilon \tau}}{i\varepsilon +\mu}=\Big[\frac{\theta(-\tau)}{1+e^{-\beta \mu}} -\frac{\theta(\tau)}{1+e^{\beta \mu}}\Big] e^{\mu \tau }.
\end{equation}
Similarly, for the determinant $\mathcal{Z}^{[\varphi]}$, the 0th term in the expansion is $\mathcal{Z}^{[\varphi]}|_{t=0}$ which factorizes into a product over sites, each contributing a factor $\left(1+ e^{\beta(\mu+U/2) }\right)^2|_{\mu\rightarrow \mu+U/2-iT\tilde{\varphi}_{0}}$.

A typical contribution at order $(n+m)$ (where $n$ is the order of the term in the expansion of $\mathcal{G}^{[\varphi]}_{\tau,\tau_0}(\bm{r},\bm{r}_0)$, while $m$ is the order of the term in the expansion of $\mathcal{Z}^{[\varphi]}$) before integration over HS fields takes the form:
\begin{equation}\frac{s^{2(n+m)}}{m} \sum\limits_{\{\bm{r}_j\}}\int\limits_{\{\tau_j\}}\prod_{j=0}^{n-1} t(s|\bm{r}_{j+1}{-}\bm{r}_{j}|)g^{[\varphi]}_{\tau_{j+1},\tau_{j}}(\bm{r}_j)\sum\limits_{\{\bm{\rho}_i\}}\int\limits_{\{\tilde{\tau}_j\}}\Big( -\prod_{i=0}^{m-1} t(s|\bm{\rho}_{i+1}{-}\bm{\rho}_{i}|)g^{[\varphi]}_{\tilde{\tau}_{i+1},\tilde{\tau}_{i}}(\bm{\rho}_i)+\overline{\prod_{i=0}^{m-1} t(s|\bm{\rho}_{i+1}{-}\bm{\rho}_{i}|)g^{[\varphi]}_{\tilde{\tau}_{i+1},\tilde{\tau}_{i}}(\bm{\rho}_i) }\Big) \label{eq:G_pert_first} \end{equation}
where $\bm{r}_n \equiv \bm{r}$, $\tau_n \equiv \tau$, and we impose $\tilde{\tau}_0 = \tilde{\tau}_{m}$, $\bm{\rho}_0 = \bm{\rho}_m$. The average $\overline{X^{[\varphi]}} = \int \mathcal{D}\varphi e^{-S_\varphi} \mathcal{Z}^{[\varphi]}|_{t=0} X^{[\varphi]}/\int \mathcal{D}\varphi e^{-S_\varphi} \mathcal{Z}^{[\varphi]}|_{t=0}$ denotes the local functional integral over $\varphi$. The product over $\{r_j\}$ sites corresponds to a connected path that connects $\bm{r}_0$ and $\bm{r}$, whereas the product over $\{\rho_i\}$ describes a closed loop originating from the expansion of $\mathcal{Z}^{[\varphi]}$. The subtraction comes from the overall normalization factor $1/Z$, ensuring that only connected diagrams contribute.

Next, we reorganize these terms based on the number of intersections in the corresponding paths, i.e., how often any given site is revisited. The simplest class of contributions corresponds to paths where all sites are distinct, i.e. no intersections occur. In this case, all functional averages over $\varphi$ decouple since each on-site Green's function depends only on the local field configuration. Moreover, only the contribution with $m=0$ survives (i.e. the expansion of $\mathcal{Z}^{[\varphi]}$ is canceled by the expansion of $1/Z$ for such processes). Diagrammatically, this situation is depicted in the second line of  Fig.~\ref{fig:figApp1}(a). We find
\begin{equation}
s^{2n}\sum\limits_{\substack{\bm{r}_0\neq \bm{r}_1\neq ... \neq \bm{r}}}\int_{\{\tau_j\}} \prod_{j=0}^{n-1} \overline{ g^{[\varphi]}_{\tau_{j+1}{-}\tau_{j}}} t(s|\bm{r}_{j+1}{-}\bm{r}_{j}|)\;.
\label{eq:G_1}  \end{equation}
The presence of the restricted lattice summations formally prevents us from directly summing over intermediate sites by going to the momentum space. This issue can be bypassed by adding and subtracting contributions involving site repetitions (e.g., $\bm{r}_1 = \bm{r}_3$ with all other sites are distinct, etc.). Clearly, the subtracted terms are sub-leading from the perspective of the $s$-counting, and so we will group them with higher-order processes involving path intersections. The remaining expression is then just Eq.~\eqref{eq:G_1} but with unrestricted summations, i.e. it is translationally invariant in both space and imaginary time, and thus can be readily Fourier-transformed, leading to $\xi^n(k/s)  g^{n+1}_{\varepsilon}$, where $g_{\varepsilon}=\int_0^\beta d\tau e^{i\ep \tau}\overline{ g^{[\varphi]}_\tau }$, and $\ep$ is the fermionic Matsubara frequency. Finally, after summing over all $n$, we obtain
\begin{equation}\label{eq:G_2_corr}
\mathcal{G}^{(0)}_{\varepsilon}(k) = \frac{1}{g^{-1}_{\varepsilon} - t(k,s)}\;.
\end{equation}
Using Eq.~\eqref{eq:t_approx_small_s_main} we can take the limit $s\ll 1$ while keeping external momentum fixed with respect to $s$, which amounts to replacing $t(k,s)\equiv s^2\sum_{\bm{r}_j\neq 0}t(s|\bm{r}_j|)e^{i\bm{r}_j\bm{k}}$ with $\xi(k/s)$, up to $s^2$ corrections. Importantly, the resulting expression does not contain small powers of $s^2$ in the denominator: these have been effectively compensated by the unrestricted summation over the (large) coordination volume, provided that the ratio $k/s$ remains fixed. From this perspective, Eq.~\eqref{eq:G_2_corr} represents the $0$-th order result in our expansion. In real space, this Green's function consists of a local part and a $\sim s^2$ long-ranged tail that propagates to distances of the order $\sim 1/s$. For the Hubbard case, we find the standard expressions for the single-site quantities
\begin{equation}
g^{-1}_{\varepsilon}=i\ep+\mu-\frac{n_0U}{2}-\frac{n_0(2-n_0)U^2/4}{i\ep+\mu+U(n_0/2-1)}\;,\label{eq:G_0_loc}
\end{equation}
and $n_0= 2(e^{\mu\beta}{+}e^{(2\mu-U)\beta})/(1{+}2e^{\mu\beta}{+}e^{(2\mu-U)\beta})$ is the averaged density. 

\begin{figure}[t!]
    \centering
\includegraphics[width=0.65\linewidth]{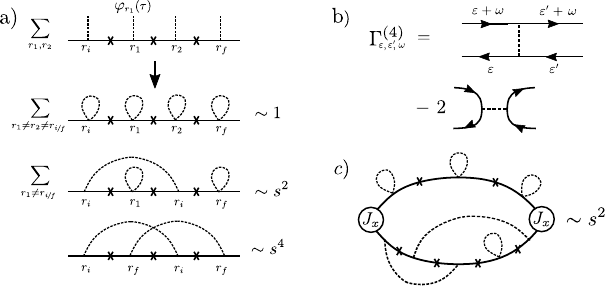}
    \caption{a) Diagrammatic expansion of the non-local Green's function in powers of $s^2$ in the Hubbard-Stratonovich representation. The cross represents a hopping amplitude. Solid lines with attached dashed segments represent free local Green’s functions in the presence of $\varphi$, and averaging over $\varphi$-fluctuations is denoted by dashed-line contractions (the power of $s^2$ associated with a diagram depends only on the total number of contractions, and is independent of whether the diagram is planar or non-planar). b) Diagrammatic representation of the four point cumulant. c) Representative $\sim s^2$ correction to the current-current correlator involving $\Gamma^{(4)}$.}
    \label{fig:figApp1}
\end{figure}

Next, we return to Eq.~\eqref{eq:G_pert_first} and consider contributions with a single repeated site: either with a self-intersection in the main path or with an intersection between the main path and a closed loop originating from the partition function. In such cases, the averaging over $\varphi$ becomes nontrivial, as the products of the form $\overline{ g^{[\varphi]}_{\tau_1,\tau_2} g^{[\varphi]}_{\tau_3,\tau_4} }$ do not factorize. Such averages are shown diagrammatically in the last two lines of Fig.~\ref{fig:figApp1}(a). Moreover, for each such average we need to subtract $\overline{ g^{[\varphi]}_{\tau_1,\tau_2}} \;\overline{g^{[\varphi]}_{\tau_3,\tau_4} }$, which originates from the terms with repeated site visits that were added to Eq.~\eqref{eq:G_1} by hand in order to make it translation-invariant. This naturally leads to the appearance of the four-point vertex $\Gamma^{(4)}$ (summed over two of its internal spin indices)
\begin{equation}
\begin{aligned}
   \Gamma^{(4)}_{\tau,\tau_2,\tau_1,\tau_0}  = \overline{ g^{[\varphi]}_{\tau_1,\tau_0}     g^{[\varphi]}_{\tau,\tau_2}}-\overline{ g^{[\varphi]}_{\tau_1,\tau_0}}\;\;\overline{     g^{[\varphi]}_{\tau,\tau_2}} - 2\overline{ g^{[\varphi]}_{\tau,\tau_0}     g^{[\varphi]}_{\tau_1,\tau_2}} +2\overline{ g^{[\varphi]}_{\tau,\tau_0}} \;\;\overline{ g^{[\varphi]}_{\tau_1,\tau_2}}\;,\label{eq:Gamma^4_def}
   \end{aligned}
\end{equation}
The extra factor of $-2$ originates from the spin summation and a fermion loop. By comparing Eq.~\eqref{eq:Gamma^4_def} with the formalism discussed in the main text, one can easily recognize $\Gamma^{(4)}$ as the local cumulant Eq.~\eqref{eq:Gamma_def_main} given in terms of operator expectation values. The $s^2$ correction associated with a repeated site visit in Fig.~\ref{fig:figApp1}(a) can be interpreted as a corrections to the single-site Green's function $g_{\ep}$ appearing in Eq.~\eqref{eq:G_2_corr}. Specifically, we find
\begin{equation}
\delta g_\ep \equiv g_\ep^2\Sigma^{(1)}_\ep  = T\sum\limits_{\varepsilon
'} \Gamma^{(4)}_{\ep,\ep'; 0}    \sum_{k \in \rm{BZ}} t^2(k,s) \mathcal{G}^{(0)}_{\ep'}(k)  \approx s^2 T\sum\limits_{\varepsilon
'} \Gamma^{(4)}_{\ep,\ep'; 0}    \int_k \xi^2(k) \mathcal{G}^{(0)}_{\ep'}(sk)\;,\label{eq:Sigma_1_H1}
\end{equation}
which is of the general form given in Eq.~\eqref{eq:Sigma_general_k_k} (specialized for the Hubbard case in Eq.~\eqref{eq:Sigma_Hubbard_main}) of the main text. Here we defined   $\Gamma^{(4)}_{\ep,\ep';\omega}=\beta^{-1}\int_{\{\tau_i\}} e^{i (\ep+\omega) \tau{-}i \ep \tau_0{-}i (\ep'+\omega) \tau_2{+}i \ep' \tau_1} \Gamma^{(4)}_{\tau,\tau_2,\tau_1,\tau_0}$ (which is depicted diagrammatically in Fig.~\ref{fig:figApp1}(b)), and the factor $ g_\ep^2$ accounts for the external legs. The appearance of the $0$th order Green's function $ \mathcal{G}^{(0)}_{\ep'}(k) $ accounts for an infinite lattice detour which does not increase the order of the diagram. In \eqref{eq:Sigma_1_H1} we also used Eq.~\eqref{eq:t_approx_small_s_main} in order to approximate the momentum summation over the full BZ with a continuum integral over unrestricted $k$. After replacing $g_{\ep}$ with $g_{\ep}+\delta g_{\ep}$ in Eq.~\eqref{eq:G_2_corr} and expanding the denominator at small $s^2$, we finally obtain the corrected Green's function
\begin{equation}
\begin{aligned}
\mathcal{G}_{\ep}(k) &= \frac{1}{g^{-1}_{\ep}- t(k,s)-\Sigma^{(1)}_{\ep}}\;.
\label{eq:G_33_corr}
\end{aligned}
\end{equation}


Higher-order corrections will appear from paths that have more than one repeated site visits. For example, at order $s^4$ there will be a non-local (i.e. momentum-dependent) correction to the Green's function originating from paths with two sites, each visited twice in the same order (diagrammatically represented in the last line of Fig.~\ref{fig:figApp1}(a)). Specifically, the corresponding self-energy in the latter case is
\begin{equation}
\begin{aligned}
\Sigma^{(2)}_{\ep}(k/s) = -s^4 g^{-2}_{\ep} T^2\sum\limits_{\varepsilon',\omega} \Gamma^{(4)}_{\varepsilon',\varepsilon;\omega}\Gamma^{(4)}_{\varepsilon'+\omega,\varepsilon';\ep-\ep'} \int_{pq} \frac{\xi(k/s+q)\xi(p+q)\xi(p)}{(1{-}g_{\ep+\omega}\xi(k/s{+}q)) (1{-}g_{\ep'}\xi(p))(1{-}g_{\ep'+\omega}\xi(p{+}q))}\;.
\end{aligned}
\end{equation}
The appearance of the denominators in this expression can be understood as a dressing of each copy of $\xi$ by an additional detour without intersections, which does not increase the order of $s^2$. The full structure of the self-energy to all orders is of the form given in Eq.~\eqref{eq:Sigma_main_series} of the main text.

\subsubsection{Conductivity and spin susceptibility}
The counting scheme outlined above allows for straightforward generalization to other observables. For instance, the Matsubara conductivity is given by
\begin{equation}
    \sigma_\omega= \frac{1}{\omega}\Big[\Pi_{\omega}^{J_xJ_x}-K_{xx} \Big]\;, 
\end{equation}
where $\Pi^{J_xJ_x}_{\omega}=\frac{1}{V} \int_0^\beta e^{i\omega \tau} \langle T J_x(\tau)J_x\rangle$ is the current-current correlator, $V$ is the system volume, and $K_{xx}$ is the diamagnetic term. For the Hubbard case, we find
\begin{equation}
    J_x= -is^2\sum\limits_{ij}(\bm{r}_i-\bm{r}_j)_x \;t(s|\bm{r}_i-\bm{r}_j|) \sum\limits_{\sigma}c^\dagger_{i\sigma} c_{j\sigma},\quad 
    K_{xx}=s^2\sum\limits_{ij}(\bm{r}_i-\bm{r}_j)^2_x \;t(s|\bm{r}_i-\bm{r}_j|) \sum\limits_{\sigma}\langle c^\dagger_{i\sigma} c_{j\sigma}\rangle.
\end{equation}
The Ward identity implies that $ \Pi^{J_xJ_x}_{\omega=0}=K_{xx}$. At the leading order in $s^2$, we thus obtain
\begin{equation}
          \Pi^{J_xJ_x}_\omega =T\sum\limits_{\ep}\int_k \left(\xi'(k)\right)^2  \mathcal{G}^{(0)}_{\ep+\omega}(sk)\mathcal{G}^{(0)}_{\ep}(sk),\quad \quad 
          K_{xx} = -\int_k \left[\xi''(k)+\frac{\xi'(k)}{k}\right]\mathcal{G}^{(0)}_{\tau=0^-}(sk)\label{eq:Jx}\;.
\end{equation}
In these expressions we have already included the spin sum and performed the angular average for an isotropic $\xi(k)$. Remarkably, Eqs.~\eqref{eq:Jx} are independent of $s$. This is because the velocity is $\nabla_k \xi(k/s)\sim 1/s$ near $k=0$, and thus the integral with the square of the velocity over the $\sim s^2$ region in momentum space is finite in the limit $s\rightarrow 0$.

Since the $0$th order Green's function does not include any spectral broadening effects, then the real frequency conductivity obtained from Eq.~\eqref{eq:Jx} after analytic continuation remains finite only at finite frequency. In order to study DC conductivity, we need to account for the next order corrections to these expressions originating from the self-energy insertions, as diagrammatically depicted in Fig.~\ref{fig:figApp1}(c). Resumming such effects amounts to replacing the Green's functions $\mathcal{G}^{(0)}$ with Eq.~\eqref{eq:G_33_corr}. The DC conductivity is then given by Eq.~\eqref{eq:kubo_bubble_main} of the main text. The lowest-order vertex correction associated with $\Gamma^{(4)}$ vanishes (as does the entire ladder built out of such vertices) because of the oddness of the velocity vertex and locality of $\Gamma^{(4)}$ (as explained in the main text, this situation is analogous to the cancellation of impurity-induced ladder corrections to the DC conductivity in disordered metals with uncorrelated impurities \cite{Lee1985}).

The analysis of the spin susceptibility $\chi^{S_zS_z}_\omega(\bm q)
=
\frac{1}{V}\int_0^\beta d\tau\,
e^{i\omega\tau}
\left\langle
T_\tau S^z(\tau,\bm q)S^z(-\bm q)
\right\rangle$ can be carried out in the HS formulation in the same way. 
In contrast to the current-current correlator, the spin vertex is local and does not amplify the small-momentum region near $k=0$. 
As a result, the leading momentum-dependent contribution to $\chi^{S_zS_z}$ appears only at order $s^2$. 
Moreover, the corresponding $O(s^2)$ vertex corrections do not vanish by parity and play an important role in the spin-fluctuation physics, as discussed in Sec.~\ref{subsec:maintext_spinsusceptibility}. 
More specific details for the Hubbard model are given in Sec.~\ref{sec:spin_appendix}. 
Here we only note that, in the HS formulation, the relevant local object is not the spin-summed vertex $\Gamma^{(4)}$ entering the single-particle self-energy, but rather the spin-channel combination
\begin{equation}
\begin{aligned}
&\Gamma^{(4,s)}_{\tau,\tau_2,\tau_1,\tau_0} =\overline{ g^{[\varphi]}_{\tau_1,\tau_0}     g^{[\varphi]}_{\tau,\tau_2}}-\overline{ g^{[\varphi]}_{\tau_1,\tau_0}}\;\;\overline{     g^{[\varphi]}_{\tau,\tau_2}} =-\sum_{\sigma'} \tau^z_{\sigma'\sigma'} \langle c_{\sigma}(\tau)c_{\sigma'}(\tau_1) \bar{c}_{\sigma'}(\tau_2) \bar{c}_{\sigma} (\tau_0)\rangle_{0}\label{eq:Gamma^4_spin}
   \end{aligned}
\end{equation}
where $\tau^z$ is the Pauli matrix, and we identified the HS average with the local cumulant in the operator formalism.

\section{Small$-s^2$ expansion of momentum integrals}\label{app:small_s_integrals}
In this section, we provide additional comments regarding the small$-s^2$ expansion of the typical momentum integrals that appear in various loop corrections and observables. First, we recall that the Fourier transform of the single-particle hopping  
\begin{equation}
     t(\bm{k},s)= \sum\limits_{\bm{r}_i \neq 0} s^2 t(s|\bm{r}_i|) e^{i\bm{k}\bm{r}_i}
\end{equation}
with sufficiently quickly decaying $t(r)$ admits the following expansion 
\begin{equation}
    t(s\bm{k},s)\approx \xi(k)+t(*,s)\;,\label{eq:t_approx_small_s}
\end{equation}
which can be obtained from the Poisson summation formula. Here  $t(*,s)$ is given by
\begin{equation}
    t(*,s)\approx -s^2 \int_{\bm{k}}\xi(k)+\mathcal{O}(s^4)\;,\label{eq:t_*_xi}
\end{equation}
 the star $*$ symbol indicates momenta away from the Gamma point, and $\int_k=\int d^2k/(2\pi)^2$ is the unrestricted momentum integral. The function $\xi(k)$ is $s-$independent and decays to zero at large momenta. We then consider the following momentum integral
\begin{equation}
    \sum\limits_{\bm{k} \in \rm{BZ}} f(t(\bm{k},s))=  f(t(*,s))+\sum\limits_{\bm{k} \in \rm{BZ}} \left[f(t(\bm{k},s))-f(t(*,s))\right] =f(t(*,s))+s^2\sum_{\bm{k} \in \rm{BZ}/s} \left[f(t(s\bm{k},s))-f(t(*,s))\right]\;,\label{eq:int_momentum}
\end{equation}
where $f(x)$ is some smooth function, $\sum_{\bm k\in{\rm BZ}}$ denotes the normalized Brillouin-zone integral, and $\rm{BZ}/s$ denotes the Brillouin zone with $1/s$ rescaled boundaries. Our goal is to approximate Eq.~\eqref{eq:int_momentum} up to the $s^2$ order. Within this accuracy, we can use the expansion in Eq.~\eqref{eq:t_approx_small_s} and, at the same time, extend the integration boundaries to infinity because the integral converges at large $k$. This way, we find 
\begin{equation}
\begin{aligned}
    \sum\limits_{\bm{k} \in \rm{BZ}} f(t(\bm{k},s)) &\approx f(t(*,s))+s^2\int_{\bm{k}} \left[f(\xi(k)+t(*,s))-f(t(*,s))\right]+\mathcal{O}(s^4)\\
    &\approx  f(0)+f'(0)t(*,s)+s^2\int_{\bm{k}} \left[f(\xi(k))-f(0)\right]+\mathcal{O}(s^4)\\
   &=f(0)+s^2\int_{\bm{k}} \left[f(\xi(k))-f(0)-\xi(k)f'(0)\right]+\mathcal{O}(s^4)  \label{eq:approx_1_k-integral}
    \end{aligned}
\end{equation}
Here in the second line we expanded all functions at small $t(*,s)\sim s^2$, and in the third line we used Eq.~\eqref{eq:t_*_xi}.
For most cases of interest, both $f(0)=0$ and $f'(0)=0$, and thus $  \sum_{\bm{k} \in \rm{BZ}} f(t(\bm{k},s)) \approx s^2\int_{\bm{k}} f(\xi(k))$. However, for certain special quantities (e.g. the local density of states) this is not the case, and the full expression in Eq.~\eqref{eq:approx_1_k-integral} should be used. In particular,  applying this formula to $f(x)=x$ we recover the fact that $\sum_{\bm{k} \in \rm{BZ}}t(\bm{k},s) =0$. Similarly, using $f(x)=1/(g^{-1}_\ep -x)$ (which corresponds to a single-particle Hubbard-I Green's function), we find
\begin{equation}
    \sum\limits_{\bm{k} \in \rm{BZ}} \frac{e^{i\bm{k}\bm{r}}}{g_{\ep}^{-1}-t(k,s)} \approx [g_{\ep}+t(*,s) g^2_{\ep}+...] \delta_{\bm{r},0} + s^2 \int_{\bm{k}}e^{i\bm{k}\bm{r}s}\left[\frac{1}{g_{\ep}^{-1}-\xi(k)}-\frac{1}{g_{\ep}^{-1}}\right]+\mathcal{O}(s^4)\;.
\end{equation}

\subsection{Susceptibilities}

Similar formulas can be established for multiple momentum integrals. For instance

\begin{equation}
    \begin{aligned}\sum\limits_{\bm{k},\bm{q}\in \rm{BZ}} e^{i\bm{q}\bm{r}} f(t(k+q,s))h(t(k,s) ) &= -\delta_{\bm{r},0}f(t(*,s))h(t(*,s) )   + s^2 \delta_{\bm{r},0}   \sum\limits_{\bm{k}\in {\rm{BZ}}/s}  \Big[f(t(*,s))h(t(sk,s))+f(t(sk,s))h(t(*,s))\Big] \\
  &+s^4 \sum\limits_{\bm{k},\bm{q}\in {\rm{BZ}}/s} e^{is\bm{q}\bm{r}} \Big[f(t(s k+sq,s))-f(t(*,s))\Big] \Big[h(t(s k,s))-h(t(*,s) ) \Big] \;.
    \end{aligned}
\end{equation}
Crucially, the double integral over $\bm{q}$ and $\bm{k}$ in the second line now converges at large momenta, and thus we can both extend the integration boundaries to infinity and replace $t$ by its continuous form $\xi$. At the same time, the single integral over $\bm{k}$ in the first line can be approximated using Eq.~\eqref{eq:approx_1_k-integral}. 


\begin{equation}
\begin{aligned}
   \sum\limits_{\bm{k},\bm{q}\in \rm{BZ}} e^{i\bm{q}\bm{r}} f(t(k+q,s)) h(t(k,s) ) &\approx \delta_{\bm{r},0}f(0 )h(0) +\delta_{\bm{r},0}f(0)\left\{t(*,s)h'(0)+s^2\int_{\bm{k}}\Big[h(\xi(k) ) -h(0 )\Big]\right\} \\&+ \delta_{\bm{r},0}h(0)\left\{t(*,s)f'(0)+s^2\int_{\bm{k}}\Big[f(\xi(k) ) -f(0 )\Big]\right\} +\delta_{\bm{r},0}\mathcal{O}(s^4)
\\&+s^4\int_{\bm{k},\bm{q}}e^{is\bm{q}\bm{r}}  \Big[f(\xi(k+q) )-f(0)\Big]\Big[h(\xi(k) ) -h(0 )\Big]\;.\label{eq:GG_approx_int}
   \end{aligned}
\end{equation}
The first and second lines here correspond to the $s^2$ corrections to the local $\bm{r}=0$ part of the integral, whereas the third line contributes a  $\sim s^4$ non-local term that depends on $sr$. Furthermore, using Eq.~\eqref{eq:t_*_xi} we can rewrite 
\begin{equation}
    t(*,s)f'(0)+s^2\int_{\bm{k}}\Big[f(\xi(k) ) -f(0 )\Big]= s^2\int_{\bm{k}}\Big[f(\xi(k) ) -f(0 )-f'(0) \xi(k)\Big]\;,
\end{equation}
where the new term under the integral effectively accounts for the fact that $t(\bm{r}=0)=0$.

An important example of this sort of integral is a convolution of two Hubbard-I Green's functions, i.e. $f(x)=1/(g_{\ep+\omega}^{-1}-x )$ and $h(x)=1/(g_{\ep}^{-1}-x )$. The full integral of interest reads as
\begin{equation}
   \chi(\bm{r},\omega)=  T\sum\limits_{\ep}\sum\limits_{\bm{k},\bm{q}\in \rm{BZ}} \frac{e^{i\bm{q}\bm{r}} }{(g_{\ep+\omega}^{-1}-t(k+q,s) )(g_{\ep}^{-1}-t(k,s) )}\;.\label{eq:chi_0}
\end{equation}

In this case, the first line in Eq.~\eqref{eq:GG_approx_int} corresponds to a local $s^2$ correction to the chemical potential induced by the flat part of the dispersion $t(k\rightarrow *,s)$. The second line involves quantities like $g_{\ep}^2 g_{\ep+\omega}\int_{\bm{k}}\xi(k)/(g^{-1}_{\ep}-  \xi(k))$, which corresponds to a $\sim s^2$ local fermion loop with only one of the Green's functions including one or more hops. Finally, the last line corresponds to the integral $g_{\ep}^2 g^2_{\ep+\omega}\int_{\bm{k},\bm{q}}\xi(k)\xi(k+q)/[(1- g_{\ep} \xi(k))(1- g_{\ep+\omega} \xi(k+q))]$ where both Green's functions include at least a single hop. We also recall that $\xi(k)/(1- g_{\ep} \xi(k))$ is essentially a dual fermion propagator. Thus, we find the following schematic structure
\begin{equation}
      \chi(\bm{r},\omega)= \Big( \chi_{\rm loc}^{(0)}(\omega)+s^2 \chi_{\rm loc}^{(1)}(\omega)+...\Big) \delta_{\bm{r},0}+ s^4 \chi_{\rm non-loc}^{(0)}( sr,\omega) +\mathcal{O}(s^6)\;.\label{eq:chi_real_space}
\end{equation}
where $\chi_{\rm loc}^{(0)}(\omega)=T\sum_\ep g_{\ep+\omega}g_\ep$, and
\begin{equation}
    \chi_{\rm loc}^{(1)}(\omega)=T\sum_\ep g^2_{\ep+\omega}g_\ep\int_{\bm{k}}\frac{\xi^2(k)}{g^{-1}_{\ep+\omega}-\xi(k)}+T\sum_\ep g_{\ep+\omega}g^2_{\ep}\int_{\bm{k}}\frac{\xi^2(k)}{g^{-1}_{\ep}-\xi(k)}\;.
\end{equation}
The non-local term is given by
\begin{equation}
    \chi_{\rm non-loc}^{(0)}( sr,\omega) = T\sum_\ep g^2_{\ep+\omega}g_\ep^2 \int_{\bm{q}} e^{i \bm{q}\bm{r}s}\int_{\bm{k}}\frac{\xi(k+q)\xi(k)}{(1-\xi(k+q)g_{\ep+\omega})(1-\xi(k)g_{\ep})}\;.
\end{equation}
We emphasize that at this point Eq.~\eqref{eq:chi_real_space} at $\bm{r}=0$ is only determined up to the $s^2$ terms, and thus the contribution from the non-local term $s^4 \chi_{\rm non-loc}^{(0)}(0,\omega) \sim s^4$ is effectively zero within our accuracy. In principle, in order to fully capture the $\sim s^4$ local effects in this integral we simply need to retain higher-order terms in the first two lines of Eq.~\eqref{eq:GG_approx_int}. However, for the present purposes we only focus on the leading order effects ($s^2$ in the local part and $s^4$ in the non-local part).

In the momentum space, we have
\begin{equation}
      \chi(\bm{Q},\omega)=\sum_{\bm{r}}e^{-i\bm{Q}\bm{r}} \chi(\bm{r},\omega)= \Big( \chi_{\rm loc}^{(0)}(\omega)+s^2 \chi_{\rm loc}^{(1)}(\omega)+...\Big) + s^2 \chi_{\rm non-loc}^{(0)}(\bm{Q}/s,\omega) +\mathcal{O}(s^4)\;,\label{eq:chi_zz_q/s}
\end{equation}
where we assumed that $\bm{Q}/s$ is fixed. Thus, the proper $s^2$ correction to the susceptibility is given by the combination $\chi_{\rm loc}^{(1)}(\omega) +\chi_{\rm non-loc}^{(0)}(\bm{Q}/s,\omega)$, where the first term is momentum-independent. 
\begin{equation}
  \chi_{\rm loc}^{(1)}(\omega) +\chi_{\rm non-loc}^{(0)}(\bm{Q}/s,\omega)= T\sum_\ep g_{\ep+\omega}g_\ep \int_{\bm{k}}\left[\frac{\xi(k+Q/s)\xi(k)}{(g_{\ep+\omega}^{-1}-\xi(k+Q/s))(g_{\ep}^{-1}-\xi(k))}  + \frac{g_{\ep}\xi^2(k)}{g_{\ep}^{-1}-\xi(k) }+\frac{g_{\ep+\omega}\xi^2(k)}{g_{\ep+\omega}^{-1}-\xi(k) }\right]\;.\label{eq:chi_chi_chi}
\end{equation}

Let us illustrate these general formulas for a free system with $g_\ep^{-1}=i\ep+\mu$, $1>\mu>0$, and $\xi(k)=1/(1+k^2)$. In the static limit, and for $T\ll |\mu|$, we have $\chi_{\rm loc}^{(0)}(0)\rightarrow 0$. The remaining static contribution is then easily evaluated
\begin{equation}
    \chi_{\rm loc}^{(1)}(0) +\chi_{\rm non-loc}^{(0)}(Q,0) =
-\frac{1}{4\pi \mu^2}-\frac{Q^2}{120\pi}\left(10+Q^2\right)+\frac{\sqrt{Q^2+4-\frac{4}{\mu }} \left[6+\mu  \left(Q^2+4\right) \left(\mu  \left(Q^2+4\right)+2\right)\right]}{120\pi \mu^2 Q}\theta(Q-2\sqrt{1/\mu -1})\label{eq:chi_0_test}
\end{equation}
where $k_F=\sqrt{1/\mu -1}$ is the Fermi momentum, $1/4\pi \mu^2$ is the compressibility, and the overall minus sign is reflecting our convention for $\chi$. At large $Q$, Eq.~\eqref{eq:chi_0_test} approaches $-(1-\mu^2)/(4\pi \mu^2)$, and it has a global minimum at $2\sqrt{1/\mu-1}$. We also note that in principle we could arrive at Eq.~\eqref{eq:chi_0_test} in a different way by first performing the Matsubara sum in free fermion loop before taking a small-$s$ limit. This way, large momentum part of the integral is automatically eliminated, so one can immediately pass to a continuous small$-s$ limit without introducing any subtractions. This leads to a more recognizable expression
\begin{equation}
    \chi_{\rm loc}^{(1)}(0) +\chi_{\rm non-loc}^{(0)}(\bm{Q}/s,0)  = \int_{|\bm{k}|<k_F}\mathcal{P}\left(\frac{1}{\xi(k)-\xi(k+Q/s)}+\frac{1}{\xi(k)-\xi(k-Q/s)}\right)\;,
\end{equation}
which agrees with Eq.~\eqref{eq:chi_0_test}.

Moving to the interacting case, it is instructive to consider the standard RPA calculation of the spin correlations in the modified Hubbard model Eq.~\eqref{eq:Hubbard_model} at small $U$. In this case, the spin vertex is $\Gamma^{(4,s)}_{\ep,\ep';\omega}=-\sum_{\sigma'} \tau^z_{\sigma'\sigma'} \langle c_{\sigma}(\tau)c_{\sigma'}(\tau_1) \bar{c}_{\sigma'}(\tau_2) \bar{c}_{\sigma} (\tau_0)\rangle_{0}\approx  - U g_{\ep+\omega}g_{\ep'+\omega}g_{\ep}g_{\ep'}+\mathcal{O}(U^2)$, where $g^{-1}_\ep=i\ep+\mu$. The RPA-dressed momentum-dependent vertex is then given by
\begin{equation}
\Gamma^{(4,s)}_{\ep,\ep';\omega}(\bm{Q}/s)= -\frac{U g_{\ep+\omega}g_{\ep'+\omega}g_{\ep}g_{\ep'}}{1+U \chi(\bm{Q},\omega)}\;,\quad (\text{RPA})
\end{equation}
where $\chi$ is given in Eq.~\eqref{eq:chi_0}. Using Eq.~\eqref{eq:chi_zz_q/s}, we obtain
\begin{equation}
\Gamma^{(4,s)}_{\ep,\ep';\omega}(\bm{Q}/s) \approx-\frac{U g_{\ep+\omega}g_{\ep'+\omega}g_{\ep}g_{\ep'}}{1+U \chi_{\rm loc}^{(0)}(\omega) +s^2 U \Big(\chi_{\rm loc}^{(1)}(\omega) +\chi_{\rm non-loc}^{(0)}(\bm{Q}/s,\omega)\Big) }\;,\quad (\text{RPA})\label{eq:RPA_Gamma_spin}
\end{equation}
Once again using Eq.~\eqref{eq:chi_0_test} for $\xi(k)=1/(1+k^2)$, $1>\mu>0$, and $T\rightarrow 0$, we find that the static $\omega=0$ RPA-dressed spin vertex in Eq.~\eqref{eq:RPA_Gamma_spin} exhibits a maximum at $Q=2s\sqrt{1/\mu-1}$, which is twice the Fermi momentum. This indicates a weak tendency towards an incommensurate spiral state. However, the actual Stoner-like instability is beyond the applicability of our perturbative treatment since it requires $U\sim 1/s^2 \gg 1$ where RPA is not controlled.  The actual strong coupling behavior of the spin susceptibility in the modified Hubbard model is discussed in Sec.~\ref{subsec:maintext_spinsusceptibility} of the main text, while some additional details are given in the Appendix \ref{sec:spin_appendix}.

\subsection{Weak-coupling self-energy correction}
As a next illustration, let us consider a perturbative $\mathcal{O}(U^2)$ self-energy correction in the modified Hubbard model at fixed  $k/s$
\begin{equation}
    \Sigma_\ep(k) = -U^2 T^2\sum\limits_{\ep'\omega} \sum\limits_{k',q\in {\rm {BZ}}}\mathcal{G}_{\ep+\omega}(k+q)\mathcal{G}_{\ep'+\omega}(k'+q)\mathcal{G}_{\ep'}(k'),\quad \quad \mathcal{G}_{\ep}(k) =\frac{1}{i\ep +\mu-t(k,s)},\quad g_\ep=\frac{1}{i\ep+\mu}
\end{equation}
Our goal is to construct a perturbative expansion of the momentum integrals in this diagram in powers of $s^2$. To this end, $\Sigma_\ep(k)$ can be exactly rewritten as follows
\begin{equation}
    \begin{aligned}
         \Sigma_\ep(k) &=-U^2 T^2\sum\limits_{\ep'\omega} \mathcal{G}_{\ep+\omega}(*)\mathcal{G}_{\ep'+\omega}(*)\mathcal{G}_{\ep'}(*)- U^2 T^2\sum\limits_{\ep'\omega} \mathcal{G}_{\ep+\omega}(*)\mathcal{G}_{\ep'+\omega}(*) \sum\limits_{k'\in {\rm {BZ}}}\delta\mathcal{G}_{\ep'}(k')\\
         &- U^2 T^2\sum\limits_{\ep'\omega} \mathcal{G}_{\ep+\omega}(*)\mathcal{G}_{\ep'}(*) \sum\limits_{k'\in {\rm {BZ}}}\delta\mathcal{G}_{\ep'+\omega}(k')- U^2 T^2\sum\limits_{\ep'\omega} \mathcal{G}_{\ep'+\omega}(*)\mathcal{G}_{\ep'}(*) \sum\limits_{k'\in {\rm {BZ}}}\delta\mathcal{G}_{\ep+\omega}(k')\\
         &-U^2 T^2\sum\limits_{\ep'\omega} \mathcal{G}_{\ep+\omega}(*)  \sum\limits_{k',q\in {\rm {BZ}}}\delta\mathcal{G}_{\ep'+\omega}(q)\delta\mathcal{G}_{\ep'}(k')-U^2 T^2\sum\limits_{\ep'\omega}  \mathcal{G}_{\ep'+\omega}(*)\sum\limits_{k',q\in {\rm {BZ}}}\delta\mathcal{G}_{\ep+\omega}(q)\delta\mathcal{G}_{\ep'}(k')\\
         &-U^2 T^2\sum\limits_{\ep'\omega} \mathcal{G}_{\ep'}(*)\sum\limits_{k',q\in {\rm {BZ}}}\delta\mathcal{G}_{\ep+\omega}(q)\delta\mathcal{G}_{\ep'+\omega}(k')-U^2 T^2\sum\limits_{\ep'\omega} \sum\limits_{k',q\in {\rm {BZ}}}\delta\mathcal{G}_{\ep+\omega}(k+q)\delta\mathcal{G}_{\ep'+\omega}(k'+q)\delta\mathcal{G}_{\ep'}(k')\;,
    \end{aligned}
\end{equation}
where we defined the following combination
\begin{equation}
    \delta \mathcal{G}_{\ep}(k)\equiv \mathcal{G}_{\ep}(k)-\mathcal{G}_{\ep}(*),\quad \quad\mathcal{G}_{\ep}(*)=\frac{1}{i\ep +\mu-t(*,s)}\;.
\end{equation}
One can easily see that all momentum integrals involving $\delta \mathcal{G}$ are  convergent at large momenta. Therefore, we can rescale variables as $k\rightarrow sk$, approximate $t(sk,s)\approx \xi(k)$, and extend the boundaries of integration regions from "$\rm{BZ}/s$" to infinity. This sequence of steps results in 
\begin{equation}
    \begin{aligned}
         \Sigma_\ep(k) &=-U^2 T^2\sum\limits_{\ep'\omega} g_{\ep+\omega}g_{\ep'+\omega}g_{\ep'}-s^2 U^2 T^2\sum\limits_{\ep'\omega} g_{\ep+\omega}g_{\ep'+\omega} g^2_{\ep'}\int_{\bm{k}'}\frac{\xi^2(k')}{g^{-1}_{\ep'}-\xi(k') }\\
         &-s^2 U^2 T^2\sum\limits_{\ep'\omega} g_{\ep+\omega}g_{\ep'} g^2_{\ep'+\omega}\int_{\bm{k}'}\frac{\xi^2(k')}{g^{-1}_{\ep'+\omega}-\xi(k') }-s^2 U^2 T^2\sum\limits_{\ep'\omega} g_{\ep'+\omega}g_{\ep'} g^2_{\ep+\omega}\int_{\bm{k}'}\frac{\xi^2(k')}{g^{-1}_{\ep+\omega}-\xi(k') }\\
    &+\mathcal{O}(s^4)
    \end{aligned}
\end{equation}
where we only retained the $s^2$ terms and used the identity
\begin{equation}
    g_\ep^2 t(*,s)+s^2\int_k\left[\frac{1}{g^{-1}_\ep -\xi(k)}-g_{\ep}\right]= s^2\int_k\left[\frac{1}{g^{-1}_\ep -\xi(k)}-g_{\ep} - g_\ep^2\xi(k)\right]=s^2g^2_{\ep}\int_{k}\frac{\xi^2(k)}{g^{-1}_{\ep}-\xi(k) }\;.
\end{equation}
After changing the variables in the frequency sums, we find
\begin{equation}
         \Sigma_\ep(k) =-U^2 T^2\sum\limits_{\ep'\omega} g_{\ep+\omega}g_{\ep'+\omega}g_{\ep'}-s^2 U^2 T^2\sum\limits_{\ep'\omega} g_{\ep+\omega}[g_{\ep'-\omega} +2g_{\ep'+\omega}]g^2_{\ep'}\int_{\bm{k}'}\frac{\xi^2(k')}{g^{-1}_{\ep'}-\xi(k') }+\mathcal{O}(s^4)
\end{equation}
We recognize the combination $   U^2g^2_\ep g^2_{\ep'} T\sum_{\omega'} g_{\ep+\omega'}[g_{\ep'-\omega'} +2g_{\ep'+\omega'}]$ as one of the $U^2$ term in the expansion of local four-point cumulant $\Gamma^{(4)}_{\ep,\ep',\omega=0}$. Similarly, for remaining $U^2$ self-energy correction (two nested tadpoles), we find
\begin{equation}
         \Sigma_\ep(k) =U^2 T^2\sum\limits_{\ep'\ep''} g_{\ep'}^2 g_{\ep''}e^{i\ep''0^+}+s^2 U^2 T^2\sum\limits_{\ep'\ep''} g_{\ep''}[g_{\ep''}+2 g_{\ep'}e^{i\ep'' 0^+}]g^2_{\ep'}\int_{\bm{k}'}\frac{\xi^2(k')}{g^{-1}_{\ep'}-\xi(k') }+\mathcal{O}(s^4)\;.
\end{equation}
After combining these results, we find a full $\mathcal{O}(s^2U^2)$ self-energy correction
\begin{equation}
\Sigma_{\ep} \approx s^2U^2  T\sum\limits_{\ep'}g_{\ep'}^2 \Big\{-T\sum\limits_{\omega} g_{\ep+\omega}[g_{\ep'-\omega}+2g_{\ep'+\omega}] +T\sum\limits_{\ep''} g^2_{\ep''}+2 g_{\ep'}T\sum\limits_{\ep''} g_{\ep''}e^{i\ep'' 0^+}\Big\} \int_{\bm{k}'}\frac{\xi^2(k')}{g^{-1}_{\ep'}-\xi(k') }\;.\label{eq:Sigma_approx_integral}
\end{equation}

Let us now compare this with the full $s^2$ self-energy to all orders in $U$, which is given by
\begin{equation}
    \Sigma_{\ep}(k)=s^2g_{\ep}^{-2}T \sum\limits_{\ep'}\Gamma^{(4)}_{\ep,\ep',\omega=0} \int_{\bm{k}'}\frac{\xi^2(k')}{g^{-1}_{\ep'}-\xi(k') }\;,\label{eq:G_test_222}
\end{equation}
where $g_\ep$ is now a full (i.e. to all orders in $U$) two-point function in the local problem. The agreement between Eq.~\eqref{eq:G_test_222} and Eq.~\eqref{eq:Sigma_approx_integral} can be easily verified by expanding $g^{-2}_\ep$ up to the order $\mathcal{O}(U)$ using a local self-energy $U T\sum_\ep g_\ep e^{i\ep 0^+}$, and also expanding $\Gamma^{(4)}$ up to $\mathcal{O}(U^2)$ as
\begin{equation}
    \Gamma^{(4)}_{\ep,\ep',\omega=0} \equiv \sum_{\sigma'}\Gamma_{\ep,\ep';\omega=0}^{(4)\sigma \sigma'\sigma'\sigma}\approx U g_\ep^2 g_{\ep'}^2+U^2 g_\ep^2 g_{\ep'}^2 \Big\{-T\sum\limits_{\omega} g_{\ep+\omega}[g_{\ep'-\omega}+2g_{\ep'+\omega}] +T\sum\limits_{\ep''} g^2_{\ep''}+2 (g_{\ep}+g_{\ep'})T\sum\limits_{\ep''} g_{\ep''}e^{i\ep'' 0^+}\Big\}\;,
\end{equation}
where here $g_\ep$ is a free local propagator.

\subsection{Order of limits in temperature-dependent integrals}

We close this Appendix with a simple example illustrating potential non-commutativity of the small-$s^2$ expansion with the low-temperature limit, $\beta\to\infty$. To this end, consider $
f(x)=x\tanh\frac{\beta x}{2}$.
In this case the result depends on the product $\beta s^2$. First, we directly apply Eq.~\eqref{eq:approx_1_k-integral}, and obtain
\begin{equation}
\sum_{\bm{k}\in{\rm BZ}}
t(\bm{k},s)
\tanh\left(\frac{\beta t(\bm{k},s)}{2}\right)
\approx
s^2\int_{\bm{k}}
\xi(k)
\tanh\left(\frac{\beta \xi(k)}{2}\right)\approx
s^2\int_{\bm{k}}\xi(k)
-
2s^2\int_{\bm{k}}\xi(k)e^{-\beta \xi(k)} ,
\label{eq:example_appendix_k_integral_0}
\end{equation}
where in the last step we kept only the leading exponentially small correction at $\beta \gg 1$. This procedure automatically corresponds to the order of limits when we take $s\rightarrow 0$ first, and $\beta\rightarrow \infty$ second. This can be seen by treating the constant shift $t(*,s)\sim s^2$ more carefully. Using
$\sum_{\bm{k}\in{\rm BZ}}t(\bm{k},s)=0$, we may write
\begin{equation}
\begin{aligned}
    &\sum\limits_{\bm{k} \in \rm{BZ}}t(\bm{k},s) \tanh\left(\frac{\beta t(\bm{k},s)}{2}\right) \approx   \underbrace{\sum\limits_{\bm{k} \in \rm{BZ}}t(\bm{k},s) }_{=0} - 2\sum\limits_{n=1}^{+\infty}(-1)^{n-1}\sum\limits_{\bm{k} \in \rm{BZ}}t(\bm{k},s) e^{-n\beta t(\bm{k},s)}\\
    &\approx \frac{-2t(*,s) e^{-\beta t(*,s)}}{1+e^{-\beta t(*,s)}} - 2s^2\sum\limits_{n=1}(-1)^{n-1} e^{-n\beta t(*,s)} \underbrace{\int_{\bm{k}} \left[(\xi(k)+t(*,s))e^{-n\beta \xi(k)}-t(*,s)\right]}_{\approx \int_{\bm{k}} \xi(k)e^{-n\beta \xi(k)}}\;.\label{eq:example_appendix_k_integral}
    \end{aligned}
\end{equation}
 In the second line, we applied the first line of Eq.~\eqref{eq:approx_1_k-integral} to each term in the series without assuming that $s^2$ is the smallest parameter. Finally, if we assume that $\beta t(*,s)\sim \beta s^2 $ is small (which again corresponds to taking the $s\rightarrow 0$ limit first and $\beta\rightarrow \infty$ second), then after using Eq.~\eqref{eq:t_*_xi} we find that Eq.~\eqref{eq:example_appendix_k_integral} agrees with Eq.~\eqref{eq:example_appendix_k_integral_0}. By contrast, when $\beta s^2$ is held fixed or becomes large, the shift $t(*,s)$ must be retained nonperturbatively, and the naive small-$s^2$ expansion of Eq.~\eqref{eq:approx_1_k-integral} is not uniform.

\section{One-loop Hubbard self-energy at arbitrary $\mu$ and $U/T$}\label{sec:Hubbard_U-loop_full}
In this section, we provide explicit expressions for the leading order one-loop self-energy in the modified Hubbard model, valid for any finite temperature beyond the small-$s^2$ ordering window and arbitrary finite chemical potential. First, the  correction to the local Green's function from the diagram in Fig.~\ref{fig:fig1}(c) reads as
\begin{equation}
\begin{aligned}
 g^2_{\ep} \Sigma_\ep&=  \frac{ s^2 U^2 e^{\beta \mu}\left(2+e^{\beta(\mu-U)}+2e^{\beta(2\mu-U)}+3e^{\beta\mu}\right)}{\left(1+2e^{\beta \mu} +e^{\beta (2\mu-U)}\right)^2(i\ep+\mu )^2 (i \ep+\mu -U)^2 }  \int_{\bm{k}} \xi^2(k) \;  \mathcal{G}^{(0)}_\varepsilon(sk) \\
&+\frac{s^2 U^2 \left(1-e^{\beta(2\mu-U)}\right)}{\left(1+2e^{\beta \mu} +e^{\beta (2\mu-U)}\right) \left(i\ep+\mu-U\right)^2} \;T\sum\limits_{\ep'} \frac{1}{\left(i\ep+i\ep'+2\mu-U\right)\left(i\ep'+\mu-U\right)^2}  \int_{\bm{k}} \xi^2(k) \;  \mathcal{G}^{(0)}_{\varepsilon'}(sk) \\
&+\frac{s^2}{(i\ep+\mu-U)^2}\left[x_0+\frac{U x_1}{i\ep+\mu}+\frac{U^2 x_2}{(i\ep+\mu)^2}\right]
\label{eq:Gamma_full_T}
\end{aligned}
\end{equation}
where $\Sigma_\ep$ is the corresponding self-energy, and
\begin{equation}
\begin{aligned}
    x_2&=\frac{ U \left(1+e^{\beta\mu}\right)}{\left(1+2e^{\beta \mu} +e^{\beta (2\mu-U)}\right) }  T\sum\limits_{\ep} \frac{1}{\left(i\ep+\mu\right)\left(i\ep+\mu-U\right)} \int_{\bm{k}} \xi(k)^2 \;  \mathcal{G}^{(0)}_{\ep}(sk)\;, \\
      x_1&= U  T\sum\limits_{\ep} \frac{\mathcal{X}^{(1)}_\ep}{\left(i\ep+\mu\right)^2\left(i\ep+\mu-U\right)^2} \int_{\bm{k}} \xi^2(k) \;  \mathcal{G}^{(0)}_{\ep}(sk)\;,\\
        x_0&= U  T\sum\limits_{\ep} \frac{\mathcal{X}^{(0)}_\ep}{\left(i\ep+\mu\right)^2\left(i\ep+\mu-U\right)^2}\int_{\bm{k}} \xi^2(k) \;  \mathcal{G}^{(0)}_{\ep}(sk)\;.
    \end{aligned}
\end{equation}
Here we also defined $\mathcal{X}^{(0,1)}_\ep$, which are quadratic polynomials in $\ep$:
\begin{equation}
\begin{aligned}
\mathcal{X}^{(0)}_\ep =&
(i\varepsilon+\mu)^2
+
\frac{
U e^{\beta U}(1+e^{\beta\mu})
\left[U-2(i\varepsilon+\mu)\right]
}{
e^{\beta U}+2e^{\beta(U+\mu)}+e^{2\beta\mu}
}
+
\frac{
\beta U e^{\beta(U+\mu)}
\left(e^{\beta U}+2e^{\beta\mu}+e^{2\beta\mu}\right)
(i\varepsilon+\mu)
\left[U-(i\varepsilon+\mu)\right]
}{
\left(e^{\beta U}+2e^{\beta(U+\mu)}+e^{2\beta\mu}\right)^2
}\;,
\\
\mathcal{X}^{(1)}_\ep =&-\frac{
e^{\beta U}\left(1+e^{\beta\mu}\right)
\left[
(i\varepsilon+\mu)^2+(i\varepsilon+\mu-U)^2
\right]
}{
e^{\beta U}+2e^{\beta(U+\mu)}+e^{2\beta\mu}
}
-\frac{
\beta U e^{\beta(U+\mu)}
\left(e^{\beta U}+2e^{\beta\mu}+e^{2\beta\mu}\right)
(i\varepsilon+\mu)
\left(U-i\varepsilon-\mu\right)
}{
\left(
e^{\beta U}+2e^{\beta(U+\mu)}+e^{2\beta\mu}
\right)^2
}\;.
\end{aligned}
\end{equation}
The first term in Eq.~\eqref{eq:Gamma_full_T} is a  "elastic" part of the self-energy, while the second line contains one non-factorizable summation over frequencies. Both lines will contribute to the imaginary part of the self-energy after the analytic continuation is performed.

Our goal is to perform the remaining frequency summations. The $0$th order Green's function $\mathcal{G}^{(0)}_{\ep}(sk)$ has its poles at
\begin{equation}
    z_{\pm}=\frac{1}{2}\left(U+\xi(k)-2\mu \pm \sqrt{\xi(k)^2+U^2-2\xi(k)U(1-n_0)}\right)
\end{equation}
so that in the complex frequency plane we have
\begin{equation}
    \mathcal{G}^{(0)}_{\ep}(sk)\equiv  \mathcal{G}^{(0)}_{-iz}(sk)= \frac{z+\mu+U(n_0/2-1)}{(z-z_+)(z-z_-)}\;,
\end{equation}
with $i\ep=z$. This allows us to perform all Matsubara summations in a closed form. Specifically, we find  
\begin{equation}
\begin{aligned}
     x_2&=\frac{ U \left(1+e^{\beta\mu}\right)}{\left(1+2e^{\beta \mu} +e^{\beta (2\mu-U)}\right) } \int_{\bm{k}} \xi^2(k)  \left(\frac{ (1-n_0/2)n_F(-\mu)}{(\mu+z_+)(\mu+z_-)}+\frac{ n_0 \;n_F(U-\mu)}{2(U-\mu-z_+)(U-\mu-z_-)} \right.\\
     &+\left.\sum\limits_{l=\pm }\frac{(z_l+\mu +U (n_0/2-1))n_F(z_{l})}{(z_l+\mu)(z_l+\mu-U) (z_l-z_{-l})}\right)\;,
     \end{aligned}
\end{equation}
where $n_F(z)=1/(e^{\beta z}+1)$, and, as before
\begin{equation}\label{eq:n_single_site}
    n_0= \frac{2(e^{\mu\beta}+e^{(2\mu-U)\beta})}{1+2e^{\mu\beta}+e^{(2\mu-U)\beta}}\;,
\end{equation}
which is the average occupation in a single-site problem. Similarly,
\begin{equation}
    \begin{aligned}
        x_{i=0,1}
        &=U\sum\limits_{l=\pm} \int_{\bm{k}}\xi^2(k)\;\frac{(z_l+\mu+U(n_0/2-1))\mathcal{X}^{(i)}_{-iz_l}n_F(z_l)}{\left(z_l+\mu\right)^2\left(z_l+\mu-U\right)^2 (z_l-z_{-l})} +U\int_{\bm{k}}\xi^2(k) \; \frac{d}{dz}\left(\frac{\mathcal{X}^{(i)}_{-iz} n_F(z)}{\left(z+\mu-U\right)^2}  \mathcal{G}^{(0)}_{-iz}(sk)\right)_{z\rightarrow -\mu}\\&+ U\int_{\bm{k}}\xi^2(k)\; \frac{d}{dz}\left(\frac{\mathcal{X}^{(i)}_{-iz} n_F(z)} {\left(z+\mu\right)^2}   \mathcal{G}^{(0)}_{-iz}(sk)\right)_{z\rightarrow U-\mu}\;.
    \end{aligned}
\end{equation}
The remaining Matsubara sum in the second line of Eq.~\eqref{eq:Gamma_full_T} can be evaluated as follows
\begin{equation}
    \begin{aligned}
        &T\sum\limits_{\ep'} \frac{1}{\left(i\ep+i\ep'+2\mu-U\right)\left(i\ep'+\mu-U\right)^2}  \int_{\bm{k}}\xi^2(k) \;  \mathcal{G}_{\ep'}^{(0)}(sk)= -\frac{n_B(U-2\mu)}{\left(i\ep+\mu\right)^2}  \int_{\bm{k}}\xi^2(k)  \;  \mathcal{G}^{(0)}_{2i\mu-iU-\ep}(sk)+\\
        &+\sum\limits_{l=\pm} \int_{\bm{k}}\xi^2(k)\frac{(z_l+\mu+U(n_0/2-1)) n_F(z_l)}{\left(i\ep+z_l+2\mu-U\right)\left(z_l+\mu-U\right)^2 (z_l-z_{-l})}   \\
        &+\int_{\bm{k}}\xi^2(k) \left\{ -\frac{n_F(U-\mu) }{\left(i\ep+\mu\right)^2}\mathcal{G}^{(0)}_{-i(U-\mu)}(sk) +\frac{1}{\left(i\ep+\mu\right)} \frac{d}{dz}\left( n_F(z) \mathcal{G}^{(0)}_{-iz}(sk)\right)_{z\rightarrow U-\mu}\right\}\;,
    \end{aligned}
\end{equation}
where $n_B(z)=1/(e^{\beta z}-1)$ is the Bose distribution. After collecting all results, we obtain the following final expression for the correction in Eq.~\eqref{eq:Gamma_full_T}
\begin{equation}
\begin{aligned}
g^2_\ep \Sigma_{\ep} &=  \frac{ s^2 U^2 e^{\beta \mu}\left(2+e^{\beta(\mu-U)}+2e^{\beta(2\mu-U)}+3e^{\beta\mu}\right)}{\left(1+2e^{\beta \mu} +e^{\beta (2\mu-U)}\right)^2(i\ep+\mu )^2 (i \ep+\mu -U)^2 } \int_{\bm{k}}\xi^2(k) \;  \mathcal{G}^{(0)}_\ep(sk)\\
&+\frac{s^2 U^2 \left(1-e^{\beta(2\mu-U)}\right)}{\left(1+2e^{\beta \mu} +e^{\beta (2\mu-U)}\right) \left(1-e^{\beta(U-2\mu)}\right)\left(i\ep+\mu\right)^2\left(i\ep+\mu-U\right)^2} \int_{\bm{k}}\xi^2(k) \; \mathcal{G}^{(0)}_{-i(U-2\mu)-\ep} (sk)\\
&+\frac{s^2}{(i\ep+\mu-U)^2}\left[x_0+\frac{U (x_1+\eta_1)}{i\ep+\mu}+\frac{U^2 (x_2+\eta_2)}{(i\ep+\mu)^2}\right] \\
&+\frac{s^2 U^2 \left(1-e^{\beta(2\mu-U)}\right)}{\left(1+2e^{\beta \mu} +e^{\beta (2\mu-U)}\right) \left(i\ep+\mu-U\right)^2}\sum\limits_{l=\pm} \int_{\bm{k}}\xi^2(k)\frac{(z_l+\mu+U(n_0/2-1)) n_F(z_l)}{\left(i\ep+z_l+2\mu-U\right)\left(z_l+\mu-U\right)^2 (z_l-z_{-l})} 
\label{eq:Gamma_full_T_2}
\end{aligned}
\end{equation}
and 
\begin{equation}
\begin{aligned}
    \eta_1 &= \frac{U \left(1-e^{\beta(2\mu-U)}\right)}{\left(1+2e^{\beta \mu} +e^{\beta (2\mu-U)}\right) }\int_{\bm{k}}\xi^2(k) \frac{d}{dz}\left( n_F(z) \mathcal{G}^{(0)}_{-iz}(sk)\right)_{z\rightarrow U-\mu}\;,\\
     \eta_2 &= \frac{  e^{\beta(2\mu-U)}-1  }{\left(1+2e^{\beta \mu} +e^{\beta (2\mu-U)}\right)\left(1+e^{\beta(U-\mu)}\right) } \int_{\bm{k}}\xi^2(k) \mathcal{G}^{(0)}_{-i(U-\mu)} (sk)\;.
    \end{aligned}
\end{equation}
The resulting Green's function then acquires the following form 
\begin{equation}\label{eq:G_with_Upsilon}
[\mathcal{G}_\ep(k)]^{-1}=i\ep+\mu-\frac{n_0U}{2}-\frac{n_0(2-n_0)U^2/4}{i\ep+\mu+U(n_0/2-1) +s^2\Upsilon_\ep}-t(k,s)\;,
\end{equation}
where we defined
\begin{equation}
     \begin{aligned}
\Upsilon_\ep&=-\frac{4 e^{\beta \mu}\left(2+e^{\beta(\mu-U)}+2e^{\beta(2\mu-U)}+3e^{\beta\mu}\right)}{n_0(2-n_0)\left(1+2e^{\beta \mu} +e^{\beta (2\mu-U)}\right)^2} \int_{\bm{k}}\xi^2(k) \;  \mathcal{G}^{(0)}_\ep(sk)\\
&+\frac{4e^{\beta(2\mu-U)}}{n_0(2-n_0)\left(1+2e^{\beta \mu} +e^{\beta (2\mu-U)}\right) } \int_{\bm{k}}\xi^2(k) \;  \mathcal{G}^{(0)}_{-i(U-2\mu)-\ep}(sk)\\
&-\frac{ 4\left(1-e^{\beta(2\mu-U)}\right)(i\ep+\mu)^2}{n_0(2-n_0)\left(1+2e^{\beta \mu} +e^{\beta (2\mu-U)}\right) }\sum\limits_{l=\pm} \int_{\bm{k}}\xi^2(k) \frac{(z_l+\mu+U(n_0/2-1)) n_F(z_l)}{\left(i\ep+z_l+2\mu-U\right)\left(z_l+\mu-U\right)^2 (z_l-z_{-l})}\\
&-\frac{4}{n_0(2-n_0) U^2}\left[x_0(i\ep+\mu)^2+U (x_1+\eta_1)(i\ep+\mu)+U^2 (x_2+\eta_2)\right] \;.\label{eq:Upsilon_final_T}
  \end{aligned}
\end{equation}
The last line contributes only to the real part after analytic continuation to real frequencies is performed. We also note that the integration over $\bm{k}$ in the last two lines has to be performed together in order to ensure convergence.

The form of the self-energy correction in Eq.~\eqref{eq:G_with_Upsilon}, written in terms of $\Upsilon_\ep$, is motivated by the fact that $\Sigma_\ep$ contains a second-order pole, $1/(i\ep+\mu+U(n_0/2-1))^2$. It is therefore natural to regard this contribution as a perturbative shift of the first-order pole, $1/(i\ep+\mu+U(n_0/2-1))$, that appears in the denominator of the zeroth-order Green's function. We emphasize that, at the level of a strict $O(s^2)$ expansion, the two representations of the Green's function (one written in terms of $\Sigma_\ep$ and the other in terms of $\Upsilon_\ep$) are equivalent. They differ only by terms beyond the order retained, regardless of whether the higher-order corrections ultimately organize into a geometric series. The representation in Eq.~\eqref{eq:G_with_Upsilon}, however, has an important conceptual advantage: it preserves analyticity in the upper half of the complex frequency plane, with $z=i\varepsilon$.

\section{DC conductivity in the modified Hubbard model at the leading order in $1/s^2$}
\label{app:conductivity_small_s}
In this section, we evaluate the leading-order contribution to the DC conductivity. Since vertex corrections do not appear at this order, the conductivity is given by a single-loop diagram with Green’s functions dressed by self-energy corrections:
\begin{equation}
\sigma
=\pi\int_{-\infty}^{\infty} d\Omega\,
\Big(-\frac{\partial n_F(\Omega)}{\partial\Omega}\Big)\,
\int_{\bm{k}} (\xi'(k))^2\,A^2_\Omega(s\mathbf{k}) \;.
\label{eq:kubo_bubble}
\end{equation}
Here $\xi'(k)\equiv v_{k}$ is the velocity, the spectral function is
\begin{equation}
A_\Omega(\mathbf{k})\equiv- \frac{1}{\pi}\frac{\operatorname{Im}\Sigma^R_\Omega}
{\big[(g_\Omega^R)^{-1}-\xi(k/s)-\operatorname{Re}\Sigma^R_\Omega\big]^2+(\operatorname{Im}\Sigma^R_\Omega)^2}\;,
\label{eq:GR_A_def}
\end{equation}
and $g_\Omega^R$ is the retarded single-site Green's function. Since at this order the self-energy is $\bm{k}$-independent, it is natural to introduce the velocity-resolved density of states 
\begin{equation}
\Phi(\xi)\equiv \int_{\bm{k}} (\xi'(k))^2\,\delta(\xi-\xi(k)) =\frac{\alpha}{\pi} \xi (1-\xi^{1/\alpha})\theta(\xi)\theta(1-\xi),
\label{eq:transport_function}
\end{equation}
where the second equality is for $\xi(k)=1/(1+k^2)^\alpha$, $\alpha\geq 1$.  Using Eq.~\eqref{eq:transport_function}, we can re-write the momentum integral in Eq.~\eqref{eq:kubo_bubble} as $\int_{\bm{k}}(\xi'(k))^2 f(\xi(k))=\int d\xi\,\Phi(\xi)f(\xi)$.

Next, we note that $\operatorname{Im}\Sigma^R_\Omega$ is parametrically small, so the
$\xi$-integral is dominated by a narrow squared Lorentzian. Using the standard
identity $\int d\xi\;\Gamma^2/(\xi^2+\Gamma^2)^2=\pi/(2\Gamma)$ and assuming that $\Phi(\xi)$ varies smooth near $\xi=(g_\Omega^R)^{-1}-\operatorname{Re}\Sigma^R_\Omega$ on the scale of $\operatorname{Im}\Sigma^R_\Omega$ (we will revisit  this assumption later), we find
\begin{equation}
\int_{\bm{k}} (\xi'(k))^2\,A^2_\Omega(s\mathbf{k})\approx -\frac{1}{2\pi \operatorname{Im}\Sigma^R_\Omega} \;\Phi\left((g_\Omega^R)^{-1}-\operatorname{Re}\Sigma^R_\Omega\right)\theta(|\operatorname{Im}\Sigma^R_\Omega|)\;.
\end{equation}
Substituting into Eq.~\eqref{eq:kubo_bubble} yields 
\begin{equation}
\sigma_{}
\approx
\frac{1}{2}\int_{-\infty}^{\infty} d\Omega\,
\frac{\partial n_F(\Omega)}{\partial\Omega}\,
\frac{\Phi\left((g_\Omega^R)^{-1}-\operatorname{Re}\Sigma^R_\Omega\right)}{\operatorname{Im}\Sigma^R_\Omega}\theta(|\operatorname{Im}\Sigma^R_\Omega|)\;.
\label{eq:sigma_drude_like}
\end{equation}

After further expanding the smooth function $\Phi$ in the $O(s^2)$ shift originating from $\operatorname{Re}\Sigma^R_\Omega$, we obtain the leading $\sim 1/s^2$ term
\begin{equation}
\sigma_{}
\approx
\frac{1}{2}\int_{-\infty}^{\infty} d\Omega\,
\frac{\partial n_F(\Omega)}{\partial\Omega}\,
\frac{\Phi\left((g_\Omega^R)^{-1}\right)}{\operatorname{Im}\Sigma^R_\Omega}\theta(|\operatorname{Im}\Sigma^R_\Omega|) +\;O(s^0).
\label{eq:sigma_leading_fixed_mu}
\end{equation}
When the density is fixed, the chemical potential should be determined from the integrated Green's function.

\subsection{Chemical potential at fixed density}\label{sec:appendix_chemical_fixed}
The physical average density $n_{\rm phys}$ is expressed through the Green's function as
\begin{equation}
    n_{\rm phys} = 2s^2T\sum_\ep e^{i\ep 0^+}\int_{\bm{k}} \left[\mathcal{G}_\ep(sk)-\mathcal{G}_\ep(k\rightarrow *) \right] +2T\sum_\ep e^{i\ep 0^+}\mathcal{G}_\ep(k\rightarrow *) \;.
\end{equation}
After performing analytic continuation, we find
\begin{equation}
    n_{\rm phys} = 2s^2 \int d\Omega\;  n_F(\Omega)\int_{\bm{k}} \left[A_\Omega(sk)-A_\Omega(k\rightarrow *) \right] +2\int d\Omega A_\Omega(k\rightarrow *) n_F(\Omega)\;.\label{eq:n_phys}
\end{equation}
At the $0$th order in $s$, $n_{\rm phys}$ coincides with the single-site expression in Eq.~\eqref{eq:n_single_site}. Beyond leading order, the density--chemical potential relation admits an expansion
\begin{equation}
n_{\rm phys}=n_0(\mu)+s^2 n_1(\mu)+O(s^4)\;,
\label{eq:n_expand}
\end{equation}
where $n_0(\mu)$ is given in Eq.~\eqref{eq:n_single_site}. In general, there are three effects contributing to $n_1(\mu)$: (i) the first term in Eq.~\eqref{eq:n_phys} which is already of the order of $s^2$ (and thus only the $0$th order spectral function should be used in the integral); (ii) a constant shift in $\xi(k)$, i.e. $\xi(k)\rightarrow \xi(k/s)-s^2 \int_{\bm{p}}\xi(p)$, in the second term in Eq.~\eqref{eq:n_phys} (see discussion in Sec.~\ref{app:small_s_integrals}); and (iii) a $s^2$ self-energy correction to the second term in Eq.~\eqref{eq:n_phys}.

Solving perturbatively at fixed $n_{\rm phys}$ gives
\begin{equation}
\mu=\mu_0(n_{\rm phys})+s^2\mu_1(n_{\rm phys})+O(s^4),
\qquad
\mu_1=-\frac{n_1(\mu_0)}{\partial_\mu n_0(\mu_0)}\;,
\label{eq:mu_fixed_n}
\end{equation}
where $\mu_0$ is the solution of Eq.~\eqref{eq:n_single_site} with $n_0$ replaced by $n_{\rm phys}$.

Crucially, Eq.~\eqref{eq:mu_fixed_n} can eventually break down once the first term $n_0(\mu)$ is small (e.g. at low temperatures), and so the $s^2$ correction in Eq.~\eqref{eq:n_expand} becomes the main effect. In this case, both terms must be retained when performing the inversion. Physically, this corresponds to the fact that the chemical potential eventually saturates to a constant value at low temperatures which is sensitive to the $s^2$ effects (this becomes particularly important at small densities of doped carriers of the order of $s^2$ and below). Finally, if the second term in Eq.~\eqref{eq:n_expand} becomes of the order of unity, then the full expression in Eq.~\eqref{eq:n_phys} must be used.


\section{Large $U$ limit in the modified Hubbard model}\label{sec:large-U}
In this section we provide additional details regarding the $\mathcal{O}(s^2)$ self-energy in the modified Hubbard model in the large $U$ limit.

First, let us consider the hole doped case. The corresponding Green's function can be obtained as the $U\rightarrow \infty$ limit of the general expression found in Eqs.\eqref{eq:G_with_Upsilon},\eqref{eq:Upsilon_final_T}, while keeping the chemical potential $\mu$ fixed. In this case, Eq.~\eqref{eq:G_with_Upsilon} simplifies to
\begin{equation}
    [\mathcal{G}_\ep(k)]^{-1}=(i\ep+\mu)\left(1+\frac{1}{1+e^{-\beta \mu }}\right)-t(k,s)+\frac{s^2 \Upsilon^{}_\ep}{1+e^{-\beta \mu }}\;,\label{eq:G_s^2_large_U}
\end{equation}
or equivalently
\begin{equation}
    \mathcal{G}_\ep(k)=\frac{1-n_0/2}{i\ep+\mu -(1-n_0/2)t(k,s) +s^2\Upsilon_\ep  n_0/2}\;,\quad \quad n_0=\frac{2e^{\beta \mu}}{1+2e^{\beta \mu}}\;.\label{eq:G_LHB}
\end{equation}
The full expression for $\Upsilon_\ep$ in Eq.~\eqref{eq:Upsilon_final_T} is also reduced dramatically: the second and the third line are both exponentially suppressed as $e^{-\beta U}\ll 1$, and thus we find
\begin{equation}
     \begin{aligned}
   & \Upsilon_\ep =-\frac{2+3e^{\beta\mu}}{1+e^{\beta\mu}} \int_{\bm{k}}\frac{\xi^2(k)}{(i\ep+\mu)\big(1+\frac{1}{1+e^{-\beta \mu }}\big)-\xi(k)}  +(i\ep+\mu )\Upsilon_1+\Upsilon_2\;,
   \label{eq:Upsilon_large_U_1}
  \end{aligned}
\end{equation}
where we defined two functions of $\beta$ and $\mu$
\begin{equation}
    \begin{aligned}
        \Upsilon_1&= \frac{1}{(1{+}e^{\beta  \mu })^2}\int_{\bm{k}}\left( [1+2 (\beta  \xi(k) {+}2) (1{+}e^{\beta \mu})e^{\beta\mu} ] e^{\frac{-\beta  \xi(k)  (e^{\beta  \mu }+1)}{2 e^{\beta  \mu }+1}}{-}e^{\beta  \mu } (4{-}\beta  \xi(k)){-}4 e^{2 \beta  \mu }{+}\beta  \xi(k) {-}1\right)n_F\left(\mu{-} \frac{\xi(k)  (e^{\beta  \mu }{+}1)}{2 e^{\beta  \mu }{+}1}\right)\\
        \Upsilon_2&=\frac{2 e^{\beta  \mu }+1}{e^{\beta  \mu }+1}\int_{\bm{k}} \xi(k)\Big(e^{\frac{-\beta  \xi(k) (e^{\beta  \mu }+1)}{2 e^{\beta  \mu }+1}}-1\Big) n_F\left(\mu - \frac{\xi(k)  (e^{\beta  \mu }+1)}{2 e^{\beta  \mu }+1}\right)\;.\label{eq:Upsilon_hole_doped_12}
    \end{aligned}
\end{equation}

Only the first term in Eq.~\eqref{eq:Upsilon_large_U_1} contributes to the imaginary part after the analytic continuation is performed. The second and third terms only provide a temperature-dependent shift of the quasiparticle residue and the chemical potential. Eq.~\eqref{eq:G_LHB} can also be rewritten in the form of Eq.~\eqref{eq:G_LHB_main} with the proper self-energy $\Sigma_\ep \equiv -s^2n_0 \Upsilon_\ep/(2-n_0)$. The momentum integral in Eq.~\eqref{eq:Upsilon_large_U_1} can be evaluated in a closed form for $\xi(k)=1/(1+k^2)^\alpha$ using
\begin{equation}
    \int_{\bm{k}}\frac{\xi^2(k)}{i\ep+\mu-(1-n_0/2)\xi(k)}= 
\frac{1}{4\pi(2\alpha-1) (i\ep+\mu)}\,
{}_2F_1\!\left(
1,\,
2-\frac{1}{\alpha};\,
3-\frac{1}{\alpha};\,
\frac{1-n_0/2}{i\ep+\mu}
\right).\label{eq:int_Sigma_full}
\end{equation}
where ${}_2F_1$ is the hypergeometric function. We also note that for $\alpha=3/2$ this expression can be expressed in terms of elementary functions.

As was already emphasized in Sec.~\ref{sec:Hubbard}, the same results could  be obtained by first imposing the Gutzwiller no-double-occupancy projection directly at the level of the Hamiltonian, Eq.~\eqref{eq:Uinf_holes_main}, which is of the form Eq.~\eqref{eq:model_general} with $\gamma_{i\sigma}=(1-n_{i\bar{\sigma}}) c_{i\sigma }$. In particular, $g_\ep=(1-n_0/2)/(i\ep+\mu)$, and the 4-point vertex in the projected theory simplifies to
\begin{equation}
\begin{aligned}\Gamma_{\ep,\ep';\omega=0}^{(4)}&\equiv \sum_{\sigma'}\Gamma_{\ep,\ep';\omega=0}^{(4)\sigma \sigma'\sigma'\sigma} = \frac{n_0(4-n_0)\beta \delta_{\ep,\ep'}}{4(i\ep+\mu)^2}\\ &+\frac{\mu  (n_0 (\beta  \mu  n_0+4)-8)
+\beta  (n_0-2) n_0 (i \ep'+\mu) (i \ep+\mu)+i (\ep'+\ep) (n_0 (\beta  \mu  n_0+2)-4) -\beta  n_0^2 \ep' \ep}{4 (i\ep'+\mu )^2 (i\ep+\mu )^2}\;.   \label{eq:Gamma_U_inft}
    \end{aligned}
\end{equation}
The first term is the static local-moment contribution, proportional to
$\beta n_0\delta_{\varepsilon,\varepsilon'}$, while the remaining terms are regular in the low-temperature local-moment regime. Substituting Eq.~\eqref{eq:Gamma_U_inft} into the general one-loop expression Eq.~\eqref{eq:Sigma_Hubbard_main} gives the $O(s^2)$ self-energy discussed in the main text.



Next, we consider the $U\rightarrow \infty$ limit on the electron-doped side of the modified Hubbard model. To this end, we shift the chemical potential as 
$\mu=U-\mu'$, where $\mu'$ is fixed, and take the large $U$ limit. Using Eq.~\eqref{eq:G_with_Upsilon} we obtain the following Green's function 
\begin{equation}
    \mathcal{G}_\ep(k)=\frac{n_0/2}{i\ep-\mu'+ t(k,s) n_0/2+s^2(1-n_0/2)\Upsilon^{}_\ep},\quad \quad n_0= 2- \frac{2e^{\beta \mu'}}{1+2e^{\beta \mu'}}\;.\label{eq:G_electron_side}
\end{equation}
We note that $\mu'<0$ corresponds to doping away from the flat part of the band. The self-energy in this limit can be directly obtained either from Eq.~\eqref{eq:Upsilon_final_T}, or using the particle-hole transformation resulting in the following rule
\begin{equation}
     \Upsilon_{\ep} \equiv \lim\limits_{U\rightarrow \infty}  \Upsilon_{\ep}|_{\mu=U-\mu'} = -  \Upsilon_{-\ep}^{(\rm hole)}|_{\substack{\xi(k)\rightarrow -\xi(k)\\\mu\rightarrow \mu'}}\;.
\end{equation}
Here our notation $\xi(k)\rightarrow -\xi(k)$ implies that we should change the sign of the dispersion in all momentum integrals. Explicitly, we obtain
\begin{equation}
    \Upsilon_{\ep}= -\frac{2+3e^{\beta \mu'}}{1+2e^{\beta \mu'}}\int_{\bm{k}}\frac{\xi^2(k)}{i\ep -\mu'+(n_0/2)\xi(k)} +(i\ep-\mu')\Upsilon_1|_{\substack{\xi(k)\rightarrow -\xi(k)\\\mu\rightarrow \mu'}}-\Upsilon_2|_{\substack{\xi(k)\rightarrow -\xi(k)\\\mu\rightarrow \mu'}}\;.
\end{equation}
Here $\Upsilon_{1,2}$ were defined in Eq.~\eqref{eq:Upsilon_hole_doped_12}. After defining the proper self-energy $\Sigma_\ep$, we arrive at Eq.~\eqref{eq:Upsilon_electron_large_U_1_main} of the main text. The associated physical density is then given in Eq.~\eqref{eq:n_for_fixed_mu_UHB_main}.

\section{Spin susceptibility in the modified Hubbard model}\label{sec:spin_appendix}

In this section we discuss the treatment of the spin correlation function in the modified Hubbard model that does not require the smallness of $U$. To this end, we note that in most cases of interest, the spin vertex at $s=0$ has the following factorizable structure
\begin{equation}
\Gamma^{(4,s)}_{\ep,\ep';\omega} = \beta \delta_{\ep,\ep'} \bar{\gamma}^{(0)}_{\ep ;\omega}+\sum\limits_{i}\bar{\gamma}^{(i)}_{\ep;\omega} \gamma_{\ep';\omega}^{(i)} \;,
\end{equation}
where the sum runs over a finite number of terms, and $\gamma_{\ep;\omega}, \bar{\gamma}_{\ep;\omega}$ are some frequency-dependent functions. At the leading order in $s^2$ this vertex acquires non-local corrections, which could be resummed by means of the following ladder equation
\begin{equation}
   \Gamma^{(4,s)}_{\ep,\ep'';\omega}(\bm{Q})=\Gamma^{(4,s)}_{\ep,\ep'';\omega}+ s^2T\sum_{\ep'}\Gamma^{(4,s)}_{\ep,\ep';\omega}  \Pi_{\ep';\omega}(\bm{Q}) \Gamma^{(4,s)}_{\ep',\ep'';\omega}(\bm{Q})\;.
\end{equation}
Here $\Gamma^{(4,s)}_{\ep,\ep'';\omega}(\bm{Q})$ is the non-local vertex dressed by $s^2$ corrections, and $ \Pi$ is 
\begin{equation}
   \Pi_{\ep;\omega}(\bm{Q})= \int_{\bm{k}}\frac{\xi(k+Q/s)\xi(k)}{(1-g_{\ep+\omega}\xi(k+Q/s))(1-g_{\ep}\xi(k))} \;.
\end{equation}
In terms of $\gamma_\ep, \bar{\gamma}_\ep$, the ladder equation is reduced to the following coupled equations 
\begin{equation}  \Gamma^{(4,s)}_{\ep,\ep'';\omega}(\bm{Q}) =
   \Gamma^{(4,s)}_{\ep,\ep'';\omega} + s^2\Gamma^{(4,s)}_{\ep,\ep'';\omega}\Pi_{\ep;\omega}(\bm{Q})\bar{\gamma}_{\ep;\omega}^{(0)}+s^2\sum\limits_{i} \bar{\gamma}_{\ep;\omega}^{(i)} A^{(i)}_{\ep'';\omega}(\bm{Q})\;,\quad\;A^{(i)}_{\ep'';\omega}(\bm{Q})=T\sum\limits_\ep  \gamma^{(i)}_{\ep;\omega} \Pi_{\ep;\omega}(\bm{Q})\Gamma^{(4,s)}_{\ep,\ep'';\omega} \;.\label{eq:Fhi_ep_1}
\end{equation}
 The solution of Eq.~\eqref{eq:Fhi_ep_1} acquires a simple form
\begin{equation}
   \Gamma^{(4,s)}_{\ep,\ep'';\omega}(\bm{Q})= \frac{  1 }{1-s^2\Pi_{\ep;\omega}(\bm{Q}) \bar{\gamma}_{\ep;\omega}^{(0)}}\left(\Gamma^{(4,s)}_{\ep,\ep'';\omega}+s^2\sum_{i} \bar{\gamma}_{\ep;\omega}^{(i)} A^{(i)}_{\ep'';\omega}(\bm{Q})\right).\label{eq:F_sol}
\end{equation}
The coefficients $A^{(i)}$ can be determined via the following algebraic system of equations
\begin{equation}
     A^{(i)}=T\sum\limits_\ep  \frac{  \gamma^{(i)}_\ep \Pi_{\ep}\Gamma_{\ep} }{1-s^2\Pi_\ep \bar{\gamma}_\ep^{(0)}}+\sum_{j} T\sum\limits_\ep  \frac{ s^2 \gamma^{(i)}_\ep \Pi_{\ep}\bar{\gamma}_\ep^{(j)} }{1-s^2\Pi_\ep \bar{\gamma}_\ep^{(0)}} A^{(j)}  \;.
\end{equation}
Here we suppressed all unimportant indices for clarity. The determinant of this system can be expanded at the leading order in $s^2$ as
\begin{equation}
    \det(...) \approx 1 - s^2 T\sum\limits_\ep  \sum_i \gamma^{(i)}_\ep \Pi_{\ep}(\bm{Q})\bar{\gamma}_\ep^{(i)} +\mathcal{O}(s^4)\;.
\end{equation}

\subsection{Hole doped case in the large-$U$ limit}\label{sec:spin_Hubbard_U_inf}
Let us now apply this general formula to the large $U$ limit. We find 
\begin{equation}
    \Gamma^{(4,s)}_{\ep,\ep';\omega} = -\frac{n_0 \beta \delta_{\omega,0}}{2(i\ep+\mu)(i\ep'+\mu)} -\frac{n_0^2 \beta \delta_{\ep,\ep'}}{4(i\ep+\mu)(i\ep+i\omega+\mu)} +
    \frac{(2-n_0)(i\ep+i\ep'+i\omega+2\mu)}{2(i\ep+\mu)(i\ep'+\mu)(i\ep+i\omega+\mu)(i\ep'+i\omega+\mu)}\;,\label{eq:Gamma_s_s112}
\end{equation}
where $n_0=2e^{\beta\mu}/(1+2e^{\beta\mu})$, and $g_{\ep}= (1-n_0/2)/(i\ep+\mu)$ (cf. Eq.~\eqref{eq:G_s^2_large_U}).We are interested in a static limit $\omega=0$ and  temperatures much lower than the bandwidth $\beta\gg 1$.

The spin susceptibility $\chi^{S_zS_z}\equiv \chi$ of a single site problem is $\chi=n_0\beta/4$. The $s^2$ correction from the 6-point vertex $\Gamma^{(6,s)}$ (depicted diagrammatically in Fig.~\ref{fig:spin_main}(b)) becomes
\begin{equation}
\delta\chi^{(1)}=s^2T\sum\limits_{\ep'}\Big[ \frac{\beta ^2 (1-n_0) n_0}{4 (i \ep'+\mu )}+\frac{\beta  n_0}{2 (i\ep'+\mu  )^2}+\frac{2 (n_0-2)}{4 (i \ep'+\mu )^3}\Big] \int_k \xi^2(k) \mathcal{G}^{(0)}_{\ep'}(sk)\;.
\end{equation}
Performing the Matsubara sums, we find
\begin{equation}
\begin{aligned}
\delta\chi^{(1)}
&=
s^2\int_k
\Bigg[
\left(
\frac{\beta^2 n_0(1-n_0)}{4}\,\xi(k)
+\frac{\beta n_0}{2-n_0}
-\frac{2}{(2-n_0)\,\xi(k)}
\right)\Big(n_F\!\left((1-n_0/2)\,\xi(k)-\mu\right)-n_F(-\mu)\Big)
\\
&\qquad\qquad
-\frac{2\beta n_0(1-n_0)}{(2-n_0)^2}
+\frac{\beta^2 n_0(1-n_0)}{2(2-n_0)}\,\xi(k)
\Bigg],\label{eq:Chi_s2_01}
\end{aligned}
\end{equation}
For $1/2>\mu>0$ and $T\rightarrow 0$ we have $n_0\rightarrow 1$ up to exponential corrections $\sim e^{-\beta\mu}$, and thus the leading behavior is accompanied by the $\beta\gg 1$ factor
\begin{equation}
\delta\chi^{(1)}
=
-\beta s^2\int_k \theta\!\left(\xi(k)/2-\mu\right)
+2s^2\int_k \frac{1}{\xi(k)}\,\theta\!\left(\xi(k)/2-\mu\right)
+O\!\left(e^{-\beta\mu}\right).
\end{equation}
For the dispersion $\xi=1/(1+k^2)^\alpha$ we find
\begin{equation}
\delta\chi^{(1)}
=
-\frac{\beta s^2}{4\pi}\Big((2\mu)^{-1/\alpha}-1\Big)
+\frac{s^2}{2\pi(\alpha+1)}
\Big((2\mu)^{-(\alpha+1)/\alpha}-1\Big)
+O\!\left(e^{-\beta\mu}\right),
\qquad 0<\mu<\frac12.
\end{equation}
Similarly, the contribution from two $\Gamma^{(4)}$ vertices (depicted diagrammatically in Fig.~\ref{fig:spin_main}(a)) for $1/2>\mu>0$ and $T\rightarrow 0$ is given by
\begin{equation}
\begin{aligned}
    \delta\chi^{(2)}(Q)&\approx -\frac{s^2\beta^2}{4}
\int_{\bm k}
\frac{\xi(k+Q/s)\,\xi(k)}{\xi(k+Q/s)-\xi(k)}
\Big[
\theta\!\left(\mu-\xi(k+Q/s)/2\right)
-
\theta\!\left(\mu-\xi(k)/2\right)
\Big]\\
&+\frac{s^2\beta}{2}
\int_{\bm k}\Big[1+
\frac{\xi(k)\theta\!\left(\mu-\xi(k+Q/s)/2\right)
-
\xi(k+Q/s)\theta\!\left(\mu-\xi(k)/2\right)}{\xi(k+Q/s)-\xi(k)}
\Big]\;.\label{eq:chi_2_T=0_q}
\end{aligned}
\end{equation}

The first line here is $=\frac{s^2\beta^2}{4} \Theta_{\rm W}(Q)$.
At $Q=0$, the momentum integral $\Theta_{\rm W}(q)$ yields $\Theta_{\rm W}(Q=0)=\Gamma(2\mu)$, where $\Gamma(x)=\int_{k}\xi^2(k) \delta(x-\xi(k))$ was defined in Eq.~\eqref{eq:broadening_prof}. The scaling of these expressions with $\beta$ suggests that $\Gamma^{(6)}$ renormalizes the magnetic moment, while the non-local contribution from two  $\Gamma^{(4)}$ vertices introduces a finite Curie-Weiss temperature scale:
\begin{equation}
    \chi(Q) \approx 
    \frac{1}{T- s^2\Theta_{\rm W}(Q)+...} \times \left(\frac{1}{4}-s^2 \int_k \theta(\xi(k)-2\mu)+...\right)\;,
\end{equation}
where  $1/4$ originates from $\langle S_z^2\rangle_0=1/4$ in a single site problem at $T=0$. 

First, we  observe that the first term in Eq.~\eqref{eq:Gamma_s_s112} is much larger than the rest due to the extra $\beta$ factor. Focusing on this term only, and using Eq.~\eqref{eq:F_sol} for $\bar{\gamma}^{(1)}=-\beta n_0 /(2(i\ep+\mu))$ and $\gamma^{(1)}= 1/(i\ep+\mu)$,
we can resum the ladder corrections and find
\begin{equation}
\begin{aligned}
\Gamma^{(4,s)}_{\ep,\ep'';0}(\bm{Q})&\approx \Gamma^{(4,s)}_{\ep,\ep'';0}-\frac{s^2 n_{0}\beta }{1+\frac{s^2 n_{0}\beta }{2}T\sum_{\ep'}(i\ep'+\mu)^{-2}\Pi_{\ep';0}(\bm{Q})} \times \frac{  T\sum_{\ep'}(i\ep'+\mu)^{-1}\Pi_{\ep';0}(\bm{Q})\Gamma^{(4,s)}_{\ep',\ep'';0}}{2 (i\ep+\mu)}\;\\
&=-\frac{n_0 \beta }{2(i\ep+\mu)(i\ep''+\mu)} \times \frac{1}{1+\frac{s^2 n_{0}\beta }{2}T\sum_{\ep'}(i\ep'+\mu)^{-2}\Pi_{\ep';0}(\bm{Q})}\;.\label{eq:spin_dressed0}
\end{aligned}
\end{equation}
Within this leading low-temperature approximation, the spin
susceptibility is then expressed as
\begin{equation}
    \chi(\bm{Q})\approx -\frac{T}{2}\sum_\ep g_\ep^2 - \frac{T^2}{2}\sum_{\ep,\ep''} \Gamma^{(4,s)}_{\ep,\ep'';0}(\bm{Q})  \approx \frac{ n_0^3}{4(2-n_0)^2} \times \frac{1}{T+\frac{s^2 n_{0} }{2}T\sum_{\ep'}(i\ep'+\mu)^{-2}\Pi_{\ep';0}(\bm{Q})}\;.
\end{equation}

The sum in the denominator can be easily evaluated as
\begin{equation}
\begin{aligned}
T\sum_{\ep'}\frac{\Pi_{\ep';0}(\mathbf Q)}{(i\ep'+\mu)^2}
&=
\frac{1}{1-n_0/2}
\int_{\mathbf k}
\frac{\xi(\mathbf k+\mathbf Q/s)\,\xi(\mathbf k)}
{\xi(\mathbf k+\mathbf Q/s)-\xi(\mathbf k)}
\\
&\quad\times
\left[
n_F\!\left(\left(1-\frac{n_0}{2}\right)\xi(\mathbf k+\mathbf Q/s)-\mu\right)
-
n_F\!\left(\left(1-\frac{n_0}{2}\right)\xi(\mathbf k)-\mu\right)
\right],
\end{aligned}
\end{equation}
which leads to Eq.~\eqref{eq:Theta_main} of the main text.

\section{Local cumulants in the correlated hopping models at half-filling}\label{app:Im_correlated_hopping_model}
The self-energy in the correlated hopping model involves the following four-point local cumulant in the electron-trion basis (the trion operator is defined as $F_\sigma=(2n_{\bar{\sigma}}-1)c_\sigma$)
\begin{equation}
\Gamma_{\ep,\ep',0}^{(4),as'sb}= \int_{\tau,\tau_{1,2}} e^{i\ep\tau-i\ep'(\tau_2-\tau_1)} \sum_{\sigma_{\rm int}}\Gamma_{\tau\tau_2\tau_1 \tau_0=0}^{(4),as'sb}\;,\label{eq:def_Gamma_topology}
\end{equation}
where 
\begin{equation}
\begin{aligned}
\sum_{\sigma_{\rm int}}\Gamma_{\tau\tau_2\tau_1 \tau_0=0}^{(4),as'sb} &\equiv -\sum_{\sigma}\langle [2n_{\downarrow}(\tau)-1]^ac_{\uparrow} (\tau)  [2n_{\bar{\sigma}}(\tau_1)-1]^{s}c_{\sigma}(\tau_1)  [2n_{\bar{\sigma}}(\tau_2)-1]^{s'}\bar{c}_{\sigma}(\tau_2) [2n_{\downarrow}-1]^b\bar{c}_{\uparrow} \rangle_{0} \\
&+2g_\tau^{ab}g_{\tau_1{-}\tau_2}^{s,s'}- g_{\tau{-}\tau_2}^{as'}g^{sb}_{\tau_1}\;.
\end{aligned}
\end{equation}
Here $a,b,s,s'=0,1$, and $g^{ab}$ is given in Eq.~\eqref{eq:g_ep_corr_hop}. Our notation here is such that $[2n_{\downarrow}(\tau)-1]^0\equiv 1$ is an identity operator, etc.

Our goal is to evaluate these local expectation values at half-filling, $\mu=U/2$. Instead of working directly in the electron-trion basis $\{c_\sigma, \;F_\sigma\}\equiv \{c_\sigma, \;(2n_{\bar{\sigma}}-1)c_\sigma\}$, it will be convenient to summarize the structure of this cumulant in the rotated basis $\{c_\sigma, \;n_{\bar{\sigma}}c_\sigma\}$. All matrices in this new basis will be denoted with tildes, e.g. $\tilde{g}^{ab}_\tau  =-\langle n_{\bar{\sigma}}^a(\tau)c_{\sigma} (\tau) n_{\bar{\sigma}}^b\bar{c}_{\sigma} \rangle_0$, and
\begin{equation}
 \sum\limits_{\sigma_{\rm int}}\tilde{\Gamma}_{\ep\ep',\omega=0}^{(4),as'sb}= -\int_{\tau,\tau_{1,2}} e^{i\ep\tau-i\ep'(\tau_2-\tau_1)} \sum\limits_{\sigma}\langle n_{\downarrow}^a(\tau)c_{\uparrow} (\tau)  n_{\bar{\sigma}}^{s}(\tau_1)c_{\sigma}(\tau_1) n_{\bar{\sigma}}^{s'}(\tau_2)\bar{c}_{\sigma}(\tau_2)  n_{\downarrow}^b\bar{c}_{\uparrow} \rangle_{0} + 2\tilde g_\ep^{ab}\tilde g_{\ep'}^{s,s'}- \beta  \delta_{\ep \ep'}\tilde g_{\ep}^{as'}\tilde g^{sb}_{\ep}\,.
\end{equation}
The inverse transformation back to the electron-trion basis follows straightforwardly from the identity $F_\sigma=(2n_{\bar{\sigma}}-1)c_{\sigma} = 2n_{\bar{\sigma}}c_{\sigma} -c_{\sigma}$.

First, we consider the simplest correlator that does not involve additional density operators at all, i.e. $a=b=s=s'=0$
\begin{equation}
 \begin{aligned}
\sum\limits_{\sigma_{\rm int}}\tilde{\Gamma}_{\ep\ep',0}^{(4),0000}&=\frac{\beta U^2\left(1+\frac{1}{2}\tanh \frac{U\beta}{4}\right) }{2\left(\ep^2+\frac{U^2}{4}\right)^2}\delta_{\ep, \ep'}-\frac{\beta U^2 \delta_{\ep,-\ep'}}{2\left(1+e^{\frac{U\beta}{2}}\right)\left(\ep^2+\frac{U^2}{4}\right)^2}\\ &+\frac{\beta U^2\left(\tanh \frac{U\beta}{4}-1\right) }{4\left(\ep'^2+\frac{U^2}{4}\right)\left(\ep^2+\frac{U^2}{4}\right)}-
 \frac{ U \left(3 U^4+4 U^2 \left(\ep'^2+\ep^2\right)-16 \ep'^2 \ep^2\right)}{16\left(\frac{U^2}{4}+ \ep'^2\right)^2 \left(\frac{U^2}{4}+ \ep^2\right)^2 }.
 \end{aligned}
\end{equation}
Next, we evaluate more involved correlators in the sector $a=b=0$, but with additional intermediate density operator insertions $s=s'=1$:
\begin{equation}
\sum\limits_{\sigma_{\rm int}}\tilde{\Gamma}_{\ep\ep',0}^{(4),0110}=\frac{\beta \left(1+\frac{1}{2}\tanh \frac{U\beta}{4}\right) }{2\left(i\ep-\frac{U}{2}\right)^2}\delta_{\ep,\ep'}-\frac{\beta \delta_{\ep,-\ep'}}{2(1{+}e^{\frac{U\beta}{2}})\left(i\ep+\frac{U}{2}\right)^2}-\frac{i \varepsilon' \left( \varepsilon^2{-}U^2/4\right)+U^3/4}{ \left(i \varepsilon'-\frac{U}{2}\right)^2 \left(\varepsilon^2+\frac{U^2}{4}\right)^2}+\frac{\beta  U  }{2 (1{+}e^{\frac{\beta  U}{2}})\left(i \varepsilon'-\frac{U}{2}\right) \left(\varepsilon^2+\frac{U^2}{4}\right)}
\end{equation}
and $\sum\limits_{\sigma_{\rm int}}\tilde{\Gamma}_{\ep\ep',0}^{(4),0100}=\sum\limits_{\sigma_{\rm int}}\tilde{\Gamma}_{\ep\ep',0}^{(4),0010}$, with
\begin{equation}
 \begin{aligned}
\sum\limits_{\sigma_{\rm int}}\tilde{\Gamma}_{\ep\ep',0}^{(4),0010}&=-\frac{\beta  U \left(1+\frac{1}{2} \tanh \frac{\beta  U}{4}\right) \delta _{\varepsilon,\varepsilon'}}{2 \left(i \varepsilon -\frac{U}{2}\right) \left(\ep^2 +\frac{U^2}{4}\right)}+\frac{\beta  U \delta _{\varepsilon,-\varepsilon'}}{2 \left(1+e^{\frac{\beta  U}{2}}\right) \left(i \varepsilon -\frac{U}{2}\right) \left(\frac{U}{2}+i \varepsilon \right)^2}\\
&+\frac{\beta  U }{2 (1+e^{\frac{\beta  U}{2}})\left(i \varepsilon'-\frac{U}{2}\right) \left(\varepsilon ^2+\frac{U^2}{4}\right)}+\frac{U^2 \left(\varepsilon '^2+\varepsilon ^2\right)-4 \varepsilon'^2 \varepsilon^2+3 U^4/4}{4 \left(i \varepsilon'-\frac{U}{2}\right) ( \varepsilon'^2+\frac{U^2}{4})\left(\varepsilon ^2+\frac{U^2}{4}\right)^2}.
 \end{aligned}
\end{equation}
The remaining sectors could be evaluated in a similar way. For instance, we consider $a=1$, $b=0$ and find
\begin{equation}
\sum\limits_{\sigma_{\rm int}}\tilde{\Gamma}_{\ep\ep',0}^{(4),1110}=\frac{\beta   \left(1+\frac{1}{2} \tanh \frac{\beta  U}{4}\right) \delta _{\varepsilon,\varepsilon'}}{2 \left(i \varepsilon -\frac{U}{2}\right)^2}- \frac{\beta  }{2(1+e^{\frac{\beta  U}{2}}) \left(i \varepsilon '-\frac{U}{2}\right) \left(i \varepsilon -\frac{U}{2}\right)}+ \frac{i \left(\varepsilon '+\varepsilon \right)-U}{2 \left(i \varepsilon '-\frac{U}{2}\right)^2 \left(i \varepsilon -\frac{U}{2}\right)^2}
\end{equation}
and
\begin{equation}
 \begin{aligned}
\sum\limits_{\sigma_{\rm int}}\tilde{\Gamma}_{\ep\ep',0}^{(4),1100}&=\frac{\beta  U \left(1+\frac{1}{2} \tanh \frac{\beta  U}{4}\right) \delta _{\ep,\ep'}}{2 \left(i \varepsilon-\frac{U}{2}\right)^2 \left(i \varepsilon+\frac{U}{2}\right)}-\frac{\beta  \delta _{\varepsilon ,-\varepsilon'}}{2 (1{+}e^{\frac{\beta  U}{2}}) \left(\varepsilon ^2+\frac{U^2}{4}\right)} -\frac{\beta  }{2(1{+}e^{\frac{\beta  U}{2}})  \left(i \varepsilon '{-}\frac{U}{2}\right) \left(i \varepsilon {-}\frac{U}{2}\right)}\\
&+\frac{i U^2 \left(\varepsilon '+\varepsilon \right)-2 U \left(\varepsilon -\varepsilon '\right)^2-4 i \varepsilon ' \varepsilon  \left(\varepsilon '+\varepsilon \right)-2 U^3}{8 \left(i \varepsilon '-\frac{U}{2}\right)^2 \left(i \varepsilon -\frac{U}{2}\right)^2  \left(\frac{U}{2}+i \varepsilon '\right) \left(\frac{U}{2}+i \varepsilon \right)}. 
 \end{aligned}
\end{equation}
We also find 
\begin{equation}
\sum\limits_{\sigma_{\rm int}}\tilde{\Gamma}_{\ep\ep',0}^{(4),1010}=\frac{\beta  \left(\frac{1}{2} \tanh \frac{\beta  U}{4}+1\right) \delta _{\varepsilon ,\varepsilon '}}{2 \left(\frac{U}{2}-i \varepsilon \right)^2}+\frac{i \left(\varepsilon '+\varepsilon \right)-U}{2 \left(\frac{U}{2}-i \varepsilon '\right)^2 \left(\frac{U}{2}-i \varepsilon \right)^2}-\frac{\beta  }{2 (1+e^{\frac{\beta  U}{2}})\left(i \varepsilon '-\frac{U}{2}\right) \left(i \varepsilon -\frac{U}{2}\right)}.
\end{equation} 
Finally,
\begin{equation}
 \begin{aligned}
\sum\limits_{\sigma_{\rm int}}\tilde{\Gamma}_{\ep\ep',0}^{(4),1000}&=\frac{\beta  U \left(1+\frac{1}{2} \tanh \frac{\beta  U}{4}\right) \delta _{\varepsilon ,\varepsilon '}}{2 \left(i \varepsilon -\frac{U}{2}\right)^2 \left(\frac{U}{2}+i \varepsilon \right)}-\frac{\beta  U \delta _{\varepsilon ,-\varepsilon '}}{2 \left(1+e^{\frac{\beta  U}{2}}\right) \left(i \varepsilon -\frac{U}{2}\right)^2 \left(i \varepsilon +\frac{U}{2}\right)}\\
&+\frac{ \beta  U }{2 \left(1+e^{\frac{\beta  U}{2}}\right)\left(\varepsilon '^2+\frac{U^2}{4}\right) \left(i \varepsilon -\frac{U}{2}\right)}-\frac{\varepsilon '^2 \varepsilon ^2- U^2 \left(\varepsilon '^2+\varepsilon ^2\right)/4-3 U^4/16}{ \left( \varepsilon^2 +\frac{U^2}{4}\right)\left(\varepsilon '^2+\frac{U^2}{4}\right)^2 \left(i \varepsilon -\frac{U}{2}\right)}
 \end{aligned}
\end{equation}
and
\begin{equation}
 \begin{aligned}
&\sum\limits_{\sigma_{\rm int}}\tilde{\Gamma}_{\ep\ep',0}^{(4),1001}=\frac{\beta  \left(1{+}\frac{1}{2} \tanh \frac{\beta  U}{4}\right) \delta _{\varepsilon ,\varepsilon '}}{2 \left(i \varepsilon -\frac{U}{2}\right)^2}-\frac{\beta  \delta _{\varepsilon ,-\varepsilon '}}{2 (1{+}e^{\frac{\beta  U}{2}}) \left(i \varepsilon -\frac{U}{2}\right)^2}+\frac{i\varepsilon(U^2 {-}4  \varepsilon '^2)  -U^3}{4 \left(\varepsilon '^2+\frac{U^2}{4}\right)^2 \left(i \varepsilon -\frac{U}{2}\right)^2}+\frac{\beta  U }{2 (1{+}e^{\frac{\beta  U}{2}})\left(\varepsilon '^2{+}\frac{U^2}{4}\right) \left(i \varepsilon {-}\frac{U}{2}\right)}.
 \end{aligned}
\end{equation}
For the remaining matrix elements, we find the following symmetry relations
\begin{gather}
(0111)=(1110),\qquad (0001)=(1000),\qquad (0011)=(1100),\notag\\
(0101)=(1010),\qquad (1111)=(1110),\qquad (1011)=(1101)=(1010),
\end{gather}
where our notation here is a compact form of the identities such as  $\sum\limits_{\sigma_{\rm int}}\tilde{\Gamma}_{\ep\ep',0}^{(4),0111}=\sum\limits_{\sigma_{\rm int}}\tilde{\Gamma}_{\ep\ep',0}^{(4),1110}$, etc.

The calculation of the self-energy in the topological model also involves triangular vertices defined as
\begin{equation}
\begin{aligned}
\Gamma^{(3),a b}_{\varepsilon\varepsilon'}
=
\int_{\tau,\tau'}
\Big\{&e^{i\varepsilon \tau+i\varepsilon' \tau'}
\left\langle
[2n_{\downarrow}(\tau)-1]^a c_\uparrow(\tau)\,
[2n_{\uparrow}(\tau')-1]^b c_\downarrow(\tau') \bar{\Delta}
\right\rangle_{0,c}\\
&-e^{i\varepsilon \tau-i\varepsilon' \tau'}\sum_\sigma
\left\langle
[2n_{\downarrow}(\tau)-1]^a c_\uparrow(\tau)\,
[2n_{\bar{\sigma}}(\tau')-1]^b \bar c_\sigma(\tau')
\bar{c}_\uparrow c_\sigma\right\rangle_{0,c}\Big\},\label{eq:Gamma3_def}
\end{aligned}
\end{equation}
where $\Delta=c_{\downarrow}c_{\uparrow}$. These cumulants can be evaluated in the same way. Working directly in the electron-trion basis and at $T\ll U$, we find
\begin{equation}
\begin{aligned}
\Gamma^{(3),a b}_{\varepsilon\varepsilon'}
&=
-\frac{3}{2}\,
\beta\,\delta_{\varepsilon,\varepsilon'}\,
g_\varepsilon \tau^x -g_\varepsilon
\begin{pmatrix}
3 & 0\\
0 & 1
\end{pmatrix}
g_{\ep'} 
\\
&=
\frac{3}{2}\,\beta\,\delta_{\varepsilon,\varepsilon'}\,
\frac{1}{\varepsilon^2+U^2/4}
\begin{pmatrix}
U/2 & i\varepsilon\\
i\varepsilon & U/2
\end{pmatrix} -\frac{1}{
\left[\varepsilon^2+U^2/4\right]
\left[\varepsilon'^2+U^2/4\right]
}
\begin{pmatrix}
U^2/4-3\varepsilon\varepsilon'
&
\frac{U}{2}(3i\varepsilon+i\varepsilon')
\\[0.4em]
\frac{U}{2}(i\varepsilon+3i\varepsilon')
&
3U^2/4-\varepsilon\varepsilon'
\end{pmatrix}.
\end{aligned}\label{eq:Gamma_3_result}
\end{equation}

\section{Derivation of the self-energy for the model with concentrated topology}
\label{sec:topological_model_appendix}

In this Appendix, we give the details needed to derive the
$O(s^2)$ self-energy in Eq.~\eqref{eq:self-energy_topological_model}.
Our starting point is Eqs.~\eqref{eq:H_model_main},
\eqref{eq:n_def_corr}, and \eqref{eq:a_beta_def}.  The terms
$\mathcal H_{\rm loc}$, $\mathcal H_{\rm 2-site}$, and
$\mathcal H_{\rm 3-site}$ are given in
Eqs.~\eqref{eq:H_2_3_site}, \eqref{eq:U_def_corr}, and
\eqref{eq:def_x_M}.  After normal ordering, the full Hamiltonian
can be written as
\begin{equation}
    \mathcal H
    =
    \mathcal H_{\rm loc}
    +
    \mathcal H_{\rm 2-site}
    +
    \mathcal H_{\rm 3-site}
    +
    \mathcal H_{\rm 4-site}
    +
    \mathcal H_{\rm 2-site\;pair}.
\end{equation}
The four-site term is
\begin{equation}
    \mathcal H_{\rm 4-site}
    =
    -\frac{s^6}{2}
    \sum_{\substack{i,j,k,l\\ {\rm all\ distinct}}}
    \sum_{\sigma,\sigma'}
    Z\!\left(
    s(\bm r_i-\bm r_l),
    s(\bm r_j-\bm r_l),
    s(\bm r_k-\bm r_l)
    \right)
    c^\dagger_{i\sigma}
    c^\dagger_{k\sigma'}
    c_{j\sigma}
    c_{l\sigma'} ,
\label{eq:H_full_topology}
\end{equation}
where
\begin{equation}
s^6
Z\!\left(
s(\bm r_i-\bm r_l),
s(\bm r_j-\bm r_l),
s(\bm r_k-\bm r_l)
\right)
 =
\sum_m
a(s|\bm r_i-\bm r_m|)
a(s|\bm r_j-\bm r_m|)
a(s|\bm r_k-\bm r_m|)
a(s|\bm r_l-\bm r_m|).
\end{equation}
Unlike the kernels $X$ and $M$, the leading part of $Z$ is not
obtained solely from the terms in which $m$ coincides with one of
the external sites. The fully nonlocal part with
$m\neq i,j,k,l$ contributes at the same order: it contains four
nonlocal $\sim s^2$ tails, but is enhanced by the sum over
$O(s^{-2})$ sites. We will not need the explicit leading form of
$Z$ below, since all $Z$-containing one-particle self-energy
diagrams either vanish or start beyond $O(s^2)$.

The remaining two-site term contains pair hopping, exchange, and
density-density interactions,
\begin{equation}
\mathcal H_{\rm 2-site\;pair}
=
\frac{s^4}{2}
\sum_{i\neq j,\sigma}
\mathcal P(s|\bm r_i-\bm r_j|)
\Bigg[
\left(n_{i\sigma}-\frac12\right)
\left(n_{j\bar\sigma}-\frac12\right)
-
c^\dagger_{i\sigma}c_{i\bar\sigma}
c^\dagger_{j\bar\sigma}c_{j\sigma}
-
c^\dagger_{i\sigma}c^\dagger_{i\bar\sigma}
c_{j\sigma}c_{j\bar\sigma}
\Bigg],
\end{equation}
where
\begin{equation}
    s^4\mathcal P(s|\bm r_i-\bm r_j|)
    =
    \sum_m
    a^2(s|\bm r_i-\bm r_m|)
    a^2(s|\bm r_j-\bm r_m|).
\label{eq:P_appendix}
\end{equation}
The Hartree part of $\mathcal H_{\rm 2-site\;pair}$ is already
absorbed into the convention used for $\mathcal H_{\rm loc}$.  For
$i\neq j$, the leading part of Eq.~\eqref{eq:P_appendix} comes from
$m=i$ and $m=j$, giving
\begin{equation}
    \mathcal P(s|\bm r_i-\bm r_j|)
    =
    2a_0^2\beta^2(s|\bm r_i-\bm r_j|)
    +O(s^2)
    =
    2\beta^2(s|\bm r_i-\bm r_j|)
    +O(s^2),
\end{equation}
where in the last equality we used $a_0=1+O(s^2)$.  After the
Hartree subtraction, a single $\mathcal P$ insertion has no
nonvanishing local one-particle contraction at half filling, while
diagrams involving additional nonlocal propagation start beyond the
$O(s^2)$ accuracy considered here.

We work in the electron--trion basis
\begin{equation}
    \gamma_{i\sigma}
    =
    \begin{pmatrix}
    c_{i\sigma}\\
    F_{i\sigma}
    \end{pmatrix},
    \qquad
    F_{i\sigma}=(2n_{i\bar\sigma}-1)c_{i\sigma}.
\end{equation}
With the normalization used in the main text, the leading two-site
hopping vertex is proportional to
$\frac{\bar U}{2}[\lambda(k)-1]\tau^x$, while the local Hubbard term
supplies the remaining constant piece.  In the scaling limit this
combines into the zeroth-order inverse Green's function Eq.~\eqref{eq:G_0th_order_main} given in
the main text, with off-diagonal matrix element
$-\bar U\lambda(k/s)/2$.  We now evaluate the nonvanishing $O(s^2)$ self-energy corrections.

\subsection{Self-energy from the $X^2$ processes}
First, let us consider a familiar self-energy contribution originating from two insertions of $\mathcal H_{\rm 2-site}$ with an intermediate lattice detour. It can be
obtained from the general formula in Eq.~\eqref{eq:Sigma_general_k_k} using the local four-point cumulant in the electron--trion
basis. The corresponding expression is
\begin{equation}
\Sigma_{X^2,\ep}^{ab}
=
\frac{T}{4}\sum_{\ep'}
\sum_{\{a_i,b_i\}}
(g_\ep^{-1})_{aa_1}
\Gamma_{\ep,\ep',0}^{(4),a_1a_2b_2b_1}
(g_\ep^{-1})_{b_1b}
\sum_{k\in \rm{BZ}}
X(k,s)^2
\left[
\tau^x
\mathcal G^{(0)}_{\ep'}(k)
\tau^x
\right]^{a_2b_2},
\end{equation}
where $\Gamma_{\ep,\ep',0}^{(4),as'sb}$ is defined in
Eq.~\eqref{eq:def_Gamma_topology}.  Passing to the continuum scaling
limit as in Eq.~\eqref{eq:X_smooth} and using the explicit local cumulants calculated in Sec.~\ref{app:Im_correlated_hopping_model} gives
\begin{equation}
\Sigma_{X^2,\ep}
=
\frac{3s^2\bar U^2}{4}
\begin{pmatrix}
\mathcal I^+_\ep & \mathcal I^-_\ep\\
\mathcal I^-_\ep & \mathcal I^+_\ep
\end{pmatrix}.
\label{eq:Sigma_X2_topology}
\end{equation}
Here $\mathcal I^{\pm}_\ep $ is defined in Eq.~\eqref{eq:I_eps_222}. We emphasize that this self-energy is momentum-independent, and all its matrix components are finite.

\subsection{Self-energy from the $M^2$ processes}

We next consider the self-energy generated by two insertions of the
three-site vertex $\mathcal H_{\rm 3-site}$. The relevant process is
shown schematically in Fig.~\ref{fig:topological_model}(a). An
electron at site $i$ is converted by the first $M$ vertex into a
local particle-hole or pair excitation at an intermediate site, while
the remaining fermion propagates through a nonlocal hopping leg. A
second $M$ vertex then recombines this excitation and produces an
electron at site $f$. Since both external legs attach to electron
components of the three-site vertex, this contribution is
proportional to $\delta_{a0}\delta_{b0}$ in the electron--trion
basis.

It is useful to keep the structure of the three-site kernel explicit:
the first argument specifies the displacement from the central site
of the vertex to one external leg, and the second argument specifies
the displacement to the other leg. Around the atomic limit, the
nonlocal part of the self-energy is
\begin{equation}
\begin{aligned}
    \Sigma_{M^2,\ep}^{ab}(\bm r_i-\bm r_f\neq 0)
    &=
    4\delta_{a0}\delta_{b0}\,
    T\sum_{\omega}
    \sum_{l,m}^{\prime}
    M\!\left(s(\bm r_l-\bm r_i),s(\bm r_l-\bm r_m)\right)
    M\!\left(s(\bm r_l-\bm r_m),s(\bm r_l-\bm r_f)\right)
\\
&\qquad\qquad \times
    \left[
    g_{\ep-\omega}^{00}\,
    \mathcal C_{Q,\omega}
    +
    g_{\omega-\ep}^{00}\,
    \mathcal C_{P,\omega}
    \right].
\end{aligned}
\label{eq:Sigma_M2_atomic}
\end{equation}
Here the prime indicates that all site indices entering each
three-site vertex are pairwise distinct. The local cumulants are
\begin{equation}
    \mathcal C_{Q,\omega}
    \equiv
    \sum_{\sigma'}
    \int_0^\beta d\tau\,
    e^{i\omega\tau}
    \left\langle
    \bar c_{\sigma'}(\tau)c_{\uparrow}(\tau)
    \bar c_{\uparrow}c_{\sigma'}
    \right\rangle_{0,c},
    \qquad
    \mathcal C_{P,\omega}
    \equiv
    \int_0^\beta d\tau\,
    e^{i\omega\tau}
    \left\langle
    \Delta(\tau)\bar\Delta(0)
    \right\rangle_{0},
    \quad
    \Delta=c_{\downarrow}c_{\uparrow}.
\end{equation}
Using the leading form $M\!\left(s(\bm r_l-\bm r_i),s(\bm r_l-\bm r_m)\right)
\simeq
\frac{\bar U}{2}\,
\beta(s|\bm r_l-\bm r_i|)
\beta(s|\bm r_l-\bm r_m|)$,
each three-site vertex supplies two nonlocal tails. One of the two
internal site sums becomes the external convolution producing
$[\lambda(k/s)-1]^2$, while the other gives the internal momentum
integral over $[\lambda(p)-1]^2$. Therefore,
\begin{equation}
\begin{aligned}
&\sum_R e^{i\bm k\bm R}
\sum_{l,m}^{\prime}
M\!\left(s(\bm r_l-\bm r_i),s(\bm r_l-\bm r_m)\right)
M\!\left(s(\bm r_l-\bm r_m),s(\bm r_l-\bm r_f)\right)
\\
&\qquad =
\frac{s^2\bar U^2}{4}
\left[\lambda(k/s)-1\right]^2
\int_{\bm p}\left[\lambda(p)-1\right]^2
+O(s^4),
\qquad
\bm R=\bm r_i-\bm r_f .
\end{aligned}
\end{equation}
Replacing the internal local propagator by the resummed
$\mathcal G^{(0)}$ then gives
\begin{equation}
    \Sigma_{M^2,\ep}^{ab}(k)
    =
    s^2\bar U^2
    \delta_{a0}\delta_{b0}
    \left[\lambda(k/s)-1\right]^2
    T\sum_{\omega}
    \int_{\bm p}\left[\lambda(p)-1\right]^2
    \left[
    \mathcal G_{\ep-\omega}^{(0),00}(sp)\,
    \mathcal C_{Q,\omega}
    +
    \mathcal G_{\omega-\ep}^{(0),00}(sp)\,
    \mathcal C_{P,\omega}
    \right]
    +O(s^4).
\label{eq:Sigma_M2_continuum}
\end{equation}
At $T\ll \bar U$, the local cumulants reduce to
$\mathcal C_{Q,\omega}=3\beta\delta_{\omega,0}/4$ and
$\mathcal C_{P,\omega}=0$. Thus
\begin{equation}
\Sigma_{M^2,\ep}(k)
=
\frac{3s^2\bar U^2}{4}
\left[\lambda(k/s)-1\right]^2
\begin{pmatrix}
\mathcal I^+_\ep & 0\\
0 & 0
\end{pmatrix}.
\label{eq:Sigma_MM}
\end{equation}

\subsection{Self-energy from the mixed $XM$ processes}

We now consider the mixed process involving one two-site vertex
$\mathcal H_{\rm 2-site}$ and one three-site vertex
$\mathcal H_{\rm 3-site}$. This process is shown diagrammatically in
Fig.~\ref{fig:topological_model}(b). One end of the diagram is
attached to the electron leg of the $M$ vertex, while the remaining
nonlocal leg is connected to the $X$ vertex. The local part of the
diagram is therefore a triangular cumulant, defined in
Eq.~\eqref{eq:Gamma3_def}.

Keeping the structure of the three-site kernel explicit, the
real-space perturbative correction to the Green's function near the atomic limit is
\begin{equation}
\begin{aligned}
\delta \mathcal G^{ab}_{\tau}&(\bm r_i-\bm r_f\neq 0)
=
\sum_u\sum_{s,s'=0,1}\int_{\tau_{1,2}}
M\!\left(s(\bm r_f-\bm r_i),s(\bm r_f-\bm r_u)\right)
X(s|\bm r_f-\bm r_u|)
\langle
\gamma_\uparrow^a(\tau)
\bar\Delta(\tau_2)
\gamma_\downarrow^{s'}(\tau_1)
\rangle_c
(\tau^x)^{s's}
g_{\tau_2-\tau_1}^{s0}
g_{\tau_2}^{0b}
\\
&\quad
+
\sum_u\sum_{s,s'=0,1}\int_{\tau_{1,2}}
M\!\left(s(\bm r_i-\bm r_f),s(\bm r_i-\bm r_u)\right)
X(s|\bm r_i-\bm r_u|)
g_{\tau-\tau_2}^{a0}
g_{\tau_1-\tau_2}^{0s'}
(\tau^x)^{s's}
\left\langle
\bar\gamma_\downarrow^{s}(\tau_1)
\Delta(\tau_2)
\bar\gamma_\uparrow^b
\right\rangle_c
\\
&\quad
-
\sum_u\sum_{s,s'=0,1}\int_{\tau_{1,2}}
M\!\left(s(\bm r_f-\bm r_i),s(\bm r_f-\bm r_u)\right)
X(s|\bm r_f-\bm r_u|)
\sum_\sigma
\left\langle
\gamma_\uparrow^a(\tau)
\bar c_\uparrow(\tau_2)c_\sigma(\tau_2)
\bar\gamma_\sigma^{s}(\tau_1)
\right\rangle_c
(\tau^x)^{ss'}
g_{\tau_1-\tau_2}^{s'0}
g_{\tau_2}^{0b}
\\
&\quad
-
\sum_u\sum_{s,s'=0,1}\int_{\tau_{1,2}}
M\!\left(s(\bm r_i-\bm r_f),s(\bm r_i-\bm r_u)\right)
X(s|\bm r_i-\bm r_u|)
\sum_\sigma
g_{\tau-\tau_2}^{a0}
g_{\tau_2-\tau_1}^{0s}
(\tau^x)^{ss'}
\langle
\gamma_\sigma^{s'}(\tau_1)
\bar c_\sigma(\tau_2)c_\uparrow(\tau_2)
\bar\gamma_\uparrow^b
\rangle_c .
\end{aligned}
\end{equation}
After Fourier transforming, the four local correlators combine into
the triangular vertex $\Gamma^{(3)}$, and we obtain
\begin{equation}
\begin{aligned}
\delta \mathcal G_{\ep}^{ab}(\bm r_i-\bm r_f\neq 0)
&=
\sum_u
M\!\left(s(\bm r_f-\bm r_i),s(\bm r_f-\bm r_u)\right)
X(s|\bm r_f-\bm r_u|)
\,T\sum_{\ep'}
\left[
\Gamma^{(3)}_{\ep\ep'}\tau^x g_{\ep'}
\right]^{a0}
g_\ep^{0b}
\\
&\quad
+
\sum_u
M\!\left(s(\bm r_i-\bm r_f),s(\bm r_i-\bm r_u)\right)
X(s|\bm r_i-\bm r_u|)
\,T\sum_{\ep'}
g_\ep^{a0}
\left[
g_{\ep'}\tau^x\Gamma^{(3)}_{\ep'\ep}
\right]^{0b}.
\end{aligned}
\label{eq:deltaG_XM_freq}
\end{equation}
Using the leading forms
\begin{equation}
M\!\left(s(\bm r_f-\bm r_i),s(\bm r_f-\bm r_u)\right)
\simeq
\frac{\bar U}{2}
\beta(s|\bm r_f-\bm r_i|)
\beta(s|\bm r_f-\bm r_u|),
\qquad
X(s|\bm r_f-\bm r_u|)
\simeq
\bar U\beta(s|\bm r_f-\bm r_u|),
\end{equation}
and similarly for the second term, we find that the corresponding 
self-energy is proportional to $\lambda(k/s)-1$.  The remaining
internal sum produces the same momentum integral that appears in the
$X^2$ and $M^2$ contributions. Within the same $\mathcal{O}(s^2)$ accuracy we can then replace the internal local propagator by the resummed
$\mathcal G^{(0)}_\ep$. This gives
\begin{equation}
\Sigma_{XM,\ep}^{ab}(k)
=
-\frac{s^2\bar U^2}{2}
\left[\lambda(k/s)-1\right]
\int_{\bm p}\left[\lambda(p)-1\right]^2
T\sum_{\ep'}\Big\{
\left[
g_\ep^{-1}\Gamma^{(3)}_{\ep\ep'}\tau^x
\mathcal G^{(0)}_{\ep'}(sp)
\right]^{a0}\delta_{b0}+\delta_{a0}
\left[
\mathcal G^{(0)}_{\ep'}(sp)\tau^x
\Gamma^{(3)}_{\ep'\ep}g_\ep^{-1}
\right]^{0b}\Big\}.
\end{equation}

At $T\ll \bar U$, we use Eq.~\eqref{eq:Gamma_3_result}. The part of
$\Gamma^{(3)}_{\ep\ep'}$ that is not diagonal in frequency gives no
contribution after the internal Matsubara summation. This follows
from the half-filling identity
\begin{equation}
    T\sum_{\ep}
    \begin{pmatrix}
        3 & 0\\
        0 & 1
    \end{pmatrix}
    g_\ep(\tau^x g_\ep)^n
    =0,
    \qquad n\geq 1.
\end{equation}
Therefore only the frequency-diagonal part of the triangular vertex
contributes, and we finally obtain
\begin{equation}
\Sigma_{XM,\ep}(k)
=
\frac{3s^2\bar U^2}{4}
\left[\lambda(k/s)-1\right]
\begin{pmatrix}
2\mathcal I^+_\ep & \mathcal I^-_\ep\\
\mathcal I^-_\ep & 0
\end{pmatrix}.
\label{eq:MXSigma}
\end{equation}
The vanishing trion-trion component is not accidental: in every
nonlocal $XM$ diagram, at least one external leg attaches to the
electron leg of the $M$ vertex, so a fully trion-trion self-energy
cannot be produced at this order.

There is also a possible local $XM$ contraction,
\begin{equation}
\delta \mathcal G^{ab}_{XM,\ep}
=
-\sum_{l,m}
M\!\left(s(\bm r_i-\bm r_l),s(\bm r_i-\bm r_m)\right)
X(s|\bm r_m-\bm r_l|)
T\sum_{\ep'}
\left[
g_{\ep'}\tau^x g_{\ep'}
\right]^{00}
\int_{\tau,\tau'}
e^{i\ep\tau}
\langle
\gamma^a_{\uparrow}(\tau)
n(\tau')
\bar\gamma^b_{\uparrow}(0)
\rangle_{0,c}.
\end{equation}
However, it vanishes because the local vertex is independent of the internal
frequency, while at half filling
\begin{equation}
    T\sum_{\ep'}
    \left[
    g_{\ep'}(\tau^x g_{\ep'})^n
    \right]^{00}
    =
    0,
    \qquad n\geq 1.
\label{eq:ggg_zero}
\end{equation}

\subsection{Absence of the remaining  $O(s^2)$ channels}
There is no
$O(s^2)$ contribution to either the local or nonlocal one-particle
self-energy from $MZ$, $Z^2$, $\mathcal P^2$, or from diagrams in
which $\mathcal P$ is mixed with any other amplitude.  After the
Hartree subtraction described above, all such diagrams either vanish
by the same half-filling identities or start at higher order in the
small-$s^2$ expansion.

There is a possible mixed $XZ$ contribution that is proportional to
\begin{equation}
    \Sigma^{ab}_{XZ,\ep}(\bm r_i-\bm r_j\neq 0)
    \propto
    \delta_{a0}\delta_{b0}
    \sum_{m,l}'
    Z\!\left(
    s(\bm r_i-\bm r_l),
    s(\bm r_j-\bm r_l),
    s(\bm r_m-\bm r_l)
    \right)
    X(s|\bm r_m-\bm r_l|)
    T\sum_{\ep'}
    \left[
    g_{\ep'}\tau^x g_{\ep'}
    \right]^{00},
\end{equation}
where the prime again indicates that the site indices entering the four-site vertex are all distinct. However, this correction vanishes by Eq.~\eqref{eq:ggg_zero} at half-filling. Therefore, after combining Eqs.~\eqref{eq:Sigma_X2_topology}, \eqref{eq:Sigma_MM}, and
\eqref{eq:MXSigma}, we finally obtain
\begin{equation}
\Sigma_\ep(k)
=
\frac{3s^2\bar U^2}{4}
\begin{pmatrix}
\lambda^2(k/s)\mathcal I^+_\ep
&
\lambda(k/s)\mathcal I^-_\ep
\\[0.3em]
\lambda(k/s)\mathcal I^-_\ep
&
\mathcal I^+_\ep
\end{pmatrix},
\label{eq:self_energy_topology_appendix}
\end{equation}
which is the self-energy quoted in the main text.

\end{widetext}


%

\end{document}